\documentclass[twocolumn,pra,nolongbibliography,floatfix,aps,superscriptaddress,superscriptaddress]{revtex4-2} %\documentclass[preprint,pra,longbibliography]{revtex4-2} %nofootinbib
\usepackage{amsthm,amsmath,amssymb,stmaryrd,amsfonts,graphicx,verbatim,xcolor,bm} 
\usepackage{ulem,hyperref,enumerate} 
\usepackage[utf8]{inputenc} 
\usepackage{siunitx,physics} 
\usepackage[english]{babel}

\usepackage{dsfont}

\usepackage{csquotes}
\MakeOuterQuote{"}

\allowdisplaybreaks

\begin{document}

\newcommand{\Kelvin}{\text{K}}
\newcommand{\bR}{\mathbf{R}}
\newcommand{\bM}{\mathbf{M}}
\newcommand{\bV}{\mathbf{V}}
\newcommand{\bQ}{\bm {Q}}
\newcommand{\bG}{\bm {G}}
\newcommand{\bK}{\bm {K}}
\newcommand{\bq}{\bm {q}}
\newcommand{\bk}{\bm {k}}
\newcommand{\bp}{\bm {p}}
\newcommand{\bL}{\mathbf{L}}
\newcommand{\bx}{\bm {x}}
\newcommand{\by}{\bm {y}}
\newcommand{\bz}{\mathbf{z}}
\newcommand{\br}{\bm {r}}
\newcommand{\eqn}[1]{(\ref{#1})}
\newcommand{\mr}{moir\'e~}
\newcommand{\Mr}{Moir\'e~}
\newcommand{\mT}{\mathcal{T}}
\newcommand{\mM}{\mathcal{M}}

\newcommand{\CommentDG}[1]{\textbf{\color{orange} [DG: #1]}}
\newcommand{\dg}[1]{{\color{orange}{#1}}}

\newcommand{\ajm}[1]{{\color{red}{#1}}}

\newcommand{\dk}[1]{{\color{teal}{#1}}}

\title{Topological superconductivity from repulsive interactions in twisted WSe$_2$}

\author{Daniele Guerci}
\email{dguerci@flatironinstitute.org}
\affiliation{Center for Computational Quantum Physics, Flatiron Institute, New York, New York 10010, USA}

\author{Daniel Kaplan}
\email{d.kaplan1@rutgers.edu}
\affiliation{Department of Physics and Astronomy, Center for Materials Theory, Rutgers University, Piscataway, New Jersey 08854, USA}

\author{Julian Ingham}
%\email{jingham@bu.edu}
\affiliation{Department of Physics, Columbia University, 538 Wst 120th Street, New York, NY 10027, USA}
\affiliation{Center for Computational Quantum Physics, Flatiron Institute, New York, New York 10010, USA}

\author{J. H. Pixley}
\affiliation{Department of Physics and Astronomy, Center for Materials Theory, Rutgers University, Piscataway, New Jersey 08854, USA}
\affiliation{Center for Computational Quantum Physics, Flatiron Institute, New York, New York 10010, USA}

\author{Andrew Millis}
\affiliation{Department of Physics, Columbia University, 538 Wst 120th Street, New York, NY 10027, USA}
\affiliation{Center for Computational Quantum Physics, Flatiron Institute, New York, New York 10010, USA}

\begin{abstract}

The recent observation of superconductivity in twisted bilayer WSe$_2$ raises intriguing questions concerning the origin and the properties of superconducting states realized in bands with non-trivial topological properties and repulsive electron-electron interactions. Using a continuum band structure model, we analyze a mechanism for Coulomb interaction-driven superconductivity in twisted bilayers of WSe$_2$. 
We discuss the symmetries and the phenomenological properties of the resulting superconducting phases and their evolution with interlayer potential difference, tunable via an out of plane electric field. The pairing strength is a non-monotonic function of interlayer potential, being larger at intermediate values due to mixing of singlet and triplet pairing. 
In contrast, at larger interlayer potential, the pairing tendency is suppressed due to enhanced Coulomb repulsion. The superconducting state is chiral in a large regime of parameters and undergoes a transition to a nodal nematic superconductor at a critical potential difference. The chiral state, characterized by an intervalley-symmetric superposition of triplet and singlet pairs, is classified as a topological superconductor within the Altland-Zirnbauer class C. At zero interlayer potential difference, the superconducting state is instead of class D, which hosts Majorana zero modes, making it a promising candidate for applications in quantum computation.

\end{abstract}

\maketitle

{\it Introduction.---} Superconductivity has been a central object of study in physics since its discovery in 1911, due to its fascinating properties of vanishing electrical resistance and macroscopic phase coherence -- which are both scientifically interesting, and provide potential utility in technological applications.
Theoretical understanding of why electronic systems superconduct has always lagged behind experimental breakthroughs; while there is a good understanding of conventional i.e. phonon mediated superconductivity~\cite{Migdal_1958,McMillan_1965,chubukov2020eliashberg}, the electronic pairing mechanisms~\cite{kohnluttingerSC_prl,chubukov_1993,maiti2013superconductivity,Raghu2010} that are generally agreed to be central to the cuprates, heavy fermions, iron pnictides, and organic superconductors, continue to defy a theoretical consensus. 

The newly discovered family of moir\'e materials -- combinations of van-der-Waals bonded atomic layers displaying tunable unit cells, correlation strengths and topologies -- has opened exciting directions in understanding interaction-driven electronic phenomena~\cite{Andrei_2020,Andrei_2021,Kennes_2021}. Superconductivity has now been observed in many members of the graphene family of moir\'e materials~\cite{Cao_2018,Yankowitz_2019,Lu_2019,arora2020superconductivity,saito2020independent,Park_2021,Hao_2021,Zhou_2021,park2022robust,burg2022emergence,su2023superconductivity}. 
% \sout{and untwisted graphene multilayers ~\cite{Zhou_2021} have been discovered to superconduct.} 
The origin of this superconductivity is debated, with proposals involving both electron-phonon and electron-electron interactions~\cite{martin_2018,mattia_2019,angeli2020jahntellercouplingmoirephonons,biao_2019,FWu_prb_2019,prl_yoreg_2021,blason_2022,gadelha2021localization,PhysRevX.8.041041,Khalaf_2021,shi2024moireopticalphononsdancing,wang2024electronphononcouplingtopological,liu2023electronkphononinteractiontwistedbilayer,Nuckolls_cdw_2023,kwan2023electronphononcouplingcompetingkekule,ingham2023quadratic}. 

In an exciting very recent development, strong evidence of superconductivity has now been reported in twisted WSe$_2$ (tWSe$_2$)~\cite{xia2024unconventional,guo2024superconductivity}, a member of the transition metal dicalcogenide (TMD) family of moir\'e materials following earlier suggestive but not conclusive findings~\cite{wang2020correlated}. 
The TMD family of materials exhibits Mott~\cite{regan2020mott,tang2020simulation,Li:2021aa,wang2020correlated,Xu_2022,mak2022semiconductor} 
% {\color{red} many of these references related to heterobilayers, which exhibit very different physics, so I'm not sure one can point to these experiments as evidence that electron phonon coupling is weak} 
and anomalous Hall~\cite{Li_2021,Cai2023,PhysRevX.13.031037,xuParkObservationFractionallyQuantized2023,zeng2023thermodynamic} physics strongly suggestive of dominant electron-electron interactions, while electron-phonon interactions are believed to be weak ~\cite{Huang_2024,Jauho_2019,Jauho_2013}, strongly suggesting that the superconductivity reported in tWSe$_2$ is mediated by electron-electron interactions.
% \sout{These experiments strongly suggest a significant role for electronically mediated pairing, further emphasized by the weak electron-phonon coupling in monolayer TMDs}
\begin{figure}
    \centering
    \includegraphics[width=\linewidth]{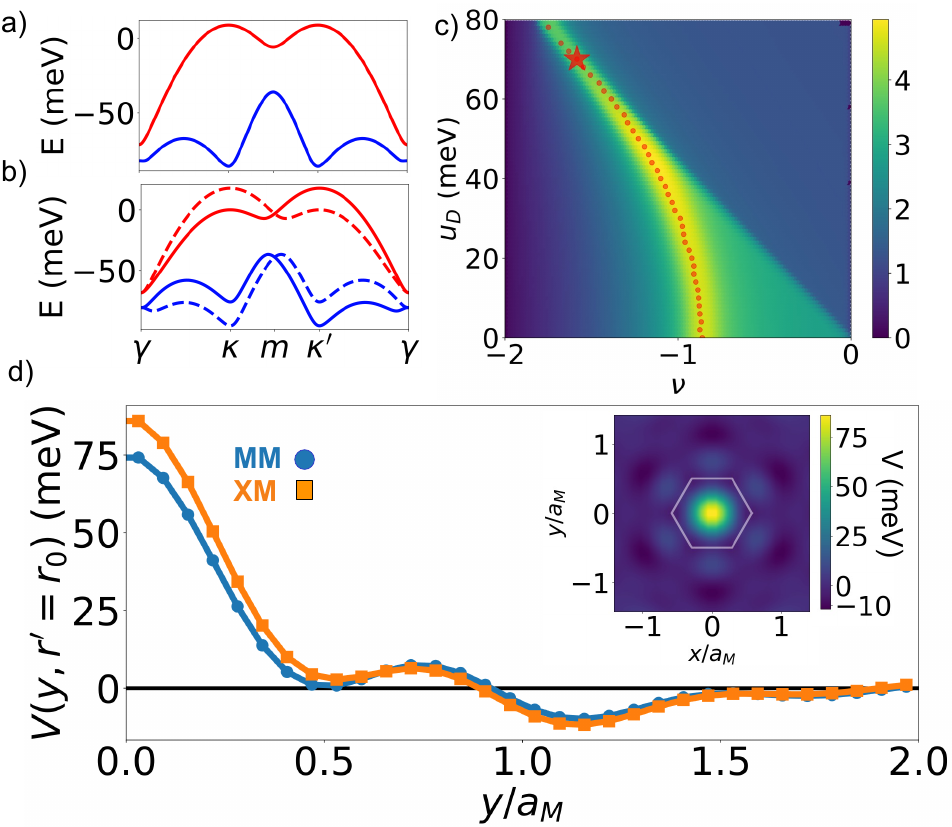}
    \caption{a) Dispersion relations of the moir\'e energy bands for vanishing displacement field and b) $u_D=20$meV. The topmost bands are topological with Chern number $1(-1)$ for spin $\uparrow(\downarrow)$. c) False color representation of the density of states as a function of $u_D$ and filling factor. The red dots show the position of the van Hove singularity, and the star denotes the location of the higher order van Hove singularity. d) Real space structure of the screened Coulomb potential $V(\br,\br')$ in the moir\'e unit cell for $\theta=5^\circ$ and $u_D=0$ obtained setting $\br'=0,(\bm a_1-\bm a_2)/3$ corresponding to the MM (blue) and XM (orange) stackings, respectively. The calculation has been performed at $\theta=5^\circ$ [bandwidth $W=80$meV, $V_{C}=e^2/(4\pi\epsilon\epsilon_0a_{\rm M})=33$meV] with model parameters from Ref.~\cite{wang2023topological} and dielectric constant $\epsilon=20$, distance from the metallic gates $d_{\rm sc}=24$nm. The inset shows the real-space evolution of the screened interaction for $\br'=0$ (MM).
    }
    \label{fig:fig1}
\end{figure}

%Topological superconductivity is a superconducting phase that exhibits nontirival topological properties, such as exotic edge states that are robust to local perturbations~\cite{Sato_2017}. 
With the rise of our understanding of topology and its role in electronic band structure, the quest to observe topological superconductivity -- a superconducting state featuring exotic zero energy states that are robust against local perturbations at the sample boundary~\cite{Sato_2017} -- intrinsically~\cite{Nandkishore_2012,crepel2023prl,Zerba_2024}, through proximity effects~\cite{fu2008}, or via twisting~\cite{Can_2021,kimtwisted2023,volkov2023prl,volkov2023prb,lucht2023topological,PhysRevB.109.184507} -- has garnered significant attention, though a definitive experimental confirmation remains elusive. 
Notably for the case of interest here, the topmost valence band of twisted homobilayer TMDs has a Chern number equal to $\pm1$ for spin up and down carriers and hosts a layer skyrmion~\cite{AllanNicolas_2024,shi2024adiabaticapproximationaharonovcasherbands,kolar2024hofstadterspectrumchernbands,Zhang_2024,dong2023,zhang2024directobservationlayerskyrmions} that generates a real-space Berry connection~\cite{FW_PRL_2019,haining2020,AllanNicolas_2024}, leading to the fundamental question of how Cooper pairs behave in the presence of this effective magnetic field~\cite{shi2006attractive,qin2019chiral,YiLi_2018}. 
% {\color{red} somewhat unclear to refer to this simply as effective magnetic field}. 
Despite the considerable progress in understanding quantum geometry within twisted homobilayer TMDs, the influence of topology on the Cooper pairing mechanism has not been explored in these systems~\cite{schrade2024nematic,YiMingWu_2023,Klebl_2023,Zegrodnik_2023,zhu2024theorysuperconductivitytwistedtransition,kim2024theorycorrelatedinsulatorssuperconductor} with only few recent studies pointing out the important role of the topological wavefunction properties~\cite{christos2024approximatesymmetriesinsulatorssuperconductivity,xie2024superconductivitytwistedwse2topologyinduced}. Furthermore, a detailed analysis of the symmetries and topology of the superconducting order is still lacking. 
% \jp{JP: This last sentence after the comment above is problematic and weakens our case completely. It looks like there is a already several papers looking into the effects of topology on cooper pairing. I don't think this is true of these papers. Also, it seems we are trying to make 2 or 3 points in one sentence, lets break into one point at a time.}

% \ajm{\bf Comment: first, while the mechanism is closely related to what Kohn and Luttinger considered, it is not quite the same: they said basically that if you look at high enough angular momentum you can find something attractive and they got a really really low Tc; second the term Kohn-Luttinger will cause some people, including Chubukov and Provokiev on one side and Kivelson on the other to get  wrong ideas about what is done here. this is why I suggest the following.} 

In this work, we study a purely electronic mechanism for superconductivity in a topological band relevant to tWSe$_2$, in which the interaction originates from the screening of the long-range Coulomb interaction~\cite{Cea_2021,Ghazaryan_2021,li2020artificial}. The model we consider has a real space pairing structure very similar to that arising from spin fluctuations, and we expect that many of our conclusions are valid beyond the specifics of the interaction (or pairing mechanism) used. 
By incorporating the interplay of van Hove singularities with quantum geometric effects, we provide a consistent theory that describes the recent superconductivity observed in ${\rm tWSe}_{2}$~\cite{xia2024unconventional,guo2024superconductivity}, and makes concrete predictions for the character of the superconducting order. 
A crucial point of the theory developed here is that the strong spin-momentum locking characteristic of TMD materials leads to opposite (in sign) Berry curvature in opposing valleys, introducing complex phases to the wavefunction which in turn generate nodes in the projected Coulomb interaction. This affects the competition between different superconducting instabilities and modifies the properties of the superconducting state. 
% {\color{red} are those two things the same? a bit unclear what the sentence means}. 
We detail the symmetries and properties of the superconducting state, with leading instabilities in the two-fold degenerate $E$ and non-degenerate $A$ irreducible representations, and characterize the evolution of the pairing in the experimentally relevant range of parameters. 
Additionally, we emphasize the significant role of the Berry phase in introducing mixing between singlet and triplet channels, as well as the competition between chiral and nematic superconductors within the two-fold degenerate $E$ subspace. The chiral superconductor breaks spontaneously time-reversal symmetry (TRS) and features an intervalley symmetric superposition of triplet ($S^z=0$) and singlet pairing belonging to the C Cartan class, while the nematic state breaks the $C_{3z}$ symmetry and displays a Bogoliubov de Gennes spectrum with a pair of Dirac nodes. Finally, we outline future experiments which can be performed to further clarify this scenario.

{\it Moir\'e band structure.---} Our model for tWSe$_2$ is given by $\hat H=\hat H_0+\hat H_{\rm int}$; we first describe the single-particle bandstructure of tWSe$_2$ which for the spin $\uparrow$ is given by the spectrum of the continuum Hamiltonian~\cite{FW_PRL_2019,haining2020}
\begin{equation}\label{continuum_model}
    H_{\uparrow}(\br)=\begin{pmatrix}
        -\frac{\bk^2}{2m} + U_+(\br)+\frac{u_D}{2} & T(\br) \\ 
        T^*(\br) & -\frac{\bk^2}{2m} + U_-(\br)-\frac{u_D}{2}
    \end{pmatrix},
\end{equation}
where the $2\times2$ matrix acts on layer space (denoted by quantum number $\ell$ in the following), $H_{\downarrow}(\br)=H_{\uparrow}(\br)^*$, $U_\pm(\br)=2v\sum_{j=1,3,5}\cos(\bm b_j\cdot\br\pm\phi)$ includes only the reciprocal lattice vectors $\bm b_j$ corresponding to the leading harmonics of the superlattice potential, $\bm a_{1/2}$ moir\'e lattice constants,  $T(\br)=w\sum_{j=1}^{3}e^{-i\bq_j\cdot\br}$, and $\bm q_j$ wavevector connecting the parabolic bands in the two layers centered around $\kappa$ and $\kappa'$ of the mini Brillouin zone (mBZ). More details are given in the SM~\cite{supplementary}. 

The continuum theory preserves TRS represented by the anti-unitary operator $\mathcal T=i \sigma^y\mathcal K$, where $\mathcal K$ is complex conjugation, $\sigma^{x,y,z}$ are Pauli matrices acting on the spin/valley degree of freedom, and $\gamma^{x,y,z}$ acts on layer space. Additionally, the theory possesses $C_{3z}$ symmetry represented by the unitary operator $C_{3z} = \exp(i\pi\sigma_z/3)$, and $C_{2y}=i\gamma^x\sigma^y$ which acts on position and spin. The model~\eqref{continuum_model} has an additional three-dimensional inversion symmetry $I$ acting as $\gamma^xH_\sigma(\bk) \gamma^x=H_\sigma(-\bk)$ which, however, is broken by higher harmonics in the moir\'e potential~\cite{wang2023topological,Jiabin_2024,PhysRevLett.132.036501,Zhang_2024} and by higher-order terms in the $\bk\cdot\bp$ expansion around the two valleys $K/K'$ of the monolayers~\cite{Korm_nyos_2015}. 
Taken together, these symmetries are those of the magnetic point group $3m'$. The displacement field $u_D$ encodes the experimentally tunable potential difference between top and bottom layers. 
A nonzero $u_D$ breaks three-dimensional inversion $I$ and two-fold rotation $C_{2y}$, reducing the symmetries to $C_{3z}$ and $\mathcal T$ and giving rise to the bands displayed in Fig.~\ref{fig:fig1}b).

%$\nu=n_0/|\bm a_1\times\bm a_2|$ ($\bm a_1,\bm a_2$)
Fig.~\ref{fig:fig1}c) shows the evolution of the single particle density of states as a function of displacement field and filling factor $\nu=n_0/\Omega$, where $\Omega$ is the area of the moir\'e unit cell. We highlight two features: a line in parameter space along which a van Hove singularity occurs at the Fermi level, starting at $\nu\approx0.86$ and evolving monotonically to larger filling factors with increasing $u_D$, shown by the red dots in Fig.~\ref{fig:fig1}c). Secondly, the displacement field introduces an energy offset between the quadratic bands of the top and bottom layer, realizing a region of low-density of states where the two-dimensional electron gas is almost layer polarized, shown in the upper right region in Fig.~\ref{fig:fig1}c). Both trends are consistent with experiments~\cite{xia2024unconventional,guo2024superconductivity}. Finally, due to $C_{2y}$ Fig.~\ref{fig:fig1}c) is symmetric under $u_D\to-u_D$. The analysis of the bandstructure for other twist angles shows similar properties (see the SM~\cite{supplementary}).

{\it Interacting Hamiltonian.---} The low-energy degrees of freedom interact via the long range Coulomb interaction: 
\begin{eqnarray}
    \hat H_{\rm int}=\frac{1}{2A}\sum_{\bq}V_0(\bq)\hat \rho_{-\bq}\hat \rho_{\bq},
    \label{eqn:Hint}
\end{eqnarray} 
where $A$ is the area of the sample, $V_{0}(\bq)=e^2\tanh(|\bq| d_{\rm sc}/2)/(2\epsilon\epsilon_0|\bq|)$ is the dual-gate screened Coulomb potential, $d_{\rm sc}$ the distance between the metallic gates, $\epsilon$ the dielectric constant and, finally, $\hat\rho_{\bq}$ is the density operator projected on the topmost band: 
\begin{equation}\label{density_operator}
    % \hat\rho_{\bq+\bm g} =\sum_{\sigma}\sum_{\bk\in{\rm mBZ}}\braket{u_{\bk\sigma}}{u_{\bk+\bq\sigma}}\hat c^\dagger_{\bk\sigma}\hat c_{\bk+\bq\sigma},    
    \hat\rho_{\bq+\bm g} =\sum_{\sigma}\sum_{\bk\in{\rm mBZ}}\Lambda^{\bk,\bk+\bq }_{\bm g\sigma}\hat c^\dagger_{\bk\sigma}\hat c_{\bk+\bq\sigma},    
\end{equation}
where $\Lambda^{\bk,\bk+\bq }_{\bm g\sigma}=\sum_{\bm g'}z^*_{\bm g'\sigma}(\bk)z_{\bm g'-\bm g\sigma}(\bk+\bq)$ with $z_{\bm g\sigma}(\bk)=\braket{\bm g}{u_{\bk \sigma}}$ are the Fourier amplitudes of the Bloch states. 
Carriers in the topmost band of ${\rm tWSe}_{2}$ shown in Fig.~\ref{fig:fig1}a) experience opposite spin/valley-dependent Chern numbers $C_{\uparrow/\downarrow}=\pm1$ and Berry phases $\Omega_{\uparrow}(\bk)=-\Omega_\downarrow(-\bk)$, which acts effectively acts as a spin dependent $\bk-$space magnetic field~\cite{AllanNicolas_2024,shi2024adiabaticapproximationaharonovcasherbands,kolar2024hofstadterspectrumchernbands,abouelkomsan2024nonabelianspinhallinsulator} and manifests also as a real-space layer winding in the Bloch orbitals $\bm \psi_{\bk\sigma}(\br)$~\cite{AllanNicolas_2024,shi2024adiabaticapproximationaharonovcasherbands,wang2021,Ledwith2020,dong2023,guerci2024layerskyrmionsidealchern}.  
This property introduces a Berry phase in the finite momentum scattering processes encoded in the overlap between Bloch states $\Lambda^{\bk,\bp}_{\bm g \sigma}$~\eqref{density_operator}, where $\bm g$ is a reciprocal lattice vector and $\bk,\bp\in{\rm mBZ}$. The requirement of $C_{3z}$ imposes $\Lambda^{C_{3z}\bk,C_{3z}\bp}_{C_{3z}\bm g \sigma}=\braket{u_{C_{3z}\bk\sigma}}{u_{C_{3z}\bp+C_{3z}\bm g\sigma}}=\Lambda^{\bk,\bp}_{\bm g \sigma}$. 
For a trivial band with vanishing Berry curvature, $\Lambda\in\mathbb R$ by a proper gauge choice. Due to the nontrivial topology of the topmost band, $\Lambda^{\bk,\bp}_{\bm g \sigma}$ is generically complex and as a result of TRS satisfies the relation $\Lambda^{\bk,\bp}_{\bm g\uparrow}=[\Lambda^{-\bk,-\bp}_{-\bm g\downarrow}]^*$. We introduce the decomposition 
\begin{eqnarray}\label{form_factors}
    \Lambda^{\bk,\bp}_{\bm g\uparrow/\downarrow}=X^{\bk,\bp}_{\bm g}\pm i Y^{\bk,\bp}_{\bm g},
\end{eqnarray}
where $X^{\bk,\bp}_{\bm g}=[\Lambda^{\bk,\bp}_{\bm g\uparrow}+\Lambda^{\bk,\bp}_{\bm g\downarrow}]/2$ and $Y^{\bk,\bp}_{\bm g}=-i[\Lambda^{\bk,\bp}_{\bm g\uparrow}-\Lambda^{\bk,\bp}_{\bm g\downarrow}]/2$. Inversion symmetry $I$ enforces $\Lambda^{\bk,\bp}_{\bm g\sigma}=\Lambda^{-\bk,-\bp}_{-\bm g\sigma}$, and therefore employing an appropriate gauge choice we have $X^{\bk,\bp}_{\bm g},Y^{\bk,\bp}_{\bm g}\in \mathbb R$, while both become complex at any nonzero displacement field. 
\begin{figure}
    \centering
    \includegraphics[width=\linewidth]{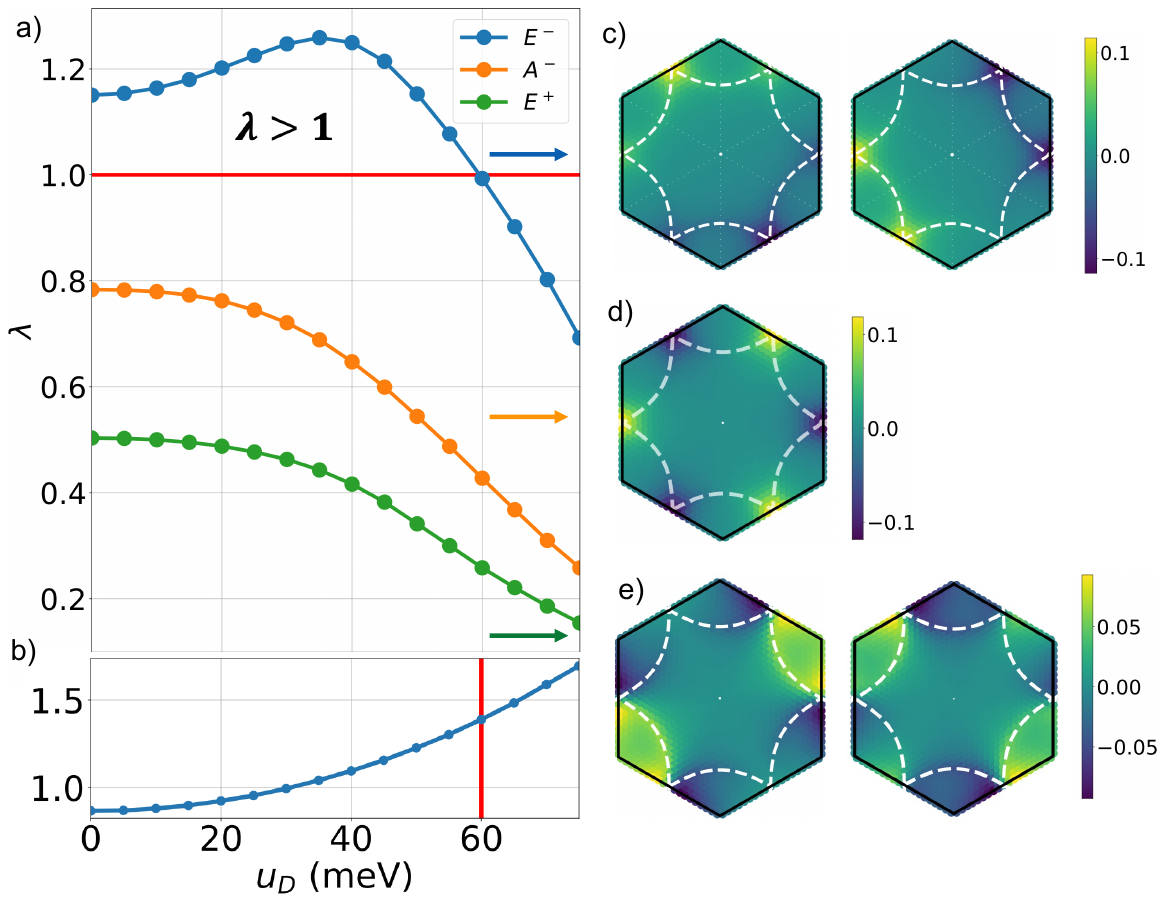}
    % \vspace{0.01cm}
    % \caption{a) Leading eigenvalues of the pairing kernel including $E^{\pm}$ and $A^-$ as a function of the displacement field tuning the chemical potential to the VHS energy. The superconducting instability takes place for $\lambda\ge1$. b) Evolution of the filling factor where the van Hove singularity is located. c) Evolution of the layer polarization $\langle\hat \gamma^z\rangle$ in unit of the filling factor as a function of the displacement field, the average is taken over the Fermi sea with chemical potential set at the van Hove filling. Panels d), e) and f) show the momentum space structure of the leading superconducting instability for $u_D=0,20,40$meV, respectively. The leading instability belongs to the two-dimensional irreducible representation $E^-$ originating from $E_u:(d^z_1,d^z_2)\sim(k_x,k_y)$ for $u_D=0$. In each panel, left and right figures show the two components spanning the two-dimensional subspace and in panels e) and f) we decompose $\phi(\bk)$ in the odd $d^z$ (upper figure) and $\psi$ even (lower figure) contributions. The calculation has been performed setting the relative dielectric constant $\epsilon=20$ [$V_C=e^2/(4\pi\epsilon\epsilon_0a_{\rm M})=33$meV], $d_{\rm sc}=24$nm, $\theta=5^\circ$, $k_BT/E_{\rm kin}=9\times 10^{-3}$, and employing the parameters in Ref.~\cite{wang2023topological}.}
    \caption{a) Leading eigenvalues of the pairing kernel including $E^{\pm}$ and $A^-$ as a function of the displacement field tuning the chemical potential to the van Hove filling. The superconducting instability takes place for $\lambda\ge1$. b) Evolution of the filling factor where the van Hove singularity is located. Panels c), d) and e) show the form factor of leading superconducting instabilities at $u_D=0$. The leading instability belongs to the two-dimensional irreducible representation $E^-$ originating from $E_u:(d^z_1,d^z_2)\sim(k_x,k_y)$ for $u_D=0$. The calculation has been performed setting the relative dielectric constant $\epsilon=20$ [$V_C=e^2/(4\pi\epsilon\epsilon_0a_{\rm M})=33$meV], $d_{\rm sc}=24$nm, $\theta=5^\circ$, $k_BT/E_{\rm kin}=9\times 10^{-3}$, and employing the parameters in Ref.~\cite{wang2023topological}.
    }
    \label{fig:fig2}
\end{figure}

{\it Pairing mechanism.---}  We study superconductivity in this system due to a purely electronic mechanism, driven by the screening of the Coulomb interaction from low-energy particle-hole excitations. Within the random phase approximation~\cite{Cea_2021,Ghazaryan_2021,li2020artificial,Long_2024,cavicchi_2024,Lewandowski_2023,li2022higher,scammell2022intrinsic,Giuliani_Vignale_2005}, this interaction takes the form: 
\begin{equation}\label{screened}
    V_{\bm g\bm g'}(\bq,i\omega)=\left[1-V_0(\bq)\Pi(\bq,i\omega)\right]^{-1}_{\bm g\bm g'}V_{0}(\bq+\bm g'),
\end{equation}
where we have introduced the particle-hole susceptibility:
\begin{equation}\label{particle_hole}
    \Pi_{\bm g\bm g'}(i\omega, \bq)=\frac{1}{A}\sum_{\bk\sigma}\frac{f(\xi_{\bk\sigma})-f(\xi_{\bk+\bq\sigma})}{i\omega+\epsilon_{\bk\sigma}-\epsilon_{\bk+\bq\sigma}}\Lambda^{\bk,\bk+\bq}_{-\bm g\sigma}\Lambda^{\bk+\bq,\bk}_{\bm g'\sigma},
\end{equation}
with $q=(\bq,i\omega)$, $\xi_{\bk\sigma}=\epsilon_{\bk\sigma}-\mu$, $f(\xi)$ the Fermi-Dirac function, and $\mu$ chemical potential. Despite the originally repulsive bare Coulomb interaction, the screening from low-energy particle-hole excitations induces an effective attraction between the carriers. Fig.~\ref{fig:fig1}c) shows the real space behavior of the screened Coulomb potential $V(\br,\br')$ which displays attractive regions around a short range repulsive core at $\br=\br'$ as opposed to the $V_0(\br)$ given below Eq.~\eqref{eqn:Hint}, which is purely repulsive and in the long distance limit $|\br|\rightarrow \infty$  goes like $V_0(\br)\sim \exp\left(-2|\br|/d_{\rm sc}\right)/|\br|$. The long range attractive regions are responsible for the pairing between holes in the bilayer. Decreasing $d_{\rm sc}$ weakens this effective attraction as demonstrated in the SM~\cite{supplementary}.   

{\it Superconducting instabilities.---} The effective pairing interaction is obtained projecting the static component $(\omega=0)$ of the screened Coulomb potential in the particle-particle channel: 
\begin{equation}\label{particle_particle_interaction}
\hat H_{\rm eff}=\sum_{\bk\bk'}\sum_{\{s_i\}}\frac{V_{s_1s_2,s_3s_4}(\bk,\bk')}{2A}{\hat c}^\dagger_{-\bk s_1}{\hat c}^\dagger_{\bk s_2}{\hat c}_{\bk' s_3}{\hat c}_{-\bk' s_4}
\end{equation}
with $s_i=\uparrow,\downarrow$ spin/valley of the carriers and the pairing potential: 
\begin{equation}
    \begin{split}
    \label{scattering_vertex_final}
    V_{s_1s_2,s_3s_4}(\bk,\bk')&=\sum_{\mu\nu=0,z}V_{\mu\nu}(\bk,\bk')\sigma^\mu_{s_1s_4}\sigma^\nu_{s_2s_3},
    \end{split}
\end{equation}
where $V_{\mu\nu}(\bk,\bk')$ represents the interaction vertices. The analysis of retardation effects $(\omega\neq 0)$ which plays a crucial role in a faithful estimate of $T_c$~\cite{Tolmachov1958,anderson1962,sham1983prb,chubukov_prb_2019,kunchen2022prb,malte_prb_2023} is beyond the scope of the present work and will be investigated in future studies. 

The presence of $\sigma^0\otimes\sigma^z$ and $\sigma^z\otimes\sigma^0$ introduce mixing between $S^z=0$ spin triplet and singlet channels. The projected interaction is directly influenced by the topological properties of the topmost topological band.  The matrix elements $V_{zz}$ and $V_{z0/0z}$ are proportional to $Y^{\bk,\bk'}_{\bm g }$ which in the small momentum transfer limit, becomes the Berry phase accumulated in the scattering process $Y^{\bk,\bk'}_{\bm g}\approx (\bk-\bk')\cdot\left[\bm A_{\uparrow}(\bk+\bk'/2)-\bm A_{\downarrow}(\bk+\bk'/2)\right]/2$ with $\bm A_{\sigma}(\bk)=-i\braket{u_{\bk\sigma}}{\nabla_{\bk}u_{\bk\sigma}}$ single particle Berry connection. Furthermore, due to TRS $Y^{\bk,\bk'}_{\bm g}=-Y^{-\bk',-\bk}_{\bm g}$ which requires the components of the pairing potential $V_{zz}(\bk,\bk')$ and $V_{z0/0z}(\bk,\bk')$ to have zeros~\cite{Murakami_2003,YiLi_2018}. Finally, we emphasize that the mixing between triplet and singlet is present only for finite $u_D$, at $u_D=0$ the inversion symmetry $I$ implies that singlet and triplet even/odd under inversion are decoupled, and the contribution of $V_{0z/z0}(\bk,\bk')$ vanishes. 

The critical temperature $T_c$ of the superconducting state is determined by solving the linearised gap equation: 
\begin{equation}\label{SC_instability}
\lambda(T_c)\phi_{\sigma\sigma'}(\bk)=\sum_{\bp}\sum_{\tau\tau'} K_{\sigma\sigma',\tau\tau'}(\bk,\bp)\phi_{\tau\tau'}(\bp)
\end{equation}
where the superconducting instability takes place when $\lambda(T_c)=1$, and $\lambda>1$ defines a stable superconducting state; the explicit expression of the pairing kernel $K(\bk,\bp)$ is given in the SM~\cite{supplementary}. Fig.~\ref{fig:fig2}a) shows the evolution of the competing superconducting instabilities as a function of the displacement field $u_D$ at the van Hove filling displayed in Fig.~\ref{fig:fig2}b). 
The leading superconducting state involves pairing of electrons in opposite valleys, for which the Cooper pairs move in bands with opposite Chern number, and due to the broken inversion symmetry, comprises a superposition of spin $S^z=0$ triplet and singlet.
Within this subspace we find three leading instabilities: two belonging to the two-dimensional irreducible representations $E^{\pm}$, and one to the one-dimensional representation $A^-$ where the superscript $\pm$ refers to the parity under inversion $I$ at $u_D=0$, e.g. $E^{\pm}$ originates from $E_{g/u}$ respectively in the limit of vanishing displacement field. Similarly, $A^-$ denotes the superconducting state originating from $A_{1u}$ state which is odd under inversion at vanishing displacement field. The form factors characterizing the three different instabilities at $u_D=0$ are given in Fig.~\ref{fig:fig2}c), d) and e).

The evolution of the eigenvalue spectrum $\lambda$~\eqref{SC_instability}, shown in Fig.~\ref{fig:fig2}a), displays a non-monotonic behavior as a function of $u_D$ --  with the leading instability belonging to the $E^-$ irreducible representation originating due to the interplay of two competing effects. At small displacement fields, up to $u_D\le40$meV, the leading $E$-channel eigenvalue $\lambda_{E^-}$ increases with increasing $u_D$. The increase is due to the broken inversion symmetry which mixes singlet and triplet channels without spoiling the Cooper log singularity in the $S^z=0$ sector ensured by TRS, $\epsilon_{\bk\uparrow}=\epsilon_{-\bk\downarrow}$. At an intermediate scale, $u_D\sim40$meV ($\nu\simeq1.04$), we find that $\lambda_{E^-}$ has a peak, highlighting the presence of an optimal $u_D$ which is consistent with both experimental reports~\cite{xia2024unconventional,guo2024superconductivity}. For larger values of $u_D$, the large layer polarization $\langle \gamma^z\rangle\ge0.5$ leads to an enhanced Coulomb repulsion, which results in the suppression of the superconducting instability. This trend contrasts with the evolution of the density of states along the van Hove line, which increases at larger $u_D$ up to the higher-order van Hove (power-law singularity in the density of states)~\cite{chamon_2017,chandrasekaran2020catastrophe,LFu_HOVHS_2019,zangj_2021,classen2024highordervanhovesingularities} at $u_D=70$meV and filling factor $\nu\simeq1.58$ displayed in Fig.~\ref{fig:fig1}b). 
% Figs.~\ref{fig:fig2}d), e) and f) show the eigenstates $ \phi(\bk)$~\eqref{SC_instability} in the mBZ for different values of the displacement field. For $u_D=0$ in Fig.~\ref{fig:fig2}d), the two degenerate eigenstates are odd under inversion symmetry $I$. As we increase $u_D$, even and odd sectors mix giving rise to superconducting instabilities without a well defined inversion symmetry. Here, the eigenstates $\phi_{1/2}(\bk)$ consists of a superposition of odd $d^z_{1/2}(\bk)=\Tr\left[\sigma^x\phi_{1/2}(\bk)\right]/2$ and even $\psi_{1/2}(\bk)=\Tr\left[-i\sigma^y\phi_{1/2}(\bk)\right]/2$ components, shown in Figs.~\ref{fig:fig2}e) and~\ref{fig:fig2}f).
\begin{figure}
    \centering
    \includegraphics[width=1.\linewidth]{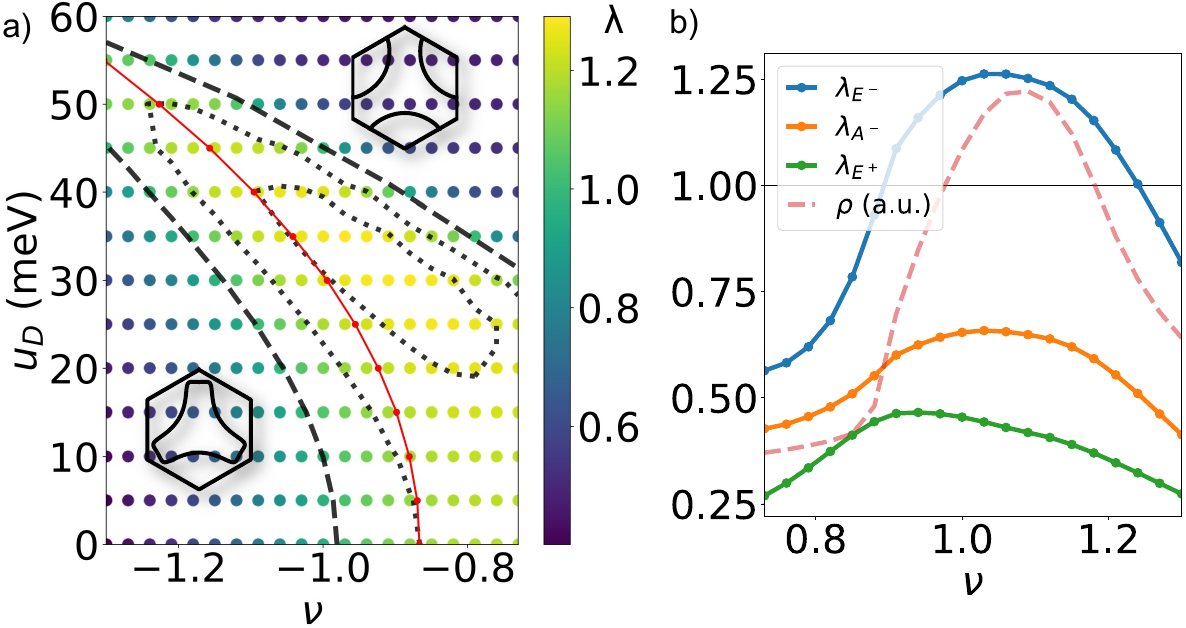}
    % \vspace{0.01cm}
    % \caption{a) Evolution of the leading instability in the relevant regime of displacement field $u_D$ and filling factor $\nu$. Black lines denote the region where the instability takes place $\lambda>1$, red one is the VHS line. b) Line cut for $u_D=40$meV, besides the leading superconducting instability $\lambda_{E^-}$ we also show the evolution of the subleading $A^-$ and $E^+$ states, and the density of states (dashed line). c) Evolution of the energy of the nematic and chiral superconductors variational states in the regime of filling factors with $\lambda_{E^-}>1$ and $u_D=40$meV. The calculation has been performed setting $\epsilon=20$ [$V_C=e^2/(4\pi\epsilon\epsilon_0a_{\rm M})=33$meV], $d_{\rm sc}=24$nm, $\theta=5^\circ$, $k_BT/E_{\rm kin}=9\times 10^{-3}$. }
     \caption{a) Evolution of the leading instability in the relevant regime of displacement field $u_D$ and filling factor $\nu$. Black dashed lines denote the region where the instability takes place $\lambda>1$, dotted ones show the contour lines for $\lambda=1.15,1.25$. Finally, the red solid line is the VHS line. We additionally present the Fermi surface of spin-up electrons in the normal state on both sides of the van Hove line. b) Line cut for $u_D=40$meV, besides the leading superconducting instability $\lambda_{E^-}$ we also show the evolution of the subleading $A^-$ and $E^+$ states, and the density of states (dashed line). The calculation has been performed setting $\epsilon=20$ [$V_C=e^2/(4\pi\epsilon\epsilon_0a_{\rm M})=33$meV], $d_{\rm sc}=24$nm, $\theta=5^\circ$, $k_BT/E_{\rm kin}=9\times 10^{-3}$. }
    \label{fig:2d_map}
\end{figure}
The evolution of the superconducting instability in the plane of filling factor $\nu$ and displacement field $u_D$ is shown in Fig.~\ref{fig:2d_map}a). The line cut in Fig.~\ref{fig:2d_map}b), taken at $u_D=40$meV, reveals that the eigenvalue $\lambda_{E^-}$ reaches its maximum at a filling factor slightly offset from the van Hove singularity.

{\it Landau theory.---} In the small displacement field regime, the Landau free-energy for the $E$-channel superconducting order parameter $\Delta(\bk)=\sum_{a=1,2}\eta_a \Delta_{a}(\bk)$ reads~\cite{sigrist_1991,Sauls_2019}: 
\begin{equation}\label{landau_ginzburg}
    \mathcal F(\bm \eta)=A\bm\eta^\dagger\cdot\bm\eta+B(\bm\eta^\dagger\cdot\bm\eta)^2+C(\bm\eta^*\times\bm\eta)^2,
\end{equation}
where $\bm\eta=(\eta_1,\eta_2)^T$ describes the decomposition of the pairing amplitudes in the two dimensional $E$ irreducible representation of the magnetic point group generated by $C_{3z}$ and $\mathcal T$, the latter mixing the $C_{3z}$ eigenstates with eigenvalues $\omega/\omega^*$, while ${\bm\eta}^* \cross \bm\eta \equiv\eta^*_1\eta_2-\eta^*_2\eta_1$. Eq.~\eqref{landau_ginzburg} is consistent with Ref.~\cite{schrade2024nematic} where  a leading $E$ superconducting instability was also found. As shown in Fig.~\ref{fig:fig2}d) and e), triplet and singlet sectors are mixed and the order parameter does not have a well defined symmetry under inversion symmetry:
\begin{equation}\label{basis_functions}
    \Delta_{a}(\bk)=i\left[\psi_{a}(\bk)+d^z_{a}(\bk)\sigma^z\right]\sigma^y,
\end{equation}
where $\Delta_{1/2}(\bk)$ are two orthogonal basis function transforming as an $E$ doublet. From Eq.~\eqref{landau_ginzburg} we find, depending on the sign of the coefficient $C$, two different superconducting ground states. For $C>0$ the state which minimizes the free-energy is a chiral combination of the two components $\bm\eta=(1,i)^T$. On the other hand, for $C<0$ we find a nematic state where $\bm \eta=(\cos\varphi,\sin\varphi)^T$ with $\varphi\in[0,2\pi)$. As we will show, in the nematic state the continuous rotational symmetry within the order parameter space $(\eta_1,\eta_2)$ is broken down to a discrete one, as including higher order corrections lowers this continuous symmetry to $C_{3}$.   

\begin{figure}
    \centering
    \includegraphics[width=1.\linewidth]{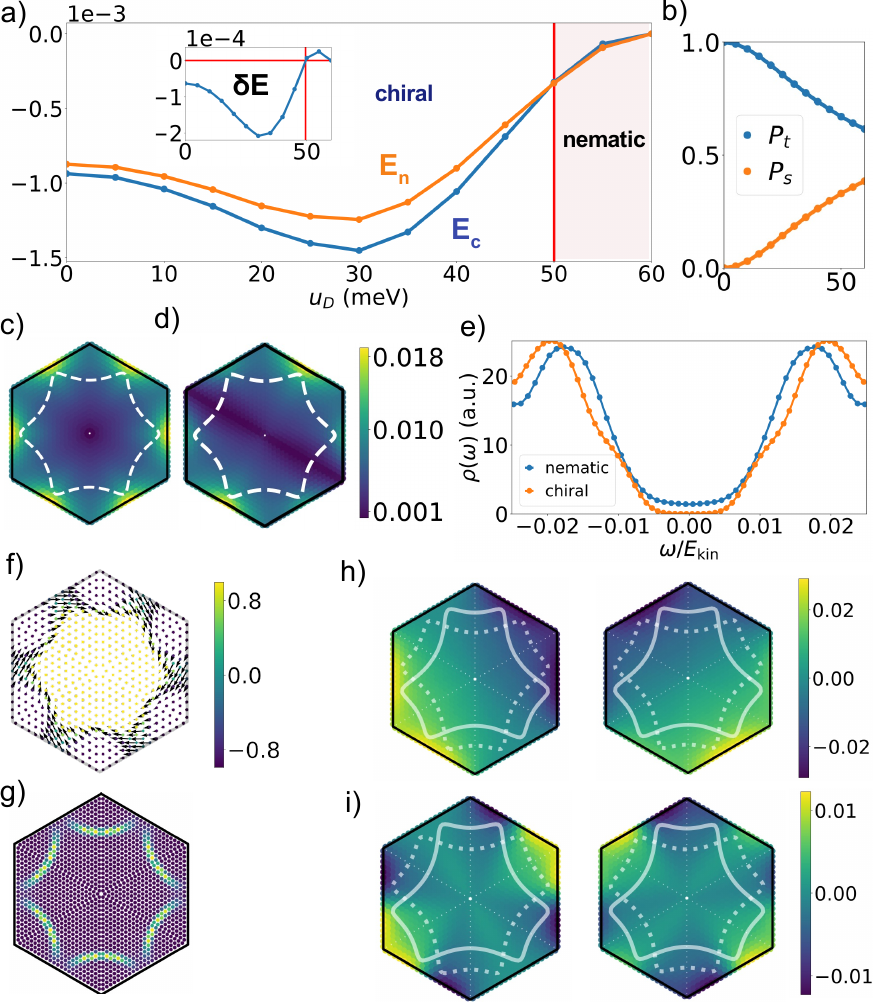}
    \vspace{0.01cm}
    \caption{a) Energy of the nematic $E_n$ and chiral $E_c$ phases measured with respect to the energy of the normal state and in unit of $E_{\rm kin}$ as a function $u_D$, and the inset shows the energy difference. b) Triplet $P_t$ and singlet $P_s$ fraction in the superconducting order parameter. Panels c) and d) show $\Tr[\Delta(\bk)\Delta^\dagger(\bk)]/2$ in unit of $E_{\rm kin}$ for the chiral and nematic states, respectively, in the zero displacement field limit, while e) the density of states of the nematic and chiral phases. f) The arrows shows the in-plane components of the pseudospin vector $[n_x(\bk),n_y(\bk)]$, and the color code show the out of plane component of the vector field $n_z(\bk)$, and g) the skyrmion number of the pseudospin $\bm n(\bk)$. Panels h) and i) show the real part (left) and the imaginariry part (right) of $d^z(\bk)$ and $\psi(\bk)$ for the saddle point solution obtained for $u_D=20$meV slightly above the VHS filling. Calculations are performed at $\theta=5^\circ$ for different values of $u_D$ in proximity of the VHS filling factor.}
    \label{fig:variational_results}
\end{figure}

{\it Analysis of competing superconducting phases.---} To characterize the properties of the $E^-$ superconducting instability we minimize the variational energy $E_\Psi=\mel{\Psi}{\hat H_0+\hat H_{\rm eff}}{\Psi}$ in the Hilbert space of the BCS wavefunction $\ket{\Psi}$ with the unconventional order parameter $ \Delta(\bk)=\eta_1\Delta_{1}(\bk)+\eta_2\Delta_{2}(\bk)$; $\hat H_{\rm eff}$ has been introduced in Eq.~\eqref{particle_particle_interaction}.
At zero displacement field Fig.~\ref{fig:variational_results}a), we find that the chiral state $\bm\eta=(1,i)^T$ is energetically favored over the nematic solution. The greater energy gain in the chiral state can be attributed to the full gap in the chiral solution shown in Fig.~\ref{fig:variational_results}c), in constrant to the nematic one, which is characterized by a line of zeros Fig.~\ref{fig:variational_results}d) which leads to gapless low energy excitations. Consequently, the density of states, shown in Fig.~\ref{fig:variational_results}e), exhibits a U-shaped gap in the chiral state, whereas it presents a V-shape in the nematic phase~\cite{vshape}. 
 
The Bogoliubov de Gennes (BdG) Hamiltonian of the chiral superconductor decouples into two blocks $\mathcal H_{\rm BdG}(\bk)=\sum_{s=\pm}(\tau^0+s\tau^z)h(\bk)/2$ in the basis $[c_{\bk\uparrow},c^\dagger_{-\bk\downarrow},c_{\bk\downarrow},c^\dagger_{-\bk\uparrow}]^T$ with $h(\bk)=\xi_{\bk}\sigma^z+d^z(\bk)\sigma^++h.c.$ and is invariant under the particle-hole symmetry $\mathcal C=-i\tau^z\sigma^x\mathcal K$ ($\mathcal C^2=+1$). 
Furthermore, $d^z(\bk)=d^z_1(\bk)+id^z_2(\bk)$ with $d^z(\bk)\sim(k_x+ik_y)$ in the continuum limit~\cite{read2000prb}, implying that this state breaks TRS $\mathcal T=i\tau^y\sigma^z\mathcal K$, and thus belongs to the D Altland-Zirnbauer class~\cite{Altland_1997,Schnyder_2008}. The $\mathbb Z$ topological invariant is determined by the winding of the pseudospin texture $\bm n(\bk)=[\Re d^z(\bk),-\Im d^z(\bk),\xi_{\bk}]/E_{\bk}$ with $E^2_{\bk}=\xi^2_{\bk}+|d^z(\bk)|^2$ displayed in Fig.~\ref{fig:variational_results}f). 
Fig.~\ref{fig:variational_results}g) shows the Pontriagyin density $Q=\bm n\cdot(\partial_{k_x}\bm n\times\partial_{k_y}\bm n)/4\pi$ whose integral in the mBZ gives a total Chern number $C=\int_{\bk}Q=1$, which also follows from the opposite sign of $n_z(\bk)$ at $\gamma$ and the boundary $\kappa/\kappa'$ of the mBZ $n_z(\gamma)=-n_z(\kappa)$~\cite{Schnyder_2008,Sato_2017}.

For nonzero displacement field, the solution becomes a superposition of $S^z=0$ triplet and singlet components~\eqref{basis_functions}. The evolution of the energy difference between the nematic and chiral phases is displayed in Fig.~\ref{fig:variational_results}a). At small displacement field, the chiral state is the lowest energy phase and is described by the BdG Hamiltonian $\mathcal H_{\rm BdG}(\bk)=\sum_{s=\pm}(\tau^0+s\tau^z) h_s(\bk)/2$ with $h_{\pm}(\bk)=\xi_{\bk\uparrow/\downarrow}\sigma^z+\left[d^z(\bk)\pm\psi(\bk)\right]\sigma^++h.c.$ with $d^z(\bk)=d^z_1(\bk)+id^z_2(\bk)$ and $\psi(\bk)=\psi_1(\bk)+i\psi_2(\bk)$ which breaks $\mathcal T$. We show the real and imaginary components of $d^z(\bk)$ and $\psi(\bk)$ composing the superconducting state in Fig.~\ref{fig:variational_results}g) and~\ref{fig:variational_results}h) for triplet and singlet, respectively. For any nonzero singlet component $\psi(\bk)$, the particle hole symmetry takes the form $\tilde{\mathcal C}=\tau^x\mathcal C=-\tau^y\sigma^x\mathcal K$ ($\tilde{\mathcal C}^2=-1$) implying that the superconducting state belongs to the C Cartan class, possessing a $\mathbb{Z}$ topological invariant with edge modes that are  described by charge-neutral complex fermionic quasiparticles~\cite{Altland_1997,Schnyder_2008}. To quantify the magnitude of the singlet component we show in Fig.~\ref{fig:variational_results}b) the coefficients $[P_{t},P_s]=\int_{\bk}[ |d^z(\bk)|^2,|\psi(\bk)|^2]/N$ with $N=\int_{\bk}\Tr[\Delta(\bk)\Delta^\dagger(\bk)]/2$. For any finite $u_D$, there is singlet-triplet mixing that, although small in the regime of small $u_D$ Fig.~\ref{fig:variational_results}b), transitions the chiral superconductor into the C topological class. 

The mixed singlet-triplet nature of the chiral superconductor has important consequences for the edge theory. In the zero displacement field case, the topological superconductor in class D hosts protected Majorana edge modes~\cite{ivanov_prl_2001,read_green_2000}. The class C case applicable to nonzero displacement field also possesses chiral modes, yet these are Dirac in nature, rather than Majorana. 
%Concretely, the edge mode is Majorana when it is mapped to itself under particle-hole conjugations evident from the block diagonal form of the particle-hole symmetry $\mathcal C$~\cite{ivanov_prl_2001}; the addition of a small singlet component to the gap function introduces two contributions to the edge which map to each other under particle-hole conjugation, meaning the edge states are not Majoranas, 
This clearly manifests in the form of the particle-hole symmetry $\tilde{\mathcal C}=\tau^x\mathcal C$ which is off-diagonal and mixes the two Kramers doublet. Despite this aspect, the edge modes -- whether they be Majorana or Dirac -- remain gapless and produce signatures in thermal transport and STM~\cite{stone2004}.

Increasing the displacement field generically enhances singlet-triplet mixing Fig.~\ref{fig:variational_results}b) which leads to a critical $u_D=u_D^c(\theta=5^\circ)\approx 50$meV beyond which the nematic solution becomes energetically favourable over the chiral one. A detailed analysis of $u^c_D$ as a function of the twist angle is left to future studies. The nematic state is time-reversal invariant and is characterized by an order parameter $\bm\eta=(\cos\varphi,\sin\varphi)^T$ with $\varphi=n\pi/3$. As shown in Fig.~\ref{fig:variational_results}d), the state is nodal and displays low-energy excitations leading to a V-shape density of states in Fig.~\ref{fig:variational_results}e). We highlight that understanding the properties in this region is more challenging, as the superconductor is in close proximity to the instability to the normal state and the variational energy has only a shallow superconducting minimum. 

{\it Conclusions.---} We have demonstrated that in tWSe$_2$, an electronic mechanism for superconductivity leads to attraction with dominant superconducting channels $E^-$ and subleading $A^-$, both originating from inversion ($I$) odd (triplet) instabilities at vanishing displacement field. 
Competing $E^-$ and $A^-$ symmetry  instabilities were also identified within a spin fluctuation mechanism~\cite{christos2024approximatesymmetriesinsulatorssuperconductivity}, while an $A^-$ is consistent with a phonon-mediated mechanism~\cite{zhu2024theorysuperconductivitytwistedtransition}.
The two-dimensional nature of the $E^-$ representation gives rise to competing chiral (gapped) and nematic (nodal) superconducting ordering, both described at finite displacement field by an intervalley symmetric superposition of $S^z=0$ triplet and singlet channels~\cite{schrade2024nematic}. 
Through variational calculations, we showed that the chiral superconductor is the lowest energy solution over a wide range of displacement fields and belongs to the C Altland-Zirnbauer class, hosting Dirac edge states that can be probed via STM and thermal transport measurements. At a critical displacement field, we find a transition from the chiral state to the nematic state, which is time-reversal symmetric and topologically trivial. This transition results in distinctive low-energy Leggett modes of the relative phase of the two-component order parameter~\cite{Leggett}. 

In both the experiments~\cite{xia2024unconventional,guo2024superconductivity}, superconductivity is seen to be enhanced at finite displacement field -- consistent with our proposal of mixed singlet-triplet pairing at finite $u_D$. In Ref.~\cite{xia2024unconventional}, superconductivity extends down to zero displacement fields, where the superconducting state hosts Majorana zero modes\cite{Sato_2017}.
We remark that the chiral superconductor despite breaking TRS is valley symmetric ($S^z=0$) and cannot be detected by probes only sensitive to valley polarization. 

Independent of the precise superconducting order parameter symmetry, unique features of this system suggest an interesting interplay with magnetic fields. The spin-valley locking implies that the in-plane Zeeman coupling is negligible, while the orbital coupling introduced by the minimal substitution  $\hat \bk_{t/b}\to\hat \bk_{t/b}\pm e\bm B_{\parallel}d/2\hbar$, with $d$ thickness of the bilayer, acts as a pair-breaking term and for large $\bm B_\parallel$ might lead to the formation of a pair density wave. An estimate of this effect can be obtained by noting that $d \approx 0.7$ nm and $a_{\rm M} \approx 4.4$ nm, which yields a momentum shift of $a_{\rm M}d{eB_\parallel/\hbar} \approx 5 \times 10^{-3}$ for a magnetic field of $1$ Tesla. 
% \ajm{\bf AJM Comment: need an estimate here--interplane distance is very small so how much flux do you really need to get an effect?}. 
On the other hand, a magnetic field $\bm B_z$ perpendicular to the bilayer strongly suppresses the intervalley symmetric superconducting state.

The prediction of a chiral superconducting ground state tunable with displacement field will allow for the observation of the superconducting diode effect (SDE) \cite{nadeem_SDE,ando_SDE}. The state predicted here realizes the minimal requirements by breaking in-plane rotational symmetries, inversion, and time reversal in the superconducting state. This ground state differs from the scenario of a Josephson diode~\cite{Davydova_JDE,Volkov_2024}, as the intrinsic effect (junction-free) probes the symmetry breaking of the superconducting ground state. Time-reversal symmetry breaking in the normal state (a necessary condition) can be achieved through via a Zeeman field\cite{scammell2022theory} or the introduction of a magnetic substrate ~\cite{narita_fieldfree}. 
The diode effect manifests in the measurement of the $j_{+}$, $j_{-}$ which stand for the forward current (along the $\bm a_1$ direction, for example) and reverse current, respectively. The diode efficiency $\eta = (|j_{+}|-|j_{-}|)/(|j_{+}|+|j_{-}|) \neq 0$ can then directly estimate the degree of TRS breaking and the size of the chiral gap that is formed. Recent proposals for the detection of the SDE without reaching critical currents may apply here as well~\cite{davydova_nonrecip_super}.

% \ajm{\bf AJM Comment: this last sentence is logically separate from the B field discussion; should have a second paragraph; also expand it a little bit--how do you test the mechanism--what is the prediction as the interlayer distance/nature of gating potential changed?? You say above that the dependence on displacement field is not monotonic, why should the dependence on the gating be monotonic. Also do we predict a change from E to A rep?}  
The importance of our purely electronic mechanism for superconductivity in tWSe$_2$ can be directly tested by controlling the Coulomb interaction via the distance from metallic gates. Increasing this distance should enhance the attractive interaction, thereby increasing $T_c$. Finally, incorporating phonons could induce attraction in the $s$-wave channel, which is $C_{3z}$-invariant, thereby enhancing the $A^-$ superconducting instability. Future research directions include a detailed analysis of retardation effects and competing instabilities.

{\it Acknowledgement.---} We acknowledge insightful discussions with Jie Wang, Kun Chen, Christophe Mora, Jen Cano, Yongxin Zeng, Olivier Gauthe, Eva Andrei, Harley Scammell, Sankar Das Sarma, Valentin Cr\'epel and Daniel Mu\~{n}oz-Segovia. 
This work has been supported in part by the NSF CAREER Grant No.~DMR 1941569 (J.H.P.).
D.K. is supported by the Abrahams Postdoctoral fellowship of the Center for Materials Theory, Rutgers University as well as the Zuckerman STEM fellowship.
A.J.M. acknowledges support from the Energy Frontier Research Center on Programmable Quantum Materials funded by the US Department of Energy office of Science, basic energy sciences, under award DESC0019443. 
This work was  performed  in part at the Aspen Center for Physics, which is supported by National Science Foundation grant PHY-2210452 (J.H.P.).
This research was supported in part by grant NSF PHY-2309135 to the Kavli Institute for Theoretical Physics (KITP).
The Flatiron Institute is a division of the Simons Foundation.

\bibliographystyle{apsrev4-2}
\bibliography{biblio.bib}

\begin{thebibliography}{131}%
\makeatletter
\providecommand \@ifxundefined [1]{%
 \@ifx{#1\undefined}
}%
\providecommand \@ifnum [1]{%
 \ifnum #1\expandafter \@firstoftwo
 \else \expandafter \@secondoftwo
 \fi
}%
\providecommand \@ifx [1]{%
 \ifx #1\expandafter \@firstoftwo
 \else \expandafter \@secondoftwo
 \fi
}%
\providecommand \natexlab [1]{#1}%
\providecommand \enquote  [1]{``#1''}%
\providecommand \bibnamefont  [1]{#1}%
\providecommand \bibfnamefont [1]{#1}%
\providecommand \citenamefont [1]{#1}%
\providecommand \href@noop [0]{\@secondoftwo}%
\providecommand \href [0]{\begingroup \@sanitize@url \@href}%
\providecommand \@href[1]{\@@startlink{#1}\@@href}%
\providecommand \@@href[1]{\endgroup#1\@@endlink}%
\providecommand \@sanitize@url [0]{\catcode `\\12\catcode `\$12\catcode
  `\&12\catcode `\#12\catcode `\^12\catcode `\_12\catcode `\%12\relax}%
\providecommand \@@startlink[1]{}%
\providecommand \@@endlink[0]{}%
\providecommand \url  [0]{\begingroup\@sanitize@url \@url }%
\providecommand \@url [1]{\endgroup\@href {#1}{\urlprefix }}%
\providecommand \urlprefix  [0]{URL }%
\providecommand \Eprint [0]{\href }%
\providecommand \doibase [0]{https://doi.org/}%
\providecommand \selectlanguage [0]{\@gobble}%
\providecommand \bibinfo  [0]{\@secondoftwo}%
\providecommand \bibfield  [0]{\@secondoftwo}%
\providecommand \translation [1]{[#1]}%
\providecommand \BibitemOpen [0]{}%
\providecommand \bibitemStop [0]{}%
\providecommand \bibitemNoStop [0]{.\EOS\space}%
\providecommand \EOS [0]{\spacefactor3000\relax}%
\providecommand \BibitemShut  [1]{\csname bibitem#1\endcsname}%
\let\auto@bib@innerbib\@empty
%</preamble>
\bibitem [{\citenamefont {Migdal}(1958)}]{Migdal_1958}%
  \BibitemOpen
  \bibfield  {author} {\bibinfo {author} {\bibfnamefont {A.~B.}\ \bibnamefont
  {Migdal}},\ }\bibfield  {journal} {\bibinfo  {journal} {JETP}\ }\href
  {https://www.osti.gov/biblio/4296936} {} (\bibinfo {year} {1958})\BibitemShut
  {NoStop}%
\bibitem [{\citenamefont {McMillan}\ and\ \citenamefont
  {Rowell}(1965)}]{McMillan_1965}%
  \BibitemOpen
  \bibfield  {author} {\bibinfo {author} {\bibfnamefont {W.~L.}\ \bibnamefont
  {McMillan}}\ and\ \bibinfo {author} {\bibfnamefont {J.~M.}\ \bibnamefont
  {Rowell}},\ }\href {https://doi.org/10.1103/PhysRevLett.14.108} {\bibfield
  {journal} {\bibinfo  {journal} {Phys. Rev. Lett.}\ }\textbf {\bibinfo
  {volume} {14}},\ \bibinfo {pages} {108} (\bibinfo {year} {1965})}\BibitemShut
  {NoStop}%
\bibitem [{\citenamefont {Chubukov}\ \emph {et~al.}(2020)\citenamefont
  {Chubukov}, \citenamefont {Abanov}, \citenamefont {Esterlis},\ and\
  \citenamefont {Kivelson}}]{chubukov2020eliashberg}%
  \BibitemOpen
  \bibfield  {author} {\bibinfo {author} {\bibfnamefont {A.~V.}\ \bibnamefont
  {Chubukov}}, \bibinfo {author} {\bibfnamefont {A.}~\bibnamefont {Abanov}},
  \bibinfo {author} {\bibfnamefont {I.}~\bibnamefont {Esterlis}},\ and\
  \bibinfo {author} {\bibfnamefont {S.~A.}\ \bibnamefont {Kivelson}},\
  }\href@noop {} {\bibfield  {journal} {\bibinfo  {journal} {Annals of
  Physics}\ }\textbf {\bibinfo {volume} {417}},\ \bibinfo {pages} {168190}
  (\bibinfo {year} {2020})}\BibitemShut {NoStop}%
\bibitem [{\citenamefont {Kohn}\ and\ \citenamefont
  {Luttinger}(1965)}]{kohnluttingerSC_prl}%
  \BibitemOpen
  \bibfield  {author} {\bibinfo {author} {\bibfnamefont {W.}~\bibnamefont
  {Kohn}}\ and\ \bibinfo {author} {\bibfnamefont {J.~M.}\ \bibnamefont
  {Luttinger}},\ }\href {https://doi.org/10.1103/PhysRevLett.15.524} {\bibfield
   {journal} {\bibinfo  {journal} {Phys. Rev. Lett.}\ }\textbf {\bibinfo
  {volume} {15}},\ \bibinfo {pages} {524} (\bibinfo {year} {1965})}\BibitemShut
  {NoStop}%
\bibitem [{\citenamefont {Chubukov}(1993)}]{chubukov_1993}%
  \BibitemOpen
  \bibfield  {author} {\bibinfo {author} {\bibfnamefont {A.~V.}\ \bibnamefont
  {Chubukov}},\ }\href {https://doi.org/10.1103/PhysRevB.48.1097} {\bibfield
  {journal} {\bibinfo  {journal} {Phys. Rev. B}\ }\textbf {\bibinfo {volume}
  {48}},\ \bibinfo {pages} {1097} (\bibinfo {year} {1993})}\BibitemShut
  {NoStop}%
\bibitem [{\citenamefont {Maiti}\ and\ \citenamefont
  {Chubukov}(2013)}]{maiti2013superconductivity}%
  \BibitemOpen
  \bibfield  {author} {\bibinfo {author} {\bibfnamefont {S.}~\bibnamefont
  {Maiti}}\ and\ \bibinfo {author} {\bibfnamefont {A.~V.}\ \bibnamefont
  {Chubukov}},\ }in\ \href@noop {} {\emph {\bibinfo {booktitle} {AIP Conference
  Proceedings}}},\ Vol.\ \bibinfo {volume} {1550}\ (\bibinfo {organization}
  {American Institute of Physics},\ \bibinfo {year} {2013})\ pp.\ \bibinfo
  {pages} {3--73}\BibitemShut {NoStop}%
\bibitem [{\citenamefont {Raghu}\ \emph {et~al.}(2010)\citenamefont {Raghu},
  \citenamefont {Kivelson},\ and\ \citenamefont {Scalapino}}]{Raghu2010}%
  \BibitemOpen
  \bibfield  {author} {\bibinfo {author} {\bibfnamefont {S.}~\bibnamefont
  {Raghu}}, \bibinfo {author} {\bibfnamefont {S.~A.}\ \bibnamefont
  {Kivelson}},\ and\ \bibinfo {author} {\bibfnamefont {D.~J.}\ \bibnamefont
  {Scalapino}},\ }\href {https://doi.org/10.1103/PhysRevB.81.224505} {\bibfield
   {journal} {\bibinfo  {journal} {Phys. Rev. B}\ }\textbf {\bibinfo {volume}
  {81}},\ \bibinfo {pages} {224505} (\bibinfo {year} {2010})}\BibitemShut
  {NoStop}%
\bibitem [{\citenamefont {Andrei}\ and\ \citenamefont
  {MacDonald}(2020)}]{Andrei_2020}%
  \BibitemOpen
  \bibfield  {author} {\bibinfo {author} {\bibfnamefont {E.~Y.}\ \bibnamefont
  {Andrei}}\ and\ \bibinfo {author} {\bibfnamefont {A.~H.}\ \bibnamefont
  {MacDonald}},\ }\href {https://doi.org/10.1038/s41563-020-00840-0} {\bibfield
   {journal} {\bibinfo  {journal} {Nature Materials}\ }\textbf {\bibinfo
  {volume} {19}},\ \bibinfo {pages} {1265–1275} (\bibinfo {year}
  {2020})}\BibitemShut {NoStop}%
\bibitem [{\citenamefont {Andrei}\ \emph {et~al.}(2021)\citenamefont {Andrei},
  \citenamefont {Efetov}, \citenamefont {Jarillo-Herrero}, \citenamefont
  {MacDonald}, \citenamefont {Mak}, \citenamefont {Senthil}, \citenamefont
  {Tutuc}, \citenamefont {Yazdani},\ and\ \citenamefont {Young}}]{Andrei_2021}%
  \BibitemOpen
  \bibfield  {author} {\bibinfo {author} {\bibfnamefont {E.~Y.}\ \bibnamefont
  {Andrei}}, \bibinfo {author} {\bibfnamefont {D.~K.}\ \bibnamefont {Efetov}},
  \bibinfo {author} {\bibfnamefont {P.}~\bibnamefont {Jarillo-Herrero}},
  \bibinfo {author} {\bibfnamefont {A.~H.}\ \bibnamefont {MacDonald}}, \bibinfo
  {author} {\bibfnamefont {K.~F.}\ \bibnamefont {Mak}}, \bibinfo {author}
  {\bibfnamefont {T.}~\bibnamefont {Senthil}}, \bibinfo {author} {\bibfnamefont
  {E.}~\bibnamefont {Tutuc}}, \bibinfo {author} {\bibfnamefont
  {A.}~\bibnamefont {Yazdani}},\ and\ \bibinfo {author} {\bibfnamefont {A.~F.}\
  \bibnamefont {Young}},\ }\href@noop {} {\bibfield  {journal} {\bibinfo
  {journal} {Nature Reviews Materials}\ }\textbf {\bibinfo {volume} {6}},\
  \bibinfo {pages} {201} (\bibinfo {year} {2021})}\BibitemShut {NoStop}%
\bibitem [{\citenamefont {Kennes}\ \emph {et~al.}(2021)\citenamefont {Kennes},
  \citenamefont {Claassen}, \citenamefont {Xian}, \citenamefont {Georges},
  \citenamefont {Millis}, \citenamefont {Hone}, \citenamefont {Dean},
  \citenamefont {Basov}, \citenamefont {Pasupathy},\ and\ \citenamefont
  {Rubio}}]{Kennes_2021}%
  \BibitemOpen
  \bibfield  {author} {\bibinfo {author} {\bibfnamefont {D.~M.}\ \bibnamefont
  {Kennes}}, \bibinfo {author} {\bibfnamefont {M.}~\bibnamefont {Claassen}},
  \bibinfo {author} {\bibfnamefont {L.}~\bibnamefont {Xian}}, \bibinfo {author}
  {\bibfnamefont {A.}~\bibnamefont {Georges}}, \bibinfo {author} {\bibfnamefont
  {A.~J.}\ \bibnamefont {Millis}}, \bibinfo {author} {\bibfnamefont
  {J.}~\bibnamefont {Hone}}, \bibinfo {author} {\bibfnamefont {C.~R.}\
  \bibnamefont {Dean}}, \bibinfo {author} {\bibfnamefont {D.~N.}\ \bibnamefont
  {Basov}}, \bibinfo {author} {\bibfnamefont {A.~N.}\ \bibnamefont
  {Pasupathy}},\ and\ \bibinfo {author} {\bibfnamefont {A.}~\bibnamefont
  {Rubio}},\ }\href {https://doi.org/10.1038/s41567-020-01154-3} {\bibfield
  {journal} {\bibinfo  {journal} {Nature Physics}\ }\textbf {\bibinfo {volume}
  {17}},\ \bibinfo {pages} {155} (\bibinfo {year} {2021})}\BibitemShut
  {NoStop}%
\bibitem [{\citenamefont {Cao}\ \emph {et~al.}(2018)\citenamefont {Cao},
  \citenamefont {Fatemi}, \citenamefont {Fang}, \citenamefont {Watanabe},
  \citenamefont {Taniguchi}, \citenamefont {Kaxiras},\ and\ \citenamefont
  {Jarillo-Herrero}}]{Cao_2018}%
  \BibitemOpen
  \bibfield  {author} {\bibinfo {author} {\bibfnamefont {Y.}~\bibnamefont
  {Cao}}, \bibinfo {author} {\bibfnamefont {V.}~\bibnamefont {Fatemi}},
  \bibinfo {author} {\bibfnamefont {S.}~\bibnamefont {Fang}}, \bibinfo {author}
  {\bibfnamefont {K.}~\bibnamefont {Watanabe}}, \bibinfo {author}
  {\bibfnamefont {T.}~\bibnamefont {Taniguchi}}, \bibinfo {author}
  {\bibfnamefont {E.}~\bibnamefont {Kaxiras}},\ and\ \bibinfo {author}
  {\bibfnamefont {P.}~\bibnamefont {Jarillo-Herrero}},\ }\href
  {https://doi.org/10.1038/nature26160} {\bibfield  {journal} {\bibinfo
  {journal} {Nature}\ }\textbf {\bibinfo {volume} {556}},\ \bibinfo {pages}
  {43} (\bibinfo {year} {2018})}\BibitemShut {NoStop}%
\bibitem [{\citenamefont {Yankowitz}\ \emph {et~al.}(2019)\citenamefont
  {Yankowitz}, \citenamefont {Chen}, \citenamefont {Polshyn}, \citenamefont
  {Zhang}, \citenamefont {Watanabe}, \citenamefont {Taniguchi}, \citenamefont
  {Graf}, \citenamefont {Young},\ and\ \citenamefont {Dean}}]{Yankowitz_2019}%
  \BibitemOpen
  \bibfield  {author} {\bibinfo {author} {\bibfnamefont {M.}~\bibnamefont
  {Yankowitz}}, \bibinfo {author} {\bibfnamefont {S.}~\bibnamefont {Chen}},
  \bibinfo {author} {\bibfnamefont {H.}~\bibnamefont {Polshyn}}, \bibinfo
  {author} {\bibfnamefont {Y.}~\bibnamefont {Zhang}}, \bibinfo {author}
  {\bibfnamefont {K.}~\bibnamefont {Watanabe}}, \bibinfo {author}
  {\bibfnamefont {T.}~\bibnamefont {Taniguchi}}, \bibinfo {author}
  {\bibfnamefont {D.}~\bibnamefont {Graf}}, \bibinfo {author} {\bibfnamefont
  {A.~F.}\ \bibnamefont {Young}},\ and\ \bibinfo {author} {\bibfnamefont
  {C.~R.}\ \bibnamefont {Dean}},\ }\href
  {https://doi.org/10.1126/science.aav1910} {\bibfield  {journal} {\bibinfo
  {journal} {Science}\ }\textbf {\bibinfo {volume} {363}},\ \bibinfo {pages}
  {1059} (\bibinfo {year} {2019})}\BibitemShut {NoStop}%
\bibitem [{\citenamefont {Lu}\ \emph {et~al.}(2019)\citenamefont {Lu},
  \citenamefont {Stepanov}, \citenamefont {Yang}, \citenamefont {Xie},
  \citenamefont {Aamir}, \citenamefont {Das}, \citenamefont {Urgell},
  \citenamefont {Watanabe}, \citenamefont {Taniguchi}, \citenamefont {Zhang},\
  and\ \citenamefont {et~al.}}]{Lu_2019}%
  \BibitemOpen
  \bibfield  {author} {\bibinfo {author} {\bibfnamefont {X.}~\bibnamefont
  {Lu}}, \bibinfo {author} {\bibfnamefont {P.}~\bibnamefont {Stepanov}},
  \bibinfo {author} {\bibfnamefont {W.}~\bibnamefont {Yang}}, \bibinfo {author}
  {\bibfnamefont {M.}~\bibnamefont {Xie}}, \bibinfo {author} {\bibfnamefont
  {M.~A.}\ \bibnamefont {Aamir}}, \bibinfo {author} {\bibfnamefont
  {I.}~\bibnamefont {Das}}, \bibinfo {author} {\bibfnamefont {C.}~\bibnamefont
  {Urgell}}, \bibinfo {author} {\bibfnamefont {K.}~\bibnamefont {Watanabe}},
  \bibinfo {author} {\bibfnamefont {T.}~\bibnamefont {Taniguchi}}, \bibinfo
  {author} {\bibfnamefont {G.}~\bibnamefont {Zhang}},\ and\ \bibinfo {author}
  {\bibnamefont {et~al.}},\ }\href {https://doi.org/10.1038/s41586-019-1695-0}
  {\bibfield  {journal} {\bibinfo  {journal} {Nature}\ }\textbf {\bibinfo
  {volume} {574}},\ \bibinfo {pages} {653} (\bibinfo {year}
  {2019})}\BibitemShut {NoStop}%
\bibitem [{\citenamefont {Arora}\ \emph {et~al.}(2020)\citenamefont {Arora},
  \citenamefont {Polski}, \citenamefont {Zhang}, \citenamefont {Thomson},
  \citenamefont {Choi}, \citenamefont {Kim}, \citenamefont {Lin}, \citenamefont
  {Wilson}, \citenamefont {Xu}, \citenamefont {Chu} \emph
  {et~al.}}]{arora2020superconductivity}%
  \BibitemOpen
  \bibfield  {author} {\bibinfo {author} {\bibfnamefont {H.~S.}\ \bibnamefont
  {Arora}}, \bibinfo {author} {\bibfnamefont {R.}~\bibnamefont {Polski}},
  \bibinfo {author} {\bibfnamefont {Y.}~\bibnamefont {Zhang}}, \bibinfo
  {author} {\bibfnamefont {A.}~\bibnamefont {Thomson}}, \bibinfo {author}
  {\bibfnamefont {Y.}~\bibnamefont {Choi}}, \bibinfo {author} {\bibfnamefont
  {H.}~\bibnamefont {Kim}}, \bibinfo {author} {\bibfnamefont {Z.}~\bibnamefont
  {Lin}}, \bibinfo {author} {\bibfnamefont {I.~Z.}\ \bibnamefont {Wilson}},
  \bibinfo {author} {\bibfnamefont {X.}~\bibnamefont {Xu}}, \bibinfo {author}
  {\bibfnamefont {J.-H.}\ \bibnamefont {Chu}}, \emph {et~al.},\ }\href@noop {}
  {\bibfield  {journal} {\bibinfo  {journal} {Nature}\ }\textbf {\bibinfo
  {volume} {583}},\ \bibinfo {pages} {379} (\bibinfo {year}
  {2020})}\BibitemShut {NoStop}%
\bibitem [{\citenamefont {Saito}\ \emph {et~al.}(2020)\citenamefont {Saito},
  \citenamefont {Ge}, \citenamefont {Watanabe}, \citenamefont {Taniguchi},\
  and\ \citenamefont {Young}}]{saito2020independent}%
  \BibitemOpen
  \bibfield  {author} {\bibinfo {author} {\bibfnamefont {Y.}~\bibnamefont
  {Saito}}, \bibinfo {author} {\bibfnamefont {J.}~\bibnamefont {Ge}}, \bibinfo
  {author} {\bibfnamefont {K.}~\bibnamefont {Watanabe}}, \bibinfo {author}
  {\bibfnamefont {T.}~\bibnamefont {Taniguchi}},\ and\ \bibinfo {author}
  {\bibfnamefont {A.~F.}\ \bibnamefont {Young}},\ }\href@noop {} {\bibfield
  {journal} {\bibinfo  {journal} {Nature Physics}\ }\textbf {\bibinfo {volume}
  {16}},\ \bibinfo {pages} {926} (\bibinfo {year} {2020})}\BibitemShut
  {NoStop}%
\bibitem [{\citenamefont {Park}\ \emph {et~al.}(2021)\citenamefont {Park},
  \citenamefont {Cao}, \citenamefont {Watanabe}, \citenamefont {Taniguchi},\
  and\ \citenamefont {Jarillo-Herrero}}]{Park_2021}%
  \BibitemOpen
  \bibfield  {author} {\bibinfo {author} {\bibfnamefont {J.~M.}\ \bibnamefont
  {Park}}, \bibinfo {author} {\bibfnamefont {Y.}~\bibnamefont {Cao}}, \bibinfo
  {author} {\bibfnamefont {K.}~\bibnamefont {Watanabe}}, \bibinfo {author}
  {\bibfnamefont {T.}~\bibnamefont {Taniguchi}},\ and\ \bibinfo {author}
  {\bibfnamefont {P.}~\bibnamefont {Jarillo-Herrero}},\ }\href
  {https://doi.org/10.1038/s41586-021-03192-0} {\bibfield  {journal} {\bibinfo
  {journal} {Nature}\ }\textbf {\bibinfo {volume} {590}},\ \bibinfo {pages}
  {249–255} (\bibinfo {year} {2021})}\BibitemShut {NoStop}%
\bibitem [{\citenamefont {Hao}\ \emph {et~al.}(2021)\citenamefont {Hao},
  \citenamefont {Zimmerman}, \citenamefont {Ledwith}, \citenamefont {Khalaf},
  \citenamefont {Najafabadi}, \citenamefont {Watanabe}, \citenamefont
  {Taniguchi}, \citenamefont {Vishwanath},\ and\ \citenamefont
  {Kim}}]{Hao_2021}%
  \BibitemOpen
  \bibfield  {author} {\bibinfo {author} {\bibfnamefont {Z.}~\bibnamefont
  {Hao}}, \bibinfo {author} {\bibfnamefont {A.~M.}\ \bibnamefont {Zimmerman}},
  \bibinfo {author} {\bibfnamefont {P.}~\bibnamefont {Ledwith}}, \bibinfo
  {author} {\bibfnamefont {E.}~\bibnamefont {Khalaf}}, \bibinfo {author}
  {\bibfnamefont {D.~H.}\ \bibnamefont {Najafabadi}}, \bibinfo {author}
  {\bibfnamefont {K.}~\bibnamefont {Watanabe}}, \bibinfo {author}
  {\bibfnamefont {T.}~\bibnamefont {Taniguchi}}, \bibinfo {author}
  {\bibfnamefont {A.}~\bibnamefont {Vishwanath}},\ and\ \bibinfo {author}
  {\bibfnamefont {P.}~\bibnamefont {Kim}},\ }\href
  {https://doi.org/10.1126/science.abg0399} {\bibfield  {journal} {\bibinfo
  {journal} {Science}\ }\textbf {\bibinfo {volume} {371}},\ \bibinfo {pages}
  {1133–1138} (\bibinfo {year} {2021})}\BibitemShut {NoStop}%
\bibitem [{\citenamefont {Zhou}\ \emph {et~al.}(2021)\citenamefont {Zhou},
  \citenamefont {Xie}, \citenamefont {Taniguchi}, \citenamefont {Watanabe},\
  and\ \citenamefont {Young}}]{Zhou_2021}%
  \BibitemOpen
  \bibfield  {author} {\bibinfo {author} {\bibfnamefont {H.}~\bibnamefont
  {Zhou}}, \bibinfo {author} {\bibfnamefont {T.}~\bibnamefont {Xie}}, \bibinfo
  {author} {\bibfnamefont {T.}~\bibnamefont {Taniguchi}}, \bibinfo {author}
  {\bibfnamefont {K.}~\bibnamefont {Watanabe}},\ and\ \bibinfo {author}
  {\bibfnamefont {A.~F.}\ \bibnamefont {Young}},\ }\href
  {https://doi.org/10.1038/s41586-021-03926-0} {\bibfield  {journal} {\bibinfo
  {journal} {Nature}\ }\textbf {\bibinfo {volume} {598}},\ \bibinfo {pages}
  {434–438} (\bibinfo {year} {2021})}\BibitemShut {NoStop}%
\bibitem [{\citenamefont {Park}\ \emph {et~al.}(2022)\citenamefont {Park},
  \citenamefont {Cao}, \citenamefont {Xia}, \citenamefont {Sun}, \citenamefont
  {Watanabe}, \citenamefont {Taniguchi},\ and\ \citenamefont
  {Jarillo-Herrero}}]{park2022robust}%
  \BibitemOpen
  \bibfield  {author} {\bibinfo {author} {\bibfnamefont {J.~M.}\ \bibnamefont
  {Park}}, \bibinfo {author} {\bibfnamefont {Y.}~\bibnamefont {Cao}}, \bibinfo
  {author} {\bibfnamefont {L.-Q.}\ \bibnamefont {Xia}}, \bibinfo {author}
  {\bibfnamefont {S.}~\bibnamefont {Sun}}, \bibinfo {author} {\bibfnamefont
  {K.}~\bibnamefont {Watanabe}}, \bibinfo {author} {\bibfnamefont
  {T.}~\bibnamefont {Taniguchi}},\ and\ \bibinfo {author} {\bibfnamefont
  {P.}~\bibnamefont {Jarillo-Herrero}},\ }\href@noop {} {\bibfield  {journal}
  {\bibinfo  {journal} {Nature Materials}\ }\textbf {\bibinfo {volume} {21}},\
  \bibinfo {pages} {877} (\bibinfo {year} {2022})}\BibitemShut {NoStop}%
\bibitem [{\citenamefont {Burg}\ \emph {et~al.}(2022)\citenamefont {Burg},
  \citenamefont {Khalaf}, \citenamefont {Wang}, \citenamefont {Watanabe},
  \citenamefont {Taniguchi},\ and\ \citenamefont {Tutuc}}]{burg2022emergence}%
  \BibitemOpen
  \bibfield  {author} {\bibinfo {author} {\bibfnamefont {G.~W.}\ \bibnamefont
  {Burg}}, \bibinfo {author} {\bibfnamefont {E.}~\bibnamefont {Khalaf}},
  \bibinfo {author} {\bibfnamefont {Y.}~\bibnamefont {Wang}}, \bibinfo {author}
  {\bibfnamefont {K.}~\bibnamefont {Watanabe}}, \bibinfo {author}
  {\bibfnamefont {T.}~\bibnamefont {Taniguchi}},\ and\ \bibinfo {author}
  {\bibfnamefont {E.}~\bibnamefont {Tutuc}},\ }\href@noop {} {\bibfield
  {journal} {\bibinfo  {journal} {Nature Materials}\ }\textbf {\bibinfo
  {volume} {21}},\ \bibinfo {pages} {884} (\bibinfo {year} {2022})}\BibitemShut
  {NoStop}%
\bibitem [{\citenamefont {Su}\ \emph {et~al.}(2023)\citenamefont {Su},
  \citenamefont {Kuiri}, \citenamefont {Watanabe}, \citenamefont {Taniguchi},\
  and\ \citenamefont {Folk}}]{su2023superconductivity}%
  \BibitemOpen
  \bibfield  {author} {\bibinfo {author} {\bibfnamefont {R.}~\bibnamefont
  {Su}}, \bibinfo {author} {\bibfnamefont {M.}~\bibnamefont {Kuiri}}, \bibinfo
  {author} {\bibfnamefont {K.}~\bibnamefont {Watanabe}}, \bibinfo {author}
  {\bibfnamefont {T.}~\bibnamefont {Taniguchi}},\ and\ \bibinfo {author}
  {\bibfnamefont {J.}~\bibnamefont {Folk}},\ }\href@noop {} {\bibfield
  {journal} {\bibinfo  {journal} {Nature Materials}\ }\textbf {\bibinfo
  {volume} {22}},\ \bibinfo {pages} {1332} (\bibinfo {year}
  {2023})}\BibitemShut {NoStop}%
\bibitem [{\citenamefont {Wu}\ \emph {et~al.}(2018)\citenamefont {Wu},
  \citenamefont {MacDonald},\ and\ \citenamefont {Martin}}]{martin_2018}%
  \BibitemOpen
  \bibfield  {author} {\bibinfo {author} {\bibfnamefont {F.}~\bibnamefont
  {Wu}}, \bibinfo {author} {\bibfnamefont {A.~H.}\ \bibnamefont {MacDonald}},\
  and\ \bibinfo {author} {\bibfnamefont {I.}~\bibnamefont {Martin}},\ }\href
  {https://doi.org/10.1103/PhysRevLett.121.257001} {\bibfield  {journal}
  {\bibinfo  {journal} {Phys. Rev. Lett.}\ }\textbf {\bibinfo {volume} {121}},\
  \bibinfo {pages} {257001} (\bibinfo {year} {2018})}\BibitemShut {NoStop}%
\bibitem [{\citenamefont {Angeli}\ \emph {et~al.}(2019)\citenamefont {Angeli},
  \citenamefont {Tosatti},\ and\ \citenamefont {Fabrizio}}]{mattia_2019}%
  \BibitemOpen
  \bibfield  {author} {\bibinfo {author} {\bibfnamefont {M.}~\bibnamefont
  {Angeli}}, \bibinfo {author} {\bibfnamefont {E.}~\bibnamefont {Tosatti}},\
  and\ \bibinfo {author} {\bibfnamefont {M.}~\bibnamefont {Fabrizio}},\ }\href
  {https://doi.org/10.1103/PhysRevX.9.041010} {\bibfield  {journal} {\bibinfo
  {journal} {Phys. Rev. X}\ }\textbf {\bibinfo {volume} {9}},\ \bibinfo {pages}
  {041010} (\bibinfo {year} {2019})}\BibitemShut {NoStop}%
\bibitem [{\citenamefont {Angeli}\ and\ \citenamefont
  {Fabrizio}(2020)}]{angeli2020jahntellercouplingmoirephonons}%
  \BibitemOpen
  \bibfield  {author} {\bibinfo {author} {\bibfnamefont {M.}~\bibnamefont
  {Angeli}}\ and\ \bibinfo {author} {\bibfnamefont {M.}~\bibnamefont
  {Fabrizio}},\ }\href {https://arxiv.org/abs/2007.01048} {\bibinfo {title}
  {Jahn-teller coupling to moir\`e phonons in the continuum model formalism for
  small angle twisted bilayer graphene}} (\bibinfo {year} {2020}),\ \Eprint
  {https://arxiv.org/abs/2007.01048} {arXiv:2007.01048 [cond-mat.supr-con]}
  \BibitemShut {NoStop}%
\bibitem [{\citenamefont {Lian}\ \emph {et~al.}(2019)\citenamefont {Lian},
  \citenamefont {Wang},\ and\ \citenamefont {Bernevig}}]{biao_2019}%
  \BibitemOpen
  \bibfield  {author} {\bibinfo {author} {\bibfnamefont {B.}~\bibnamefont
  {Lian}}, \bibinfo {author} {\bibfnamefont {Z.}~\bibnamefont {Wang}},\ and\
  \bibinfo {author} {\bibfnamefont {B.~A.}\ \bibnamefont {Bernevig}},\ }\href
  {https://doi.org/10.1103/PhysRevLett.122.257002} {\bibfield  {journal}
  {\bibinfo  {journal} {Phys. Rev. Lett.}\ }\textbf {\bibinfo {volume} {122}},\
  \bibinfo {pages} {257002} (\bibinfo {year} {2019})}\BibitemShut {NoStop}%
\bibitem [{\citenamefont {Wu}\ \emph {et~al.}(2019{\natexlab{a}})\citenamefont
  {Wu}, \citenamefont {Hwang},\ and\ \citenamefont {Das~Sarma}}]{FWu_prb_2019}%
  \BibitemOpen
  \bibfield  {author} {\bibinfo {author} {\bibfnamefont {F.}~\bibnamefont
  {Wu}}, \bibinfo {author} {\bibfnamefont {E.}~\bibnamefont {Hwang}},\ and\
  \bibinfo {author} {\bibfnamefont {S.}~\bibnamefont {Das~Sarma}},\ }\href
  {https://doi.org/10.1103/PhysRevB.99.165112} {\bibfield  {journal} {\bibinfo
  {journal} {Phys. Rev. B}\ }\textbf {\bibinfo {volume} {99}},\ \bibinfo
  {pages} {165112} (\bibinfo {year} {2019}{\natexlab{a}})}\BibitemShut
  {NoStop}%
\bibitem [{\citenamefont {Shavit}\ \emph {et~al.}(2021)\citenamefont {Shavit},
  \citenamefont {Berg}, \citenamefont {Stern},\ and\ \citenamefont
  {Oreg}}]{prl_yoreg_2021}%
  \BibitemOpen
  \bibfield  {author} {\bibinfo {author} {\bibfnamefont {G.}~\bibnamefont
  {Shavit}}, \bibinfo {author} {\bibfnamefont {E.}~\bibnamefont {Berg}},
  \bibinfo {author} {\bibfnamefont {A.}~\bibnamefont {Stern}},\ and\ \bibinfo
  {author} {\bibfnamefont {Y.}~\bibnamefont {Oreg}},\ }\href
  {https://doi.org/10.1103/PhysRevLett.127.247703} {\bibfield  {journal}
  {\bibinfo  {journal} {Phys. Rev. Lett.}\ }\textbf {\bibinfo {volume} {127}},\
  \bibinfo {pages} {247703} (\bibinfo {year} {2021})}\BibitemShut {NoStop}%
\bibitem [{\citenamefont {Blason}\ and\ \citenamefont
  {Fabrizio}(2022)}]{blason_2022}%
  \BibitemOpen
  \bibfield  {author} {\bibinfo {author} {\bibfnamefont {A.}~\bibnamefont
  {Blason}}\ and\ \bibinfo {author} {\bibfnamefont {M.}~\bibnamefont
  {Fabrizio}},\ }\href {https://doi.org/10.1103/PhysRevB.106.235112} {\bibfield
   {journal} {\bibinfo  {journal} {Phys. Rev. B}\ }\textbf {\bibinfo {volume}
  {106}},\ \bibinfo {pages} {235112} (\bibinfo {year} {2022})}\BibitemShut
  {NoStop}%
\bibitem [{\citenamefont {Gadelha}\ \emph {et~al.}(2021)\citenamefont
  {Gadelha}, \citenamefont {Ohlberg}, \citenamefont {Rabelo}, \citenamefont
  {Neto}, \citenamefont {Vasconcelos}, \citenamefont {Campos}, \citenamefont
  {Lemos}, \citenamefont {Ornelas}, \citenamefont {Miranda}, \citenamefont
  {Nadas} \emph {et~al.}}]{gadelha2021localization}%
  \BibitemOpen
  \bibfield  {author} {\bibinfo {author} {\bibfnamefont {A.~C.}\ \bibnamefont
  {Gadelha}}, \bibinfo {author} {\bibfnamefont {D.~A.}\ \bibnamefont
  {Ohlberg}}, \bibinfo {author} {\bibfnamefont {C.}~\bibnamefont {Rabelo}},
  \bibinfo {author} {\bibfnamefont {E.~G.}\ \bibnamefont {Neto}}, \bibinfo
  {author} {\bibfnamefont {T.~L.}\ \bibnamefont {Vasconcelos}}, \bibinfo
  {author} {\bibfnamefont {J.~L.}\ \bibnamefont {Campos}}, \bibinfo {author}
  {\bibfnamefont {J.~S.}\ \bibnamefont {Lemos}}, \bibinfo {author}
  {\bibfnamefont {V.}~\bibnamefont {Ornelas}}, \bibinfo {author} {\bibfnamefont
  {D.}~\bibnamefont {Miranda}}, \bibinfo {author} {\bibfnamefont
  {R.}~\bibnamefont {Nadas}}, \emph {et~al.},\ }\href@noop {} {\bibfield
  {journal} {\bibinfo  {journal} {Nature}\ }\textbf {\bibinfo {volume} {590}},\
  \bibinfo {pages} {405} (\bibinfo {year} {2021})}\BibitemShut {NoStop}%
\bibitem [{\citenamefont {Isobe}\ \emph {et~al.}(2018)\citenamefont {Isobe},
  \citenamefont {Yuan},\ and\ \citenamefont {Fu}}]{PhysRevX.8.041041}%
  \BibitemOpen
  \bibfield  {author} {\bibinfo {author} {\bibfnamefont {H.}~\bibnamefont
  {Isobe}}, \bibinfo {author} {\bibfnamefont {N.~F.~Q.}\ \bibnamefont {Yuan}},\
  and\ \bibinfo {author} {\bibfnamefont {L.}~\bibnamefont {Fu}},\ }\href
  {https://doi.org/10.1103/PhysRevX.8.041041} {\bibfield  {journal} {\bibinfo
  {journal} {Phys. Rev. X}\ }\textbf {\bibinfo {volume} {8}},\ \bibinfo {pages}
  {041041} (\bibinfo {year} {2018})}\BibitemShut {NoStop}%
\bibitem [{\citenamefont {Khalaf}\ \emph {et~al.}(2021)\citenamefont {Khalaf},
  \citenamefont {Chatterjee}, \citenamefont {Bultinck}, \citenamefont
  {Zaletel},\ and\ \citenamefont {Vishwanath}}]{Khalaf_2021}%
  \BibitemOpen
  \bibfield  {author} {\bibinfo {author} {\bibfnamefont {E.}~\bibnamefont
  {Khalaf}}, \bibinfo {author} {\bibfnamefont {S.}~\bibnamefont {Chatterjee}},
  \bibinfo {author} {\bibfnamefont {N.}~\bibnamefont {Bultinck}}, \bibinfo
  {author} {\bibfnamefont {M.~P.}\ \bibnamefont {Zaletel}},\ and\ \bibinfo
  {author} {\bibfnamefont {A.}~\bibnamefont {Vishwanath}},\ }\bibfield
  {journal} {\bibinfo  {journal} {Science Advances}\ }\textbf {\bibinfo
  {volume} {7}},\ \href {https://doi.org/10.1126/sciadv.abf5299}
  {10.1126/sciadv.abf5299} (\bibinfo {year} {2021})\BibitemShut {NoStop}%
\bibitem [{\citenamefont {Shi}\ \emph {et~al.}(2024{\natexlab{a}})\citenamefont
  {Shi}, \citenamefont {Miao},\ and\ \citenamefont
  {Dai}}]{shi2024moireopticalphononsdancing}%
  \BibitemOpen
  \bibfield  {author} {\bibinfo {author} {\bibfnamefont {H.}~\bibnamefont
  {Shi}}, \bibinfo {author} {\bibfnamefont {W.}~\bibnamefont {Miao}},\ and\
  \bibinfo {author} {\bibfnamefont {X.}~\bibnamefont {Dai}},\ }\href
  {https://arxiv.org/abs/2402.11824} {\bibinfo {title} {Moir\'{e} optical
  phonons dancing with heavy electrons in magic-angle twisted bilayer
  graphene}} (\bibinfo {year} {2024}{\natexlab{a}}),\ \Eprint
  {https://arxiv.org/abs/2402.11824} {arXiv:2402.11824 [cond-mat.mes-hall]}
  \BibitemShut {NoStop}%
\bibitem [{\citenamefont {Wang}\ \emph
  {et~al.}(2024{\natexlab{a}})\citenamefont {Wang}, \citenamefont {Zhou},
  \citenamefont {Lian},\ and\ \citenamefont
  {Song}}]{wang2024electronphononcouplingtopological}%
  \BibitemOpen
  \bibfield  {author} {\bibinfo {author} {\bibfnamefont {Y.-J.}\ \bibnamefont
  {Wang}}, \bibinfo {author} {\bibfnamefont {G.-D.}\ \bibnamefont {Zhou}},
  \bibinfo {author} {\bibfnamefont {B.}~\bibnamefont {Lian}},\ and\ \bibinfo
  {author} {\bibfnamefont {Z.-D.}\ \bibnamefont {Song}},\ }\href
  {https://arxiv.org/abs/2407.11116} {\bibinfo {title} {Electron phonon
  coupling in the topological heavy fermion model of twisted bilayer graphene}}
  (\bibinfo {year} {2024}{\natexlab{a}}),\ \Eprint
  {https://arxiv.org/abs/2407.11116} {arXiv:2407.11116 [cond-mat.str-el]}
  \BibitemShut {NoStop}%
\bibitem [{\citenamefont {Liu}\ \emph {et~al.}(2023)\citenamefont {Liu},
  \citenamefont {Chen}, \citenamefont {Yazdani},\ and\ \citenamefont
  {Bernevig}}]{liu2023electronkphononinteractiontwistedbilayer}%
  \BibitemOpen
  \bibfield  {author} {\bibinfo {author} {\bibfnamefont {C.-X.}\ \bibnamefont
  {Liu}}, \bibinfo {author} {\bibfnamefont {Y.}~\bibnamefont {Chen}}, \bibinfo
  {author} {\bibfnamefont {A.}~\bibnamefont {Yazdani}},\ and\ \bibinfo {author}
  {\bibfnamefont {B.~A.}\ \bibnamefont {Bernevig}},\ }\href
  {https://arxiv.org/abs/2303.15551} {\bibinfo {title} {Electron-k-phonon
  interaction in twisted bilayer graphene}} (\bibinfo {year} {2023}),\ \Eprint
  {https://arxiv.org/abs/2303.15551} {arXiv:2303.15551 [cond-mat.supr-con]}
  \BibitemShut {NoStop}%
\bibitem [{\citenamefont {Nuckolls}\ \emph {et~al.}(2023)\citenamefont
  {Nuckolls}, \citenamefont {Lee}, \citenamefont {Oh}, \citenamefont {Wong},
  \citenamefont {Soejima}, \citenamefont {Hong}, \citenamefont {Călugăru},
  \citenamefont {Herzog-Arbeitman}, \citenamefont {Bernevig}, \citenamefont
  {Watanabe}, \citenamefont {Taniguchi}, \citenamefont {Regnault},
  \citenamefont {Zaletel},\ and\ \citenamefont {Yazdani}}]{Nuckolls_cdw_2023}%
  \BibitemOpen
  \bibfield  {author} {\bibinfo {author} {\bibfnamefont {K.~P.}\ \bibnamefont
  {Nuckolls}}, \bibinfo {author} {\bibfnamefont {R.~L.}\ \bibnamefont {Lee}},
  \bibinfo {author} {\bibfnamefont {M.}~\bibnamefont {Oh}}, \bibinfo {author}
  {\bibfnamefont {D.}~\bibnamefont {Wong}}, \bibinfo {author} {\bibfnamefont
  {T.}~\bibnamefont {Soejima}}, \bibinfo {author} {\bibfnamefont {J.~P.}\
  \bibnamefont {Hong}}, \bibinfo {author} {\bibfnamefont {D.}~\bibnamefont
  {Călugăru}}, \bibinfo {author} {\bibfnamefont {J.}~\bibnamefont
  {Herzog-Arbeitman}}, \bibinfo {author} {\bibfnamefont {B.~A.}\ \bibnamefont
  {Bernevig}}, \bibinfo {author} {\bibfnamefont {K.}~\bibnamefont {Watanabe}},
  \bibinfo {author} {\bibfnamefont {T.}~\bibnamefont {Taniguchi}}, \bibinfo
  {author} {\bibfnamefont {N.}~\bibnamefont {Regnault}}, \bibinfo {author}
  {\bibfnamefont {M.~P.}\ \bibnamefont {Zaletel}},\ and\ \bibinfo {author}
  {\bibfnamefont {A.}~\bibnamefont {Yazdani}},\ }\href
  {https://doi.org/10.1038/s41586-023-06226-x} {\bibfield  {journal} {\bibinfo
  {journal} {Nature}\ }\textbf {\bibinfo {volume} {620}},\ \bibinfo {pages}
  {525–532} (\bibinfo {year} {2023})}\BibitemShut {NoStop}%
\bibitem [{\citenamefont {Kwan}\ \emph {et~al.}(2023)\citenamefont {Kwan},
  \citenamefont {Wagner}, \citenamefont {Bultinck}, \citenamefont {Simon},
  \citenamefont {Berg},\ and\ \citenamefont
  {Parameswaran}}]{kwan2023electronphononcouplingcompetingkekule}%
  \BibitemOpen
  \bibfield  {author} {\bibinfo {author} {\bibfnamefont {Y.~H.}\ \bibnamefont
  {Kwan}}, \bibinfo {author} {\bibfnamefont {G.}~\bibnamefont {Wagner}},
  \bibinfo {author} {\bibfnamefont {N.}~\bibnamefont {Bultinck}}, \bibinfo
  {author} {\bibfnamefont {S.~H.}\ \bibnamefont {Simon}}, \bibinfo {author}
  {\bibfnamefont {E.}~\bibnamefont {Berg}},\ and\ \bibinfo {author}
  {\bibfnamefont {S.~A.}\ \bibnamefont {Parameswaran}},\ }\href
  {https://arxiv.org/abs/2303.13602} {\bibinfo {title} {Electron-phonon
  coupling and competing kekul\'e orders in twisted bilayer graphene}}
  (\bibinfo {year} {2023}),\ \Eprint {https://arxiv.org/abs/2303.13602}
  {arXiv:2303.13602 [cond-mat.str-el]} \BibitemShut {NoStop}%
\bibitem [{\citenamefont {Ingham}\ \emph {et~al.}(2023)\citenamefont {Ingham},
  \citenamefont {Li}, \citenamefont {Scheurer},\ and\ \citenamefont
  {Scammell}}]{ingham2023quadratic}%
  \BibitemOpen
  \bibfield  {author} {\bibinfo {author} {\bibfnamefont {J.}~\bibnamefont
  {Ingham}}, \bibinfo {author} {\bibfnamefont {T.}~\bibnamefont {Li}}, \bibinfo
  {author} {\bibfnamefont {M.~S.}\ \bibnamefont {Scheurer}},\ and\ \bibinfo
  {author} {\bibfnamefont {H.~D.}\ \bibnamefont {Scammell}},\ }\href@noop {}
  {\bibinfo {title} {Quadratic dirac fermions and the competition of ordered
  states in twisted bilayer graphene}} (\bibinfo {year} {2023}),\ \Eprint
  {https://arxiv.org/abs/2308.00748} {arXiv:2308.00748} \BibitemShut {NoStop}%
\bibitem [{\citenamefont {Xia}\ \emph {et~al.}(2024)\citenamefont {Xia},
  \citenamefont {Han}, \citenamefont {Watanabe}, \citenamefont {Taniguchi},
  \citenamefont {Shan},\ and\ \citenamefont {Mak}}]{xia2024unconventional}%
  \BibitemOpen
  \bibfield  {author} {\bibinfo {author} {\bibfnamefont {Y.}~\bibnamefont
  {Xia}}, \bibinfo {author} {\bibfnamefont {Z.}~\bibnamefont {Han}}, \bibinfo
  {author} {\bibfnamefont {K.}~\bibnamefont {Watanabe}}, \bibinfo {author}
  {\bibfnamefont {T.}~\bibnamefont {Taniguchi}}, \bibinfo {author}
  {\bibfnamefont {J.}~\bibnamefont {Shan}},\ and\ \bibinfo {author}
  {\bibfnamefont {K.~F.}\ \bibnamefont {Mak}},\ }\href@noop {} {\bibinfo
  {title} {Unconventional superconductivity in twisted bilayer wse2}} (\bibinfo
  {year} {2024}),\ \Eprint {https://arxiv.org/abs/2405.14784} {arXiv:2405.14784
  [cond-mat.mes-hall]} \BibitemShut {NoStop}%
\bibitem [{\citenamefont {Guo}\ \emph {et~al.}(2024)\citenamefont {Guo},
  \citenamefont {Pack}, \citenamefont {Swann}, \citenamefont {Holtzman},
  \citenamefont {Cothrine}, \citenamefont {Watanabe}, \citenamefont
  {Taniguchi}, \citenamefont {Mandrus}, \citenamefont {Barmak}, \citenamefont
  {Hone}, \citenamefont {Millis}, \citenamefont {Pasupathy},\ and\
  \citenamefont {Dean}}]{guo2024superconductivity}%
  \BibitemOpen
  \bibfield  {author} {\bibinfo {author} {\bibfnamefont {Y.}~\bibnamefont
  {Guo}}, \bibinfo {author} {\bibfnamefont {J.}~\bibnamefont {Pack}}, \bibinfo
  {author} {\bibfnamefont {J.}~\bibnamefont {Swann}}, \bibinfo {author}
  {\bibfnamefont {L.}~\bibnamefont {Holtzman}}, \bibinfo {author}
  {\bibfnamefont {M.}~\bibnamefont {Cothrine}}, \bibinfo {author}
  {\bibfnamefont {K.}~\bibnamefont {Watanabe}}, \bibinfo {author}
  {\bibfnamefont {T.}~\bibnamefont {Taniguchi}}, \bibinfo {author}
  {\bibfnamefont {D.}~\bibnamefont {Mandrus}}, \bibinfo {author} {\bibfnamefont
  {K.}~\bibnamefont {Barmak}}, \bibinfo {author} {\bibfnamefont
  {J.}~\bibnamefont {Hone}}, \bibinfo {author} {\bibfnamefont {A.~J.}\
  \bibnamefont {Millis}}, \bibinfo {author} {\bibfnamefont {A.~N.}\
  \bibnamefont {Pasupathy}},\ and\ \bibinfo {author} {\bibfnamefont {C.~R.}\
  \bibnamefont {Dean}},\ }\href@noop {} {\bibinfo {title} {Superconductivity in
  twisted bilayer wse$_2$}} (\bibinfo {year} {2024}),\ \Eprint
  {https://arxiv.org/abs/2406.03418} {arXiv:2406.03418 [cond-mat.mes-hall]}
  \BibitemShut {NoStop}%
\bibitem [{\citenamefont {Wang}\ \emph {et~al.}(2020)\citenamefont {Wang},
  \citenamefont {Shih}, \citenamefont {Ghiotto}, \citenamefont {Xian},
  \citenamefont {Rhodes}, \citenamefont {Tan}, \citenamefont {Claassen},
  \citenamefont {Kennes}, \citenamefont {Bai}, \citenamefont {Kim} \emph
  {et~al.}}]{wang2020correlated}%
  \BibitemOpen
  \bibfield  {author} {\bibinfo {author} {\bibfnamefont {L.}~\bibnamefont
  {Wang}}, \bibinfo {author} {\bibfnamefont {E.-M.}\ \bibnamefont {Shih}},
  \bibinfo {author} {\bibfnamefont {A.}~\bibnamefont {Ghiotto}}, \bibinfo
  {author} {\bibfnamefont {L.}~\bibnamefont {Xian}}, \bibinfo {author}
  {\bibfnamefont {D.~A.}\ \bibnamefont {Rhodes}}, \bibinfo {author}
  {\bibfnamefont {C.}~\bibnamefont {Tan}}, \bibinfo {author} {\bibfnamefont
  {M.}~\bibnamefont {Claassen}}, \bibinfo {author} {\bibfnamefont {D.~M.}\
  \bibnamefont {Kennes}}, \bibinfo {author} {\bibfnamefont {Y.}~\bibnamefont
  {Bai}}, \bibinfo {author} {\bibfnamefont {B.}~\bibnamefont {Kim}}, \emph
  {et~al.},\ }\href@noop {} {\bibfield  {journal} {\bibinfo  {journal} {Nature
  materials}\ }\textbf {\bibinfo {volume} {19}},\ \bibinfo {pages} {861}
  (\bibinfo {year} {2020})}\BibitemShut {NoStop}%
\bibitem [{\citenamefont {Regan}\ \emph {et~al.}(2020)\citenamefont {Regan},
  \citenamefont {Wang}, \citenamefont {Jin}, \citenamefont {Bakti~Utama},
  \citenamefont {Gao}, \citenamefont {Wei}, \citenamefont {Zhao}, \citenamefont
  {Zhao}, \citenamefont {Zhang}, \citenamefont {Yumigeta} \emph
  {et~al.}}]{regan2020mott}%
  \BibitemOpen
  \bibfield  {author} {\bibinfo {author} {\bibfnamefont {E.~C.}\ \bibnamefont
  {Regan}}, \bibinfo {author} {\bibfnamefont {D.}~\bibnamefont {Wang}},
  \bibinfo {author} {\bibfnamefont {C.}~\bibnamefont {Jin}}, \bibinfo {author}
  {\bibfnamefont {M.~I.}\ \bibnamefont {Bakti~Utama}}, \bibinfo {author}
  {\bibfnamefont {B.}~\bibnamefont {Gao}}, \bibinfo {author} {\bibfnamefont
  {X.}~\bibnamefont {Wei}}, \bibinfo {author} {\bibfnamefont {S.}~\bibnamefont
  {Zhao}}, \bibinfo {author} {\bibfnamefont {W.}~\bibnamefont {Zhao}}, \bibinfo
  {author} {\bibfnamefont {Z.}~\bibnamefont {Zhang}}, \bibinfo {author}
  {\bibfnamefont {K.}~\bibnamefont {Yumigeta}}, \emph {et~al.},\ }\href@noop {}
  {\bibfield  {journal} {\bibinfo  {journal} {Nature}\ }\textbf {\bibinfo
  {volume} {579}},\ \bibinfo {pages} {359} (\bibinfo {year}
  {2020})}\BibitemShut {NoStop}%
\bibitem [{\citenamefont {Tang}\ \emph {et~al.}(2020)\citenamefont {Tang},
  \citenamefont {Li}, \citenamefont {Li}, \citenamefont {Xu}, \citenamefont
  {Liu}, \citenamefont {Barmak}, \citenamefont {Watanabe}, \citenamefont
  {Taniguchi}, \citenamefont {MacDonald}, \citenamefont {Shan} \emph
  {et~al.}}]{tang2020simulation}%
  \BibitemOpen
  \bibfield  {author} {\bibinfo {author} {\bibfnamefont {Y.}~\bibnamefont
  {Tang}}, \bibinfo {author} {\bibfnamefont {L.}~\bibnamefont {Li}}, \bibinfo
  {author} {\bibfnamefont {T.}~\bibnamefont {Li}}, \bibinfo {author}
  {\bibfnamefont {Y.}~\bibnamefont {Xu}}, \bibinfo {author} {\bibfnamefont
  {S.}~\bibnamefont {Liu}}, \bibinfo {author} {\bibfnamefont {K.}~\bibnamefont
  {Barmak}}, \bibinfo {author} {\bibfnamefont {K.}~\bibnamefont {Watanabe}},
  \bibinfo {author} {\bibfnamefont {T.}~\bibnamefont {Taniguchi}}, \bibinfo
  {author} {\bibfnamefont {A.~H.}\ \bibnamefont {MacDonald}}, \bibinfo {author}
  {\bibfnamefont {J.}~\bibnamefont {Shan}}, \emph {et~al.},\ }\href@noop {}
  {\bibfield  {journal} {\bibinfo  {journal} {Nature}\ }\textbf {\bibinfo
  {volume} {579}},\ \bibinfo {pages} {353} (\bibinfo {year}
  {2020})}\BibitemShut {NoStop}%
\bibitem [{\citenamefont {Li}\ \emph {et~al.}(2021{\natexlab{a}})\citenamefont
  {Li}, \citenamefont {Jiang}, \citenamefont {Li}, \citenamefont {Zhang},
  \citenamefont {Kang}, \citenamefont {Zhu}, \citenamefont {Watanabe},
  \citenamefont {Taniguchi}, \citenamefont {Chowdhury}, \citenamefont {Fu},
  \citenamefont {Shan},\ and\ \citenamefont {Mak}}]{Li:2021aa}%
  \BibitemOpen
  \bibfield  {author} {\bibinfo {author} {\bibfnamefont {T.}~\bibnamefont
  {Li}}, \bibinfo {author} {\bibfnamefont {S.}~\bibnamefont {Jiang}}, \bibinfo
  {author} {\bibfnamefont {L.}~\bibnamefont {Li}}, \bibinfo {author}
  {\bibfnamefont {Y.}~\bibnamefont {Zhang}}, \bibinfo {author} {\bibfnamefont
  {K.}~\bibnamefont {Kang}}, \bibinfo {author} {\bibfnamefont {J.}~\bibnamefont
  {Zhu}}, \bibinfo {author} {\bibfnamefont {K.}~\bibnamefont {Watanabe}},
  \bibinfo {author} {\bibfnamefont {T.}~\bibnamefont {Taniguchi}}, \bibinfo
  {author} {\bibfnamefont {D.}~\bibnamefont {Chowdhury}}, \bibinfo {author}
  {\bibfnamefont {L.}~\bibnamefont {Fu}}, \bibinfo {author} {\bibfnamefont
  {J.}~\bibnamefont {Shan}},\ and\ \bibinfo {author} {\bibfnamefont {K.~F.}\
  \bibnamefont {Mak}},\ }\href {https://doi.org/10.1038/s41586-021-03853-0}
  {\bibfield  {journal} {\bibinfo  {journal} {Nature}\ }\textbf {\bibinfo
  {volume} {597}},\ \bibinfo {pages} {350} (\bibinfo {year}
  {2021}{\natexlab{a}})}\BibitemShut {NoStop}%
\bibitem [{\citenamefont {Xu}\ \emph {et~al.}(2022)\citenamefont {Xu},
  \citenamefont {Kang}, \citenamefont {Watanabe}, \citenamefont {Taniguchi},
  \citenamefont {Mak},\ and\ \citenamefont {Shan}}]{Xu_2022}%
  \BibitemOpen
  \bibfield  {author} {\bibinfo {author} {\bibfnamefont {Y.}~\bibnamefont
  {Xu}}, \bibinfo {author} {\bibfnamefont {K.}~\bibnamefont {Kang}}, \bibinfo
  {author} {\bibfnamefont {K.}~\bibnamefont {Watanabe}}, \bibinfo {author}
  {\bibfnamefont {T.}~\bibnamefont {Taniguchi}}, \bibinfo {author}
  {\bibfnamefont {K.~F.}\ \bibnamefont {Mak}},\ and\ \bibinfo {author}
  {\bibfnamefont {J.}~\bibnamefont {Shan}},\ }\href
  {https://doi.org/10.1038/s41565-022-01180-7} {\bibfield  {journal} {\bibinfo
  {journal} {Nature Nanotechnology}\ }\textbf {\bibinfo {volume} {17}},\
  \bibinfo {pages} {934–939} (\bibinfo {year} {2022})}\BibitemShut {NoStop}%
\bibitem [{\citenamefont {Mak}\ and\ \citenamefont
  {Shan}(2022)}]{mak2022semiconductor}%
  \BibitemOpen
  \bibfield  {author} {\bibinfo {author} {\bibfnamefont {K.~F.}\ \bibnamefont
  {Mak}}\ and\ \bibinfo {author} {\bibfnamefont {J.}~\bibnamefont {Shan}},\
  }\href@noop {} {\bibfield  {journal} {\bibinfo  {journal} {Nature
  Nanotechnology}\ }\textbf {\bibinfo {volume} {17}},\ \bibinfo {pages} {686}
  (\bibinfo {year} {2022})}\BibitemShut {NoStop}%
\bibitem [{\citenamefont {Li}\ \emph {et~al.}(2021{\natexlab{b}})\citenamefont
  {Li}, \citenamefont {Jiang}, \citenamefont {Shen}, \citenamefont {Zhang},
  \citenamefont {Li}, \citenamefont {Tao}, \citenamefont {Devakul},
  \citenamefont {Watanabe}, \citenamefont {Taniguchi}, \citenamefont {Fu},
  \citenamefont {Shan},\ and\ \citenamefont {Mak}}]{Li_2021}%
  \BibitemOpen
  \bibfield  {author} {\bibinfo {author} {\bibfnamefont {T.}~\bibnamefont
  {Li}}, \bibinfo {author} {\bibfnamefont {S.}~\bibnamefont {Jiang}}, \bibinfo
  {author} {\bibfnamefont {B.}~\bibnamefont {Shen}}, \bibinfo {author}
  {\bibfnamefont {Y.}~\bibnamefont {Zhang}}, \bibinfo {author} {\bibfnamefont
  {L.}~\bibnamefont {Li}}, \bibinfo {author} {\bibfnamefont {Z.}~\bibnamefont
  {Tao}}, \bibinfo {author} {\bibfnamefont {T.}~\bibnamefont {Devakul}},
  \bibinfo {author} {\bibfnamefont {K.}~\bibnamefont {Watanabe}}, \bibinfo
  {author} {\bibfnamefont {T.}~\bibnamefont {Taniguchi}}, \bibinfo {author}
  {\bibfnamefont {L.}~\bibnamefont {Fu}}, \bibinfo {author} {\bibfnamefont
  {J.}~\bibnamefont {Shan}},\ and\ \bibinfo {author} {\bibfnamefont {K.~F.}\
  \bibnamefont {Mak}},\ }\href {https://doi.org/10.1038/s41586-021-04171-1}
  {\bibfield  {journal} {\bibinfo  {journal} {Nature}\ }\textbf {\bibinfo
  {volume} {600}},\ \bibinfo {pages} {641–646} (\bibinfo {year}
  {2021}{\natexlab{b}})}\BibitemShut {NoStop}%
\bibitem [{\citenamefont {Cai}\ \emph {et~al.}(2023)\citenamefont {Cai},
  \citenamefont {Anderson}, \citenamefont {Wang}, \citenamefont {Zhang},
  \citenamefont {Liu}, \citenamefont {Holtzmann}, \citenamefont {Zhang},
  \citenamefont {Fan}, \citenamefont {Taniguchi}, \citenamefont {Watanabe},
  \citenamefont {Ran}, \citenamefont {Cao}, \citenamefont {Fu}, \citenamefont
  {Xiao}, \citenamefont {Yao},\ and\ \citenamefont {Xu}}]{Cai2023}%
  \BibitemOpen
  \bibfield  {author} {\bibinfo {author} {\bibfnamefont {J.}~\bibnamefont
  {Cai}}, \bibinfo {author} {\bibfnamefont {E.}~\bibnamefont {Anderson}},
  \bibinfo {author} {\bibfnamefont {C.}~\bibnamefont {Wang}}, \bibinfo {author}
  {\bibfnamefont {X.}~\bibnamefont {Zhang}}, \bibinfo {author} {\bibfnamefont
  {X.}~\bibnamefont {Liu}}, \bibinfo {author} {\bibfnamefont {W.}~\bibnamefont
  {Holtzmann}}, \bibinfo {author} {\bibfnamefont {Y.}~\bibnamefont {Zhang}},
  \bibinfo {author} {\bibfnamefont {F.}~\bibnamefont {Fan}}, \bibinfo {author}
  {\bibfnamefont {T.}~\bibnamefont {Taniguchi}}, \bibinfo {author}
  {\bibfnamefont {K.}~\bibnamefont {Watanabe}}, \bibinfo {author}
  {\bibfnamefont {Y.}~\bibnamefont {Ran}}, \bibinfo {author} {\bibfnamefont
  {T.}~\bibnamefont {Cao}}, \bibinfo {author} {\bibfnamefont {L.}~\bibnamefont
  {Fu}}, \bibinfo {author} {\bibfnamefont {D.}~\bibnamefont {Xiao}}, \bibinfo
  {author} {\bibfnamefont {W.}~\bibnamefont {Yao}},\ and\ \bibinfo {author}
  {\bibfnamefont {X.}~\bibnamefont {Xu}},\ }\href
  {https://doi.org/10.1038/s41586-023-06289-w} {\bibfield  {journal} {\bibinfo
  {journal} {Nature}\ }\textbf {\bibinfo {volume} {622}},\ \bibinfo {pages}
  {63} (\bibinfo {year} {2023})}\BibitemShut {NoStop}%
\bibitem [{\citenamefont {Xu}\ \emph {et~al.}(2023)\citenamefont {Xu},
  \citenamefont {Sun}, \citenamefont {Jia}, \citenamefont {Liu}, \citenamefont
  {Xu}, \citenamefont {Li}, \citenamefont {Gu}, \citenamefont {Watanabe},
  \citenamefont {Taniguchi}, \citenamefont {Tong}, \citenamefont {Jia},
  \citenamefont {Shi}, \citenamefont {Jiang}, \citenamefont {Zhang},
  \citenamefont {Liu},\ and\ \citenamefont {Li}}]{PhysRevX.13.031037}%
  \BibitemOpen
  \bibfield  {author} {\bibinfo {author} {\bibfnamefont {F.}~\bibnamefont
  {Xu}}, \bibinfo {author} {\bibfnamefont {Z.}~\bibnamefont {Sun}}, \bibinfo
  {author} {\bibfnamefont {T.}~\bibnamefont {Jia}}, \bibinfo {author}
  {\bibfnamefont {C.}~\bibnamefont {Liu}}, \bibinfo {author} {\bibfnamefont
  {C.}~\bibnamefont {Xu}}, \bibinfo {author} {\bibfnamefont {C.}~\bibnamefont
  {Li}}, \bibinfo {author} {\bibfnamefont {Y.}~\bibnamefont {Gu}}, \bibinfo
  {author} {\bibfnamefont {K.}~\bibnamefont {Watanabe}}, \bibinfo {author}
  {\bibfnamefont {T.}~\bibnamefont {Taniguchi}}, \bibinfo {author}
  {\bibfnamefont {B.}~\bibnamefont {Tong}}, \bibinfo {author} {\bibfnamefont
  {J.}~\bibnamefont {Jia}}, \bibinfo {author} {\bibfnamefont {Z.}~\bibnamefont
  {Shi}}, \bibinfo {author} {\bibfnamefont {S.}~\bibnamefont {Jiang}}, \bibinfo
  {author} {\bibfnamefont {Y.}~\bibnamefont {Zhang}}, \bibinfo {author}
  {\bibfnamefont {X.}~\bibnamefont {Liu}},\ and\ \bibinfo {author}
  {\bibfnamefont {T.}~\bibnamefont {Li}},\ }\href
  {https://doi.org/10.1103/PhysRevX.13.031037} {\bibfield  {journal} {\bibinfo
  {journal} {Phys. Rev. X}\ }\textbf {\bibinfo {volume} {13}},\ \bibinfo
  {pages} {031037} (\bibinfo {year} {2023})}\BibitemShut {NoStop}%
\bibitem [{\citenamefont {Park}\ \emph {et~al.}(2023)\citenamefont {Park},
  \citenamefont {Cai}, \citenamefont {Anderson}, \citenamefont {Zhang},
  \citenamefont {Zhu}, \citenamefont {Liu}, \citenamefont {Wang}, \citenamefont
  {Holtzmann}, \citenamefont {Hu}, \citenamefont {Liu}, \citenamefont
  {Taniguchi}, \citenamefont {Watanabe}, \citenamefont {Chu}, \citenamefont
  {Cao}, \citenamefont {Fu}, \citenamefont {Yao}, \citenamefont {Chang},
  \citenamefont {Cobden}, \citenamefont {Xiao},\ and\ \citenamefont
  {Xu}}]{xuParkObservationFractionallyQuantized2023}%
  \BibitemOpen
  \bibfield  {author} {\bibinfo {author} {\bibfnamefont {H.}~\bibnamefont
  {Park}}, \bibinfo {author} {\bibfnamefont {J.}~\bibnamefont {Cai}}, \bibinfo
  {author} {\bibfnamefont {E.}~\bibnamefont {Anderson}}, \bibinfo {author}
  {\bibfnamefont {Y.}~\bibnamefont {Zhang}}, \bibinfo {author} {\bibfnamefont
  {J.}~\bibnamefont {Zhu}}, \bibinfo {author} {\bibfnamefont {X.}~\bibnamefont
  {Liu}}, \bibinfo {author} {\bibfnamefont {C.}~\bibnamefont {Wang}}, \bibinfo
  {author} {\bibfnamefont {W.}~\bibnamefont {Holtzmann}}, \bibinfo {author}
  {\bibfnamefont {C.}~\bibnamefont {Hu}}, \bibinfo {author} {\bibfnamefont
  {Z.}~\bibnamefont {Liu}}, \bibinfo {author} {\bibfnamefont {T.}~\bibnamefont
  {Taniguchi}}, \bibinfo {author} {\bibfnamefont {K.}~\bibnamefont {Watanabe}},
  \bibinfo {author} {\bibfnamefont {J.-H.}\ \bibnamefont {Chu}}, \bibinfo
  {author} {\bibfnamefont {T.}~\bibnamefont {Cao}}, \bibinfo {author}
  {\bibfnamefont {L.}~\bibnamefont {Fu}}, \bibinfo {author} {\bibfnamefont
  {W.}~\bibnamefont {Yao}}, \bibinfo {author} {\bibfnamefont {C.-Z.}\
  \bibnamefont {Chang}}, \bibinfo {author} {\bibfnamefont {D.}~\bibnamefont
  {Cobden}}, \bibinfo {author} {\bibfnamefont {D.}~\bibnamefont {Xiao}},\ and\
  \bibinfo {author} {\bibfnamefont {X.}~\bibnamefont {Xu}},\ }\href
  {https://doi.org/10.1038/s41586-023-06536-0} {\bibfield  {journal} {\bibinfo
  {journal} {Nature}\ }\textbf {\bibinfo {volume} {622}},\ \bibinfo {pages}
  {74} (\bibinfo {year} {2023})}\BibitemShut {NoStop}%
\bibitem [{\citenamefont {Zeng}\ \emph {et~al.}(2023)\citenamefont {Zeng},
  \citenamefont {Xia}, \citenamefont {Kang}, \citenamefont {Zhu}, \citenamefont
  {Kn{\"u}ppel}, \citenamefont {Vaswani}, \citenamefont {Watanabe},
  \citenamefont {Taniguchi}, \citenamefont {Mak},\ and\ \citenamefont
  {Shan}}]{zeng2023thermodynamic}%
  \BibitemOpen
  \bibfield  {author} {\bibinfo {author} {\bibfnamefont {Y.}~\bibnamefont
  {Zeng}}, \bibinfo {author} {\bibfnamefont {Z.}~\bibnamefont {Xia}}, \bibinfo
  {author} {\bibfnamefont {K.}~\bibnamefont {Kang}}, \bibinfo {author}
  {\bibfnamefont {J.}~\bibnamefont {Zhu}}, \bibinfo {author} {\bibfnamefont
  {P.}~\bibnamefont {Kn{\"u}ppel}}, \bibinfo {author} {\bibfnamefont
  {C.}~\bibnamefont {Vaswani}}, \bibinfo {author} {\bibfnamefont
  {K.}~\bibnamefont {Watanabe}}, \bibinfo {author} {\bibfnamefont
  {T.}~\bibnamefont {Taniguchi}}, \bibinfo {author} {\bibfnamefont {K.~F.}\
  \bibnamefont {Mak}},\ and\ \bibinfo {author} {\bibfnamefont {J.}~\bibnamefont
  {Shan}},\ }\href@noop {} {\bibfield  {journal} {\bibinfo  {journal} {Nature}\
  }\textbf {\bibinfo {volume} {622}},\ \bibinfo {pages} {69} (\bibinfo {year}
  {2023})}\BibitemShut {NoStop}%
\bibitem [{\citenamefont {Huang}\ and\ \citenamefont
  {Das~Sarma}(2024)}]{Huang_2024}%
  \BibitemOpen
  \bibfield  {author} {\bibinfo {author} {\bibfnamefont {Y.}~\bibnamefont
  {Huang}}\ and\ \bibinfo {author} {\bibfnamefont {S.}~\bibnamefont
  {Das~Sarma}},\ }\href {https://doi.org/10.1103/PhysRevB.109.245431}
  {\bibfield  {journal} {\bibinfo  {journal} {Phys. Rev. B}\ }\textbf {\bibinfo
  {volume} {109}},\ \bibinfo {pages} {245431} (\bibinfo {year}
  {2024})}\BibitemShut {NoStop}%
\bibitem [{\citenamefont {Kaasbjerg}\ \emph {et~al.}(2019)\citenamefont
  {Kaasbjerg}, \citenamefont {Low},\ and\ \citenamefont {Jauho}}]{Jauho_2019}%
  \BibitemOpen
  \bibfield  {author} {\bibinfo {author} {\bibfnamefont {K.}~\bibnamefont
  {Kaasbjerg}}, \bibinfo {author} {\bibfnamefont {T.}~\bibnamefont {Low}},\
  and\ \bibinfo {author} {\bibfnamefont {A.-P.}\ \bibnamefont {Jauho}},\ }\href
  {https://doi.org/10.1103/PhysRevB.100.115409} {\bibfield  {journal} {\bibinfo
   {journal} {Phys. Rev. B}\ }\textbf {\bibinfo {volume} {100}},\ \bibinfo
  {pages} {115409} (\bibinfo {year} {2019})}\BibitemShut {NoStop}%
\bibitem [{\citenamefont {Kaasbjerg}\ \emph {et~al.}(2013)\citenamefont
  {Kaasbjerg}, \citenamefont {Thygesen},\ and\ \citenamefont
  {Jauho}}]{Jauho_2013}%
  \BibitemOpen
  \bibfield  {author} {\bibinfo {author} {\bibfnamefont {K.}~\bibnamefont
  {Kaasbjerg}}, \bibinfo {author} {\bibfnamefont {K.~S.}\ \bibnamefont
  {Thygesen}},\ and\ \bibinfo {author} {\bibfnamefont {A.-P.}\ \bibnamefont
  {Jauho}},\ }\href {https://doi.org/10.1103/PhysRevB.87.235312} {\bibfield
  {journal} {\bibinfo  {journal} {Phys. Rev. B}\ }\textbf {\bibinfo {volume}
  {87}},\ \bibinfo {pages} {235312} (\bibinfo {year} {2013})}\BibitemShut
  {NoStop}%
\bibitem [{\citenamefont {Wang}\ \emph {et~al.}(2023)\citenamefont {Wang},
  \citenamefont {Devakul}, \citenamefont {Zaletel},\ and\ \citenamefont
  {Fu}}]{wang2023topological}%
  \BibitemOpen
  \bibfield  {author} {\bibinfo {author} {\bibfnamefont {T.}~\bibnamefont
  {Wang}}, \bibinfo {author} {\bibfnamefont {T.}~\bibnamefont {Devakul}},
  \bibinfo {author} {\bibfnamefont {M.~P.}\ \bibnamefont {Zaletel}},\ and\
  \bibinfo {author} {\bibfnamefont {L.}~\bibnamefont {Fu}},\ }\href@noop {}
  {\bibinfo {title} {Topological magnets and magnons in twisted bilayer
  mote$_2$ and wse$_2$}} (\bibinfo {year} {2023}),\ \Eprint
  {https://arxiv.org/abs/2306.02501} {arXiv:2306.02501 [cond-mat.str-el]}
  \BibitemShut {NoStop}%
\bibitem [{\citenamefont {Sato}\ and\ \citenamefont {Ando}(2017)}]{Sato_2017}%
  \BibitemOpen
  \bibfield  {author} {\bibinfo {author} {\bibfnamefont {M.}~\bibnamefont
  {Sato}}\ and\ \bibinfo {author} {\bibfnamefont {Y.}~\bibnamefont {Ando}},\
  }\href {https://doi.org/10.1088/1361-6633/aa6ac7} {\bibfield  {journal}
  {\bibinfo  {journal} {Reports on Progress in Physics}\ }\textbf {\bibinfo
  {volume} {80}},\ \bibinfo {pages} {076501} (\bibinfo {year}
  {2017})}\BibitemShut {NoStop}%
\bibitem [{\citenamefont {Nandkishore}\ \emph {et~al.}(2012)\citenamefont
  {Nandkishore}, \citenamefont {Levitov},\ and\ \citenamefont
  {Chubukov}}]{Nandkishore_2012}%
  \BibitemOpen
  \bibfield  {author} {\bibinfo {author} {\bibfnamefont {R.}~\bibnamefont
  {Nandkishore}}, \bibinfo {author} {\bibfnamefont {L.~S.}\ \bibnamefont
  {Levitov}},\ and\ \bibinfo {author} {\bibfnamefont {A.~V.}\ \bibnamefont
  {Chubukov}},\ }\href {https://doi.org/10.1038/nphys2208} {\bibfield
  {journal} {\bibinfo  {journal} {Nature Physics}\ }\textbf {\bibinfo {volume}
  {8}},\ \bibinfo {pages} {158–163} (\bibinfo {year} {2012})}\BibitemShut
  {NoStop}%
\bibitem [{\citenamefont {Cr\'epel}\ \emph {et~al.}(2023)\citenamefont
  {Cr\'epel}, \citenamefont {Guerci}, \citenamefont {Cano}, \citenamefont
  {Pixley},\ and\ \citenamefont {Millis}}]{crepel2023prl}%
  \BibitemOpen
  \bibfield  {author} {\bibinfo {author} {\bibfnamefont {V.}~\bibnamefont
  {Cr\'epel}}, \bibinfo {author} {\bibfnamefont {D.}~\bibnamefont {Guerci}},
  \bibinfo {author} {\bibfnamefont {J.}~\bibnamefont {Cano}}, \bibinfo {author}
  {\bibfnamefont {J.~H.}\ \bibnamefont {Pixley}},\ and\ \bibinfo {author}
  {\bibfnamefont {A.}~\bibnamefont {Millis}},\ }\href
  {https://doi.org/10.1103/PhysRevLett.131.056001} {\bibfield  {journal}
  {\bibinfo  {journal} {Phys. Rev. Lett.}\ }\textbf {\bibinfo {volume} {131}},\
  \bibinfo {pages} {056001} (\bibinfo {year} {2023})}\BibitemShut {NoStop}%
\bibitem [{\citenamefont {Zerba}\ \emph {et~al.}(2024)\citenamefont {Zerba},
  \citenamefont {Kuhlenkamp}, \citenamefont {Imamoğlu},\ and\ \citenamefont
  {Knap}}]{Zerba_2024}%
  \BibitemOpen
  \bibfield  {author} {\bibinfo {author} {\bibfnamefont {C.}~\bibnamefont
  {Zerba}}, \bibinfo {author} {\bibfnamefont {C.}~\bibnamefont {Kuhlenkamp}},
  \bibinfo {author} {\bibfnamefont {A.}~\bibnamefont {Imamoğlu}},\ and\
  \bibinfo {author} {\bibfnamefont {M.}~\bibnamefont {Knap}},\ }\bibfield
  {journal} {\bibinfo  {journal} {Physical Review Letters}\ }\textbf {\bibinfo
  {volume} {133}},\ \href {https://doi.org/10.1103/physrevlett.133.056902}
  {10.1103/physrevlett.133.056902} (\bibinfo {year} {2024})\BibitemShut
  {NoStop}%
\bibitem [{\citenamefont {Fu}\ and\ \citenamefont {Kane}(2008)}]{fu2008}%
  \BibitemOpen
  \bibfield  {author} {\bibinfo {author} {\bibfnamefont {L.}~\bibnamefont
  {Fu}}\ and\ \bibinfo {author} {\bibfnamefont {C.~L.}\ \bibnamefont {Kane}},\
  }\href {https://doi.org/10.1103/PhysRevLett.100.096407} {\bibfield  {journal}
  {\bibinfo  {journal} {Phys. Rev. Lett.}\ }\textbf {\bibinfo {volume} {100}},\
  \bibinfo {pages} {096407} (\bibinfo {year} {2008})}\BibitemShut {NoStop}%
\bibitem [{\citenamefont {Can}\ \emph {et~al.}(2021)\citenamefont {Can},
  \citenamefont {Tummuru}, \citenamefont {Day}, \citenamefont {Elfimov},
  \citenamefont {Damascelli},\ and\ \citenamefont {Franz}}]{Can_2021}%
  \BibitemOpen
  \bibfield  {author} {\bibinfo {author} {\bibfnamefont {O.}~\bibnamefont
  {Can}}, \bibinfo {author} {\bibfnamefont {T.}~\bibnamefont {Tummuru}},
  \bibinfo {author} {\bibfnamefont {R.~P.}\ \bibnamefont {Day}}, \bibinfo
  {author} {\bibfnamefont {I.}~\bibnamefont {Elfimov}}, \bibinfo {author}
  {\bibfnamefont {A.}~\bibnamefont {Damascelli}},\ and\ \bibinfo {author}
  {\bibfnamefont {M.}~\bibnamefont {Franz}},\ }\href
  {https://doi.org/10.1038/s41567-020-01142-7} {\bibfield  {journal} {\bibinfo
  {journal} {Nature Physics}\ }\textbf {\bibinfo {volume} {17}},\ \bibinfo
  {pages} {519–524} (\bibinfo {year} {2021})}\BibitemShut {NoStop}%
\bibitem [{\citenamefont {Zhao}\ \emph {et~al.}(2023)\citenamefont {Zhao},
  \citenamefont {Cui}, \citenamefont {Volkov}, \citenamefont {Yoo},
  \citenamefont {Lee}, \citenamefont {Gardener}, \citenamefont {Akey},
  \citenamefont {Engelke}, \citenamefont {Ronen}, \citenamefont {Zhong},
  \citenamefont {Gu}, \citenamefont {Plugge}, \citenamefont {Tummuru},
  \citenamefont {Kim}, \citenamefont {Franz}, \citenamefont {Pixley},
  \citenamefont {Poccia},\ and\ \citenamefont {Kim}}]{kimtwisted2023}%
  \BibitemOpen
  \bibfield  {author} {\bibinfo {author} {\bibfnamefont {S.~Y.~F.}\
  \bibnamefont {Zhao}}, \bibinfo {author} {\bibfnamefont {X.}~\bibnamefont
  {Cui}}, \bibinfo {author} {\bibfnamefont {P.~A.}\ \bibnamefont {Volkov}},
  \bibinfo {author} {\bibfnamefont {H.}~\bibnamefont {Yoo}}, \bibinfo {author}
  {\bibfnamefont {S.}~\bibnamefont {Lee}}, \bibinfo {author} {\bibfnamefont
  {J.~A.}\ \bibnamefont {Gardener}}, \bibinfo {author} {\bibfnamefont {A.~J.}\
  \bibnamefont {Akey}}, \bibinfo {author} {\bibfnamefont {R.}~\bibnamefont
  {Engelke}}, \bibinfo {author} {\bibfnamefont {Y.}~\bibnamefont {Ronen}},
  \bibinfo {author} {\bibfnamefont {R.}~\bibnamefont {Zhong}}, \bibinfo
  {author} {\bibfnamefont {G.}~\bibnamefont {Gu}}, \bibinfo {author}
  {\bibfnamefont {S.}~\bibnamefont {Plugge}}, \bibinfo {author} {\bibfnamefont
  {T.}~\bibnamefont {Tummuru}}, \bibinfo {author} {\bibfnamefont
  {M.}~\bibnamefont {Kim}}, \bibinfo {author} {\bibfnamefont {M.}~\bibnamefont
  {Franz}}, \bibinfo {author} {\bibfnamefont {J.~H.}\ \bibnamefont {Pixley}},
  \bibinfo {author} {\bibfnamefont {N.}~\bibnamefont {Poccia}},\ and\ \bibinfo
  {author} {\bibfnamefont {P.}~\bibnamefont {Kim}},\ }\href
  {https://doi.org/10.1126/science.abl8371} {\bibfield  {journal} {\bibinfo
  {journal} {Science}\ }\textbf {\bibinfo {volume} {382}},\ \bibinfo {pages}
  {1422} (\bibinfo {year} {2023})},\ \Eprint
  {https://arxiv.org/abs/https://www.science.org/doi/pdf/10.1126/science.abl8371}
  {https://www.science.org/doi/pdf/10.1126/science.abl8371} \BibitemShut
  {NoStop}%
\bibitem [{\citenamefont {Volkov}\ \emph
  {et~al.}(2023{\natexlab{a}})\citenamefont {Volkov}, \citenamefont {Wilson},
  \citenamefont {Lucht},\ and\ \citenamefont {Pixley}}]{volkov2023prl}%
  \BibitemOpen
  \bibfield  {author} {\bibinfo {author} {\bibfnamefont {P.~A.}\ \bibnamefont
  {Volkov}}, \bibinfo {author} {\bibfnamefont {J.~H.}\ \bibnamefont {Wilson}},
  \bibinfo {author} {\bibfnamefont {K.~P.}\ \bibnamefont {Lucht}},\ and\
  \bibinfo {author} {\bibfnamefont {J.~H.}\ \bibnamefont {Pixley}},\ }\href
  {https://doi.org/10.1103/PhysRevLett.130.186001} {\bibfield  {journal}
  {\bibinfo  {journal} {Phys. Rev. Lett.}\ }\textbf {\bibinfo {volume} {130}},\
  \bibinfo {pages} {186001} (\bibinfo {year} {2023}{\natexlab{a}})}\BibitemShut
  {NoStop}%
\bibitem [{\citenamefont {Volkov}\ \emph
  {et~al.}(2023{\natexlab{b}})\citenamefont {Volkov}, \citenamefont {Wilson},
  \citenamefont {Lucht},\ and\ \citenamefont {Pixley}}]{volkov2023prb}%
  \BibitemOpen
  \bibfield  {author} {\bibinfo {author} {\bibfnamefont {P.~A.}\ \bibnamefont
  {Volkov}}, \bibinfo {author} {\bibfnamefont {J.~H.}\ \bibnamefont {Wilson}},
  \bibinfo {author} {\bibfnamefont {K.~P.}\ \bibnamefont {Lucht}},\ and\
  \bibinfo {author} {\bibfnamefont {J.~H.}\ \bibnamefont {Pixley}},\ }\href
  {https://doi.org/10.1103/PhysRevB.107.174506} {\bibfield  {journal} {\bibinfo
   {journal} {Phys. Rev. B}\ }\textbf {\bibinfo {volume} {107}},\ \bibinfo
  {pages} {174506} (\bibinfo {year} {2023}{\natexlab{b}})}\BibitemShut
  {NoStop}%
\bibitem [{\citenamefont {Lucht}\ \emph {et~al.}(2023)\citenamefont {Lucht},
  \citenamefont {Pixley},\ and\ \citenamefont {Volkov}}]{lucht2023topological}%
  \BibitemOpen
  \bibfield  {author} {\bibinfo {author} {\bibfnamefont {K.~P.}\ \bibnamefont
  {Lucht}}, \bibinfo {author} {\bibfnamefont {J.}~\bibnamefont {Pixley}},\ and\
  \bibinfo {author} {\bibfnamefont {P.~A.}\ \bibnamefont {Volkov}},\
  }\href@noop {} {\bibfield  {journal} {\bibinfo  {journal} {arXiv preprint
  arXiv:2312.13367}\ } (\bibinfo {year} {2023})}\BibitemShut {NoStop}%
\bibitem [{\citenamefont {Lucht}\ \emph {et~al.}(2024)\citenamefont {Lucht},
  \citenamefont {Volkov},\ and\ \citenamefont {Pixley}}]{PhysRevB.109.184507}%
  \BibitemOpen
  \bibfield  {author} {\bibinfo {author} {\bibfnamefont {K.~P.}\ \bibnamefont
  {Lucht}}, \bibinfo {author} {\bibfnamefont {P.~A.}\ \bibnamefont {Volkov}},\
  and\ \bibinfo {author} {\bibfnamefont {J.~H.}\ \bibnamefont {Pixley}},\
  }\href {https://doi.org/10.1103/PhysRevB.109.184507} {\bibfield  {journal}
  {\bibinfo  {journal} {Phys. Rev. B}\ }\textbf {\bibinfo {volume} {109}},\
  \bibinfo {pages} {184507} (\bibinfo {year} {2024})}\BibitemShut {NoStop}%
\bibitem [{\citenamefont {Morales-Dur\'an}\ \emph {et~al.}(2024)\citenamefont
  {Morales-Dur\'an}, \citenamefont {Wei}, \citenamefont {Shi},\ and\
  \citenamefont {MacDonald}}]{AllanNicolas_2024}%
  \BibitemOpen
  \bibfield  {author} {\bibinfo {author} {\bibfnamefont {N.}~\bibnamefont
  {Morales-Dur\'an}}, \bibinfo {author} {\bibfnamefont {N.}~\bibnamefont
  {Wei}}, \bibinfo {author} {\bibfnamefont {J.}~\bibnamefont {Shi}},\ and\
  \bibinfo {author} {\bibfnamefont {A.~H.}\ \bibnamefont {MacDonald}},\ }\href
  {https://doi.org/10.1103/PhysRevLett.132.096602} {\bibfield  {journal}
  {\bibinfo  {journal} {Phys. Rev. Lett.}\ }\textbf {\bibinfo {volume} {132}},\
  \bibinfo {pages} {096602} (\bibinfo {year} {2024})}\BibitemShut {NoStop}%
\bibitem [{\citenamefont {Shi}\ \emph {et~al.}(2024{\natexlab{b}})\citenamefont
  {Shi}, \citenamefont {Morales-Durán}, \citenamefont {Khalaf},\ and\
  \citenamefont
  {MacDonald}}]{shi2024adiabaticapproximationaharonovcasherbands}%
  \BibitemOpen
  \bibfield  {author} {\bibinfo {author} {\bibfnamefont {J.}~\bibnamefont
  {Shi}}, \bibinfo {author} {\bibfnamefont {N.}~\bibnamefont {Morales-Durán}},
  \bibinfo {author} {\bibfnamefont {E.}~\bibnamefont {Khalaf}},\ and\ \bibinfo
  {author} {\bibfnamefont {A.~H.}\ \bibnamefont {MacDonald}},\ }\href
  {https://arxiv.org/abs/2404.13455} {\bibinfo {title} {Adiabatic approximation
  and aharonov-casher bands in twisted homobilayer tmds}} (\bibinfo {year}
  {2024}{\natexlab{b}}),\ \Eprint {https://arxiv.org/abs/2404.13455}
  {arXiv:2404.13455 [cond-mat.mes-hall]} \BibitemShut {NoStop}%
\bibitem [{\citenamefont {Kolář}\ \emph {et~al.}(2024)\citenamefont
  {Kolář}, \citenamefont {Yang}, \citenamefont {von Oppen},\ and\
  \citenamefont {Mora}}]{kolar2024hofstadterspectrumchernbands}%
  \BibitemOpen
  \bibfield  {author} {\bibinfo {author} {\bibfnamefont {K.}~\bibnamefont
  {Kolář}}, \bibinfo {author} {\bibfnamefont {K.}~\bibnamefont {Yang}},
  \bibinfo {author} {\bibfnamefont {F.}~\bibnamefont {von Oppen}},\ and\
  \bibinfo {author} {\bibfnamefont {C.}~\bibnamefont {Mora}},\ }\href
  {https://arxiv.org/abs/2406.06680} {\bibinfo {title} {Hofstadter spectrum of
  chern bands in twisted transition metal dichalcogenides}} (\bibinfo {year}
  {2024}),\ \Eprint {https://arxiv.org/abs/2406.06680} {arXiv:2406.06680
  [cond-mat.mes-hall]} \BibitemShut {NoStop}%
\bibitem [{\citenamefont {Zhang}\ \emph
  {et~al.}(2024{\natexlab{a}})\citenamefont {Zhang}, \citenamefont {Wang},
  \citenamefont {Liu}, \citenamefont {Fan}, \citenamefont {Cao},\ and\
  \citenamefont {Xiao}}]{Zhang_2024}%
  \BibitemOpen
  \bibfield  {author} {\bibinfo {author} {\bibfnamefont {X.-W.}\ \bibnamefont
  {Zhang}}, \bibinfo {author} {\bibfnamefont {C.}~\bibnamefont {Wang}},
  \bibinfo {author} {\bibfnamefont {X.}~\bibnamefont {Liu}}, \bibinfo {author}
  {\bibfnamefont {Y.}~\bibnamefont {Fan}}, \bibinfo {author} {\bibfnamefont
  {T.}~\bibnamefont {Cao}},\ and\ \bibinfo {author} {\bibfnamefont
  {D.}~\bibnamefont {Xiao}},\ }\bibfield  {journal} {\bibinfo  {journal}
  {Nature Communications}\ }\textbf {\bibinfo {volume} {15}},\ \href
  {https://doi.org/10.1038/s41467-024-48511-x} {10.1038/s41467-024-48511-x}
  (\bibinfo {year} {2024}{\natexlab{a}})\BibitemShut {NoStop}%
\bibitem [{\citenamefont {Dong}\ \emph {et~al.}(2023)\citenamefont {Dong},
  \citenamefont {Wang}, \citenamefont {Ledwith}, \citenamefont {Vishwanath},\
  and\ \citenamefont {Parker}}]{dong2023}%
  \BibitemOpen
  \bibfield  {author} {\bibinfo {author} {\bibfnamefont {J.}~\bibnamefont
  {Dong}}, \bibinfo {author} {\bibfnamefont {J.}~\bibnamefont {Wang}}, \bibinfo
  {author} {\bibfnamefont {P.~J.}\ \bibnamefont {Ledwith}}, \bibinfo {author}
  {\bibfnamefont {A.}~\bibnamefont {Vishwanath}},\ and\ \bibinfo {author}
  {\bibfnamefont {D.~E.}\ \bibnamefont {Parker}},\ }\href
  {https://doi.org/10.1103/PhysRevLett.131.136502} {\bibfield  {journal}
  {\bibinfo  {journal} {Phys. Rev. Lett.}\ }\textbf {\bibinfo {volume} {131}},\
  \bibinfo {pages} {136502} (\bibinfo {year} {2023})}\BibitemShut {NoStop}%
\bibitem [{\citenamefont {Zhang}\ \emph
  {et~al.}(2024{\natexlab{b}})\citenamefont {Zhang}, \citenamefont
  {Morales-Durán}, \citenamefont {Li}, \citenamefont {Yao}, \citenamefont
  {Su}, \citenamefont {Lin}, \citenamefont {Dong}, \citenamefont {Kim},
  \citenamefont {Robinson}, \citenamefont {Macdonald},\ and\ \citenamefont
  {Shih}}]{zhang2024directobservationlayerskyrmions}%
  \BibitemOpen
  \bibfield  {author} {\bibinfo {author} {\bibfnamefont {F.}~\bibnamefont
  {Zhang}}, \bibinfo {author} {\bibfnamefont {N.}~\bibnamefont
  {Morales-Durán}}, \bibinfo {author} {\bibfnamefont {Y.}~\bibnamefont {Li}},
  \bibinfo {author} {\bibfnamefont {W.}~\bibnamefont {Yao}}, \bibinfo {author}
  {\bibfnamefont {J.-J.}\ \bibnamefont {Su}}, \bibinfo {author} {\bibfnamefont
  {Y.-C.}\ \bibnamefont {Lin}}, \bibinfo {author} {\bibfnamefont
  {C.}~\bibnamefont {Dong}}, \bibinfo {author} {\bibfnamefont {H.}~\bibnamefont
  {Kim}}, \bibinfo {author} {\bibfnamefont {J.~A.}\ \bibnamefont {Robinson}},
  \bibinfo {author} {\bibfnamefont {A.~H.}\ \bibnamefont {Macdonald}},\ and\
  \bibinfo {author} {\bibfnamefont {C.-K.}\ \bibnamefont {Shih}},\ }\href
  {https://arxiv.org/abs/2406.20036} {\bibinfo {title} {Direct observation of
  layer skyrmions in twisted wse2 bilayers}} (\bibinfo {year}
  {2024}{\natexlab{b}}),\ \Eprint {https://arxiv.org/abs/2406.20036}
  {arXiv:2406.20036 [cond-mat.mes-hall]} \BibitemShut {NoStop}%
\bibitem [{\citenamefont {Wu}\ \emph {et~al.}(2019{\natexlab{b}})\citenamefont
  {Wu}, \citenamefont {Lovorn}, \citenamefont {Tutuc}, \citenamefont {Martin},\
  and\ \citenamefont {MacDonald}}]{FW_PRL_2019}%
  \BibitemOpen
  \bibfield  {author} {\bibinfo {author} {\bibfnamefont {F.}~\bibnamefont
  {Wu}}, \bibinfo {author} {\bibfnamefont {T.}~\bibnamefont {Lovorn}}, \bibinfo
  {author} {\bibfnamefont {E.}~\bibnamefont {Tutuc}}, \bibinfo {author}
  {\bibfnamefont {I.}~\bibnamefont {Martin}},\ and\ \bibinfo {author}
  {\bibfnamefont {A.~H.}\ \bibnamefont {MacDonald}},\ }\href
  {https://doi.org/10.1103/PhysRevLett.122.086402} {\bibfield  {journal}
  {\bibinfo  {journal} {Phys. Rev. Lett.}\ }\textbf {\bibinfo {volume} {122}},\
  \bibinfo {pages} {086402} (\bibinfo {year} {2019}{\natexlab{b}})}\BibitemShut
  {NoStop}%
\bibitem [{\citenamefont {Pan}\ \emph {et~al.}(2020)\citenamefont {Pan},
  \citenamefont {Wu},\ and\ \citenamefont {Das~Sarma}}]{haining2020}%
  \BibitemOpen
  \bibfield  {author} {\bibinfo {author} {\bibfnamefont {H.}~\bibnamefont
  {Pan}}, \bibinfo {author} {\bibfnamefont {F.}~\bibnamefont {Wu}},\ and\
  \bibinfo {author} {\bibfnamefont {S.}~\bibnamefont {Das~Sarma}},\ }\href
  {https://doi.org/10.1103/PhysRevResearch.2.033087} {\bibfield  {journal}
  {\bibinfo  {journal} {Phys. Rev. Res.}\ }\textbf {\bibinfo {volume} {2}},\
  \bibinfo {pages} {033087} (\bibinfo {year} {2020})}\BibitemShut {NoStop}%
\bibitem [{\citenamefont {Shi}\ and\ \citenamefont
  {Niu}(2006)}]{shi2006attractive}%
  \BibitemOpen
  \bibfield  {author} {\bibinfo {author} {\bibfnamefont {J.}~\bibnamefont
  {Shi}}\ and\ \bibinfo {author} {\bibfnamefont {Q.}~\bibnamefont {Niu}},\
  }\href@noop {} {\bibinfo {title} {Attractive electron-electron interaction
  induced by geometric phase in a bloch band}} (\bibinfo {year} {2006}),\
  \Eprint {https://arxiv.org/abs/cond-mat/0601531} {arXiv:cond-mat/0601531
  [cond-mat.supr-con]} \BibitemShut {NoStop}%
\bibitem [{\citenamefont {Qin}\ \emph {et~al.}(2019)\citenamefont {Qin},
  \citenamefont {Li},\ and\ \citenamefont {Zhang}}]{qin2019chiral}%
  \BibitemOpen
  \bibfield  {author} {\bibinfo {author} {\bibfnamefont {W.}~\bibnamefont
  {Qin}}, \bibinfo {author} {\bibfnamefont {L.}~\bibnamefont {Li}},\ and\
  \bibinfo {author} {\bibfnamefont {Z.}~\bibnamefont {Zhang}},\ }\href@noop {}
  {\bibfield  {journal} {\bibinfo  {journal} {Nature Physics}\ }\textbf
  {\bibinfo {volume} {15}},\ \bibinfo {pages} {796} (\bibinfo {year}
  {2019})}\BibitemShut {NoStop}%
\bibitem [{\citenamefont {Li}\ and\ \citenamefont {Haldane}(2018)}]{YiLi_2018}%
  \BibitemOpen
  \bibfield  {author} {\bibinfo {author} {\bibfnamefont {Y.}~\bibnamefont
  {Li}}\ and\ \bibinfo {author} {\bibfnamefont {F.~D.~M.}\ \bibnamefont
  {Haldane}},\ }\href {https://doi.org/10.1103/PhysRevLett.120.067003}
  {\bibfield  {journal} {\bibinfo  {journal} {Phys. Rev. Lett.}\ }\textbf
  {\bibinfo {volume} {120}},\ \bibinfo {pages} {067003} (\bibinfo {year}
  {2018})}\BibitemShut {NoStop}%
\bibitem [{\citenamefont {Schrade}\ and\ \citenamefont
  {Fu}(2024)}]{schrade2024nematic}%
  \BibitemOpen
  \bibfield  {author} {\bibinfo {author} {\bibfnamefont {C.}~\bibnamefont
  {Schrade}}\ and\ \bibinfo {author} {\bibfnamefont {L.}~\bibnamefont {Fu}},\
  }\href@noop {} {\bibinfo {title} {Nematic, chiral and topological
  superconductivity in transition metal dichalcogenides}} (\bibinfo {year}
  {2024}),\ \Eprint {https://arxiv.org/abs/2110.10172} {arXiv:2110.10172}
  \BibitemShut {NoStop}%
\bibitem [{\citenamefont {Wu}\ \emph {et~al.}(2023)\citenamefont {Wu},
  \citenamefont {Wu},\ and\ \citenamefont {Yao}}]{YiMingWu_2023}%
  \BibitemOpen
  \bibfield  {author} {\bibinfo {author} {\bibfnamefont {Y.-M.}\ \bibnamefont
  {Wu}}, \bibinfo {author} {\bibfnamefont {Z.}~\bibnamefont {Wu}},\ and\
  \bibinfo {author} {\bibfnamefont {H.}~\bibnamefont {Yao}},\ }\href
  {https://doi.org/10.1103/PhysRevLett.130.126001} {\bibfield  {journal}
  {\bibinfo  {journal} {Phys. Rev. Lett.}\ }\textbf {\bibinfo {volume} {130}},\
  \bibinfo {pages} {126001} (\bibinfo {year} {2023})}\BibitemShut {NoStop}%
\bibitem [{\citenamefont {Klebl}\ \emph {et~al.}(2023)\citenamefont {Klebl},
  \citenamefont {Fischer}, \citenamefont {Classen}, \citenamefont {Scherer},\
  and\ \citenamefont {Kennes}}]{Klebl_2023}%
  \BibitemOpen
  \bibfield  {author} {\bibinfo {author} {\bibfnamefont {L.}~\bibnamefont
  {Klebl}}, \bibinfo {author} {\bibfnamefont {A.}~\bibnamefont {Fischer}},
  \bibinfo {author} {\bibfnamefont {L.}~\bibnamefont {Classen}}, \bibinfo
  {author} {\bibfnamefont {M.~M.}\ \bibnamefont {Scherer}},\ and\ \bibinfo
  {author} {\bibfnamefont {D.~M.}\ \bibnamefont {Kennes}},\ }\href
  {https://doi.org/10.1103/PhysRevResearch.5.L012034} {\bibfield  {journal}
  {\bibinfo  {journal} {Phys. Rev. Res.}\ }\textbf {\bibinfo {volume} {5}},\
  \bibinfo {pages} {L012034} (\bibinfo {year} {2023})}\BibitemShut {NoStop}%
\bibitem [{\citenamefont {Zegrodnik}\ and\ \citenamefont
  {Biborski}(2023)}]{Zegrodnik_2023}%
  \BibitemOpen
  \bibfield  {author} {\bibinfo {author} {\bibfnamefont {M.}~\bibnamefont
  {Zegrodnik}}\ and\ \bibinfo {author} {\bibfnamefont {A.}~\bibnamefont
  {Biborski}},\ }\href {https://doi.org/10.1103/PhysRevB.108.064506} {\bibfield
   {journal} {\bibinfo  {journal} {Phys. Rev. B}\ }\textbf {\bibinfo {volume}
  {108}},\ \bibinfo {pages} {064506} (\bibinfo {year} {2023})}\BibitemShut
  {NoStop}%
\bibitem [{\citenamefont {Zhu}\ \emph {et~al.}(2024)\citenamefont {Zhu},
  \citenamefont {Chou}, \citenamefont {Xie},\ and\ \citenamefont
  {Sarma}}]{zhu2024theorysuperconductivitytwistedtransition}%
  \BibitemOpen
  \bibfield  {author} {\bibinfo {author} {\bibfnamefont {J.}~\bibnamefont
  {Zhu}}, \bibinfo {author} {\bibfnamefont {Y.-Z.}\ \bibnamefont {Chou}},
  \bibinfo {author} {\bibfnamefont {M.}~\bibnamefont {Xie}},\ and\ \bibinfo
  {author} {\bibfnamefont {S.~D.}\ \bibnamefont {Sarma}},\ }\href
  {https://arxiv.org/abs/2406.19348} {\bibinfo {title} {Theory of
  superconductivity in twisted transition metal dichalcogenide homobilayers}}
  (\bibinfo {year} {2024}),\ \Eprint {https://arxiv.org/abs/2406.19348}
  {arXiv:2406.19348 [cond-mat.supr-con]} \BibitemShut {NoStop}%
\bibitem [{\citenamefont {Kim}\ \emph {et~al.}(2024)\citenamefont {Kim},
  \citenamefont {Mendez-Valderrama}, \citenamefont {Wang},\ and\ \citenamefont
  {Chowdhury}}]{kim2024theorycorrelatedinsulatorssuperconductor}%
  \BibitemOpen
  \bibfield  {author} {\bibinfo {author} {\bibfnamefont {S.}~\bibnamefont
  {Kim}}, \bibinfo {author} {\bibfnamefont {J.~F.}\ \bibnamefont
  {Mendez-Valderrama}}, \bibinfo {author} {\bibfnamefont {X.}~\bibnamefont
  {Wang}},\ and\ \bibinfo {author} {\bibfnamefont {D.}~\bibnamefont
  {Chowdhury}},\ }\href {https://arxiv.org/abs/2406.03525} {\bibinfo {title}
  {Theory of correlated insulator(s) and superconductor at $\nu=1$ in twisted
  wse$_2$}} (\bibinfo {year} {2024}),\ \Eprint
  {https://arxiv.org/abs/2406.03525} {arXiv:2406.03525 [cond-mat.str-el]}
  \BibitemShut {NoStop}%
\bibitem [{\citenamefont {Christos}\ \emph {et~al.}(2024)\citenamefont
  {Christos}, \citenamefont {Bonetti},\ and\ \citenamefont
  {Scheurer}}]{christos2024approximatesymmetriesinsulatorssuperconductivity}%
  \BibitemOpen
  \bibfield  {author} {\bibinfo {author} {\bibfnamefont {M.}~\bibnamefont
  {Christos}}, \bibinfo {author} {\bibfnamefont {P.~M.}\ \bibnamefont
  {Bonetti}},\ and\ \bibinfo {author} {\bibfnamefont {M.~S.}\ \bibnamefont
  {Scheurer}},\ }\href {https://arxiv.org/abs/2407.02393} {\bibinfo {title}
  {Approximate symmetries, insulators, and superconductivity in continuum-model
  description of twisted wse$_2$}} (\bibinfo {year} {2024}),\ \Eprint
  {https://arxiv.org/abs/2407.02393} {arXiv:2407.02393 [cond-mat.supr-con]}
  \BibitemShut {NoStop}%
\bibitem [{\citenamefont {Xie}\ \emph {et~al.}(2024)\citenamefont {Xie},
  \citenamefont {Chen}, \citenamefont {Sur}, \citenamefont {Fang},
  \citenamefont {Cano},\ and\ \citenamefont
  {Si}}]{xie2024superconductivitytwistedwse2topologyinduced}%
  \BibitemOpen
  \bibfield  {author} {\bibinfo {author} {\bibfnamefont {F.}~\bibnamefont
  {Xie}}, \bibinfo {author} {\bibfnamefont {L.}~\bibnamefont {Chen}}, \bibinfo
  {author} {\bibfnamefont {S.}~\bibnamefont {Sur}}, \bibinfo {author}
  {\bibfnamefont {Y.}~\bibnamefont {Fang}}, \bibinfo {author} {\bibfnamefont
  {J.}~\bibnamefont {Cano}},\ and\ \bibinfo {author} {\bibfnamefont
  {Q.}~\bibnamefont {Si}},\ }\href {https://arxiv.org/abs/2408.10185} {\bibinfo
  {title} {Superconductivity in twisted wse$_2$ from topology-induced quantum
  fluctuations}} (\bibinfo {year} {2024}),\ \Eprint
  {https://arxiv.org/abs/2408.10185} {arXiv:2408.10185 [cond-mat.str-el]}
  \BibitemShut {NoStop}%
\bibitem [{\citenamefont {Cea}\ and\ \citenamefont {Guinea}(2021)}]{Cea_2021}%
  \BibitemOpen
  \bibfield  {author} {\bibinfo {author} {\bibfnamefont {T.}~\bibnamefont
  {Cea}}\ and\ \bibinfo {author} {\bibfnamefont {F.}~\bibnamefont {Guinea}},\
  }\bibfield  {journal} {\bibinfo  {journal} {Proceedings of the National
  Academy of Sciences}\ }\textbf {\bibinfo {volume} {118}},\ \href
  {https://doi.org/10.1073/pnas.2107874118} {10.1073/pnas.2107874118} (\bibinfo
  {year} {2021})\BibitemShut {NoStop}%
\bibitem [{\citenamefont {Ghazaryan}\ \emph {et~al.}(2021)\citenamefont
  {Ghazaryan}, \citenamefont {Holder}, \citenamefont {Serbyn},\ and\
  \citenamefont {Berg}}]{Ghazaryan_2021}%
  \BibitemOpen
  \bibfield  {author} {\bibinfo {author} {\bibfnamefont {A.}~\bibnamefont
  {Ghazaryan}}, \bibinfo {author} {\bibfnamefont {T.}~\bibnamefont {Holder}},
  \bibinfo {author} {\bibfnamefont {M.}~\bibnamefont {Serbyn}},\ and\ \bibinfo
  {author} {\bibfnamefont {E.}~\bibnamefont {Berg}},\ }\bibfield  {journal}
  {\bibinfo  {journal} {Physical Review Letters}\ }\textbf {\bibinfo {volume}
  {127}},\ \href {https://doi.org/10.1103/physrevlett.127.247001}
  {10.1103/physrevlett.127.247001} (\bibinfo {year} {2021})\BibitemShut
  {NoStop}%
\bibitem [{\citenamefont {Li}\ \emph {et~al.}(2020)\citenamefont {Li},
  \citenamefont {Ingham},\ and\ \citenamefont {Scammell}}]{li2020artificial}%
  \BibitemOpen
  \bibfield  {author} {\bibinfo {author} {\bibfnamefont {T.}~\bibnamefont
  {Li}}, \bibinfo {author} {\bibfnamefont {J.}~\bibnamefont {Ingham}},\ and\
  \bibinfo {author} {\bibfnamefont {H.~D.}\ \bibnamefont {Scammell}},\ }\href
  {https://doi.org/10.1103/PhysRevResearch.2.043155} {\bibfield  {journal}
  {\bibinfo  {journal} {Phys. Rev. Research}\ }\textbf {\bibinfo {volume}
  {2}},\ \bibinfo {pages} {043155} (\bibinfo {year} {2020})}\BibitemShut
  {NoStop}%
\bibitem [{sup()}]{supplementary}%
  \BibitemOpen
  \href@noop {} {}\bibinfo {note} {See Supplementary Material at url ... for
  details on the continuum model, the derivation of the screened Coulomb
  interaction, the linearised gap equation and the variational
  calculation.}\BibitemShut {Stop}%
\bibitem [{\citenamefont {Yu}\ \emph {et~al.}(2024)\citenamefont {Yu},
  \citenamefont {Herzog-Arbeitman}, \citenamefont {Wang}, \citenamefont
  {Vafek}, \citenamefont {Bernevig},\ and\ \citenamefont
  {Regnault}}]{Jiabin_2024}%
  \BibitemOpen
  \bibfield  {author} {\bibinfo {author} {\bibfnamefont {J.}~\bibnamefont
  {Yu}}, \bibinfo {author} {\bibfnamefont {J.}~\bibnamefont
  {Herzog-Arbeitman}}, \bibinfo {author} {\bibfnamefont {M.}~\bibnamefont
  {Wang}}, \bibinfo {author} {\bibfnamefont {O.}~\bibnamefont {Vafek}},
  \bibinfo {author} {\bibfnamefont {B.~A.}\ \bibnamefont {Bernevig}},\ and\
  \bibinfo {author} {\bibfnamefont {N.}~\bibnamefont {Regnault}},\ }\href
  {https://doi.org/10.1103/PhysRevB.109.045147} {\bibfield  {journal} {\bibinfo
   {journal} {Phys. Rev. B}\ }\textbf {\bibinfo {volume} {109}},\ \bibinfo
  {pages} {045147} (\bibinfo {year} {2024})}\BibitemShut {NoStop}%
\bibitem [{\citenamefont {Wang}\ \emph
  {et~al.}(2024{\natexlab{b}})\citenamefont {Wang}, \citenamefont {Zhang},
  \citenamefont {Liu}, \citenamefont {He}, \citenamefont {Xu}, \citenamefont
  {Ran}, \citenamefont {Cao},\ and\ \citenamefont
  {Xiao}}]{PhysRevLett.132.036501}%
  \BibitemOpen
  \bibfield  {author} {\bibinfo {author} {\bibfnamefont {C.}~\bibnamefont
  {Wang}}, \bibinfo {author} {\bibfnamefont {X.-W.}\ \bibnamefont {Zhang}},
  \bibinfo {author} {\bibfnamefont {X.}~\bibnamefont {Liu}}, \bibinfo {author}
  {\bibfnamefont {Y.}~\bibnamefont {He}}, \bibinfo {author} {\bibfnamefont
  {X.}~\bibnamefont {Xu}}, \bibinfo {author} {\bibfnamefont {Y.}~\bibnamefont
  {Ran}}, \bibinfo {author} {\bibfnamefont {T.}~\bibnamefont {Cao}},\ and\
  \bibinfo {author} {\bibfnamefont {D.}~\bibnamefont {Xiao}},\ }\href
  {https://doi.org/10.1103/PhysRevLett.132.036501} {\bibfield  {journal}
  {\bibinfo  {journal} {Phys. Rev. Lett.}\ }\textbf {\bibinfo {volume} {132}},\
  \bibinfo {pages} {036501} (\bibinfo {year} {2024}{\natexlab{b}})}\BibitemShut
  {NoStop}%
\bibitem [{\citenamefont {Kormányos}\ \emph {et~al.}(2015)\citenamefont
  {Kormányos}, \citenamefont {Burkard}, \citenamefont {Gmitra}, \citenamefont
  {Fabian}, \citenamefont {Zólyomi}, \citenamefont {Drummond},\ and\
  \citenamefont {Fal’ko}}]{Korm_nyos_2015}%
  \BibitemOpen
  \bibfield  {author} {\bibinfo {author} {\bibfnamefont {A.}~\bibnamefont
  {Kormányos}}, \bibinfo {author} {\bibfnamefont {G.}~\bibnamefont {Burkard}},
  \bibinfo {author} {\bibfnamefont {M.}~\bibnamefont {Gmitra}}, \bibinfo
  {author} {\bibfnamefont {J.}~\bibnamefont {Fabian}}, \bibinfo {author}
  {\bibfnamefont {V.}~\bibnamefont {Zólyomi}}, \bibinfo {author}
  {\bibfnamefont {N.~D.}\ \bibnamefont {Drummond}},\ and\ \bibinfo {author}
  {\bibfnamefont {V.}~\bibnamefont {Fal’ko}},\ }\href
  {https://doi.org/10.1088/2053-1583/2/2/022001} {\bibfield  {journal}
  {\bibinfo  {journal} {2D Materials}\ }\textbf {\bibinfo {volume} {2}},\
  \bibinfo {pages} {022001} (\bibinfo {year} {2015})}\BibitemShut {NoStop}%
\bibitem [{\citenamefont {Abouelkomsan}\ and\ \citenamefont
  {Fu}(2024)}]{abouelkomsan2024nonabelianspinhallinsulator}%
  \BibitemOpen
  \bibfield  {author} {\bibinfo {author} {\bibfnamefont {A.}~\bibnamefont
  {Abouelkomsan}}\ and\ \bibinfo {author} {\bibfnamefont {L.}~\bibnamefont
  {Fu}},\ }\href {https://arxiv.org/abs/2406.14617} {\bibinfo {title}
  {Non-abelian spin hall insulator}} (\bibinfo {year} {2024}),\ \Eprint
  {https://arxiv.org/abs/2406.14617} {arXiv:2406.14617 [cond-mat.mes-hall]}
  \BibitemShut {NoStop}%
\bibitem [{\citenamefont {Wang}\ \emph {et~al.}(2021)\citenamefont {Wang},
  \citenamefont {Cano}, \citenamefont {Millis}, \citenamefont {Liu},\ and\
  \citenamefont {Yang}}]{wang2021}%
  \BibitemOpen
  \bibfield  {author} {\bibinfo {author} {\bibfnamefont {J.}~\bibnamefont
  {Wang}}, \bibinfo {author} {\bibfnamefont {J.}~\bibnamefont {Cano}}, \bibinfo
  {author} {\bibfnamefont {A.~J.}\ \bibnamefont {Millis}}, \bibinfo {author}
  {\bibfnamefont {Z.}~\bibnamefont {Liu}},\ and\ \bibinfo {author}
  {\bibfnamefont {B.}~\bibnamefont {Yang}},\ }\href
  {https://doi.org/10.1103/PhysRevLett.127.246403} {\bibfield  {journal}
  {\bibinfo  {journal} {Phys. Rev. Lett.}\ }\textbf {\bibinfo {volume} {127}},\
  \bibinfo {pages} {246403} (\bibinfo {year} {2021})}\BibitemShut {NoStop}%
\bibitem [{\citenamefont {Ledwith}\ \emph {et~al.}(2020)\citenamefont
  {Ledwith}, \citenamefont {Tarnopolsky}, \citenamefont {Khalaf},\ and\
  \citenamefont {Vishwanath}}]{Ledwith2020}%
  \BibitemOpen
  \bibfield  {author} {\bibinfo {author} {\bibfnamefont {P.~J.}\ \bibnamefont
  {Ledwith}}, \bibinfo {author} {\bibfnamefont {G.}~\bibnamefont
  {Tarnopolsky}}, \bibinfo {author} {\bibfnamefont {E.}~\bibnamefont
  {Khalaf}},\ and\ \bibinfo {author} {\bibfnamefont {A.}~\bibnamefont
  {Vishwanath}},\ }\href {https://doi.org/10.1103/PhysRevResearch.2.023237}
  {\bibfield  {journal} {\bibinfo  {journal} {Phys. Rev. Res.}\ }\textbf
  {\bibinfo {volume} {2}},\ \bibinfo {pages} {023237} (\bibinfo {year}
  {2020})}\BibitemShut {NoStop}%
\bibitem [{\citenamefont {Guerci}\ \emph {et~al.}(2024)\citenamefont {Guerci},
  \citenamefont {Wang},\ and\ \citenamefont
  {Mora}}]{guerci2024layerskyrmionsidealchern}%
  \BibitemOpen
  \bibfield  {author} {\bibinfo {author} {\bibfnamefont {D.}~\bibnamefont
  {Guerci}}, \bibinfo {author} {\bibfnamefont {J.}~\bibnamefont {Wang}},\ and\
  \bibinfo {author} {\bibfnamefont {C.}~\bibnamefont {Mora}},\ }\href
  {https://arxiv.org/abs/2408.12652} {\bibinfo {title} {Layer skyrmions for
  ideal chern bands and twisted bilayer graphene}} (\bibinfo {year} {2024}),\
  \Eprint {https://arxiv.org/abs/2408.12652} {arXiv:2408.12652
  [cond-mat.mes-hall]} \BibitemShut {NoStop}%
\bibitem [{\citenamefont {Long}\ \emph {et~al.}(2024)\citenamefont {Long},
  \citenamefont {Jimeno-Pozo}, \citenamefont {Sainz-Cruz}, \citenamefont
  {Pantaleón},\ and\ \citenamefont {Guinea}}]{Long_2024}%
  \BibitemOpen
  \bibfield  {author} {\bibinfo {author} {\bibfnamefont {M.}~\bibnamefont
  {Long}}, \bibinfo {author} {\bibfnamefont {A.}~\bibnamefont {Jimeno-Pozo}},
  \bibinfo {author} {\bibfnamefont {H.}~\bibnamefont {Sainz-Cruz}}, \bibinfo
  {author} {\bibfnamefont {P.~A.}\ \bibnamefont {Pantaleón}},\ and\ \bibinfo
  {author} {\bibfnamefont {F.}~\bibnamefont {Guinea}},\ }\bibfield  {journal}
  {\bibinfo  {journal} {Proceedings of the National Academy of Sciences}\
  }\textbf {\bibinfo {volume} {121}},\ \href
  {https://doi.org/10.1073/pnas.2405259121} {10.1073/pnas.2405259121} (\bibinfo
  {year} {2024})\BibitemShut {NoStop}%
\bibitem [{\citenamefont {Cavicchi}\ \emph {et~al.}(2024)\citenamefont
  {Cavicchi}, \citenamefont {Torre}, \citenamefont {Jarillo-Herrero},
  \citenamefont {Koppens},\ and\ \citenamefont {Polini}}]{cavicchi_2024}%
  \BibitemOpen
  \bibfield  {author} {\bibinfo {author} {\bibfnamefont {L.}~\bibnamefont
  {Cavicchi}}, \bibinfo {author} {\bibfnamefont {I.}~\bibnamefont {Torre}},
  \bibinfo {author} {\bibfnamefont {P.}~\bibnamefont {Jarillo-Herrero}},
  \bibinfo {author} {\bibfnamefont {F.~H.~L.}\ \bibnamefont {Koppens}},\ and\
  \bibinfo {author} {\bibfnamefont {M.}~\bibnamefont {Polini}},\ }\href
  {https://doi.org/10.1103/PhysRevB.110.045431} {\bibfield  {journal} {\bibinfo
   {journal} {Phys. Rev. B}\ }\textbf {\bibinfo {volume} {110}},\ \bibinfo
  {pages} {045431} (\bibinfo {year} {2024})}\BibitemShut {NoStop}%
\bibitem [{\citenamefont {Papaj}\ and\ \citenamefont
  {Lewandowski}(2023)}]{Lewandowski_2023}%
  \BibitemOpen
  \bibfield  {author} {\bibinfo {author} {\bibfnamefont {M.}~\bibnamefont
  {Papaj}}\ and\ \bibinfo {author} {\bibfnamefont {C.}~\bibnamefont
  {Lewandowski}},\ }\href {https://doi.org/10.1126/sciadv.adg3262} {\bibfield
  {journal} {\bibinfo  {journal} {Science Advances}\ }\textbf {\bibinfo
  {volume} {9}},\ \bibinfo {pages} {eadg3262} (\bibinfo {year} {2023})},\
  \Eprint
  {https://arxiv.org/abs/https://www.science.org/doi/pdf/10.1126/sciadv.adg3262}
  {https://www.science.org/doi/pdf/10.1126/sciadv.adg3262} \BibitemShut
  {NoStop}%
\bibitem [{\citenamefont {Li}\ \emph {et~al.}(2022)\citenamefont {Li},
  \citenamefont {Geier}, \citenamefont {Ingham},\ and\ \citenamefont
  {Scammell}}]{li2022higher}%
  \BibitemOpen
  \bibfield  {author} {\bibinfo {author} {\bibfnamefont {T.}~\bibnamefont
  {Li}}, \bibinfo {author} {\bibfnamefont {M.}~\bibnamefont {Geier}}, \bibinfo
  {author} {\bibfnamefont {J.}~\bibnamefont {Ingham}},\ and\ \bibinfo {author}
  {\bibfnamefont {H.~D.}\ \bibnamefont {Scammell}},\ }\href
  {https://doi.org/10.1088/2053-1583/ac4060} {\bibfield  {journal} {\bibinfo
  {journal} {2D Mater.}\ }\textbf {\bibinfo {volume} {9}},\ \bibinfo {pages}
  {015031} (\bibinfo {year} {2022})}\BibitemShut {NoStop}%
\bibitem [{\citenamefont {Scammell}\ \emph
  {et~al.}(2022{\natexlab{a}})\citenamefont {Scammell}, \citenamefont {Ingham},
  \citenamefont {Geier},\ and\ \citenamefont {Li}}]{scammell2022intrinsic}%
  \BibitemOpen
  \bibfield  {author} {\bibinfo {author} {\bibfnamefont {H.~D.}\ \bibnamefont
  {Scammell}}, \bibinfo {author} {\bibfnamefont {J.}~\bibnamefont {Ingham}},
  \bibinfo {author} {\bibfnamefont {M.}~\bibnamefont {Geier}},\ and\ \bibinfo
  {author} {\bibfnamefont {T.}~\bibnamefont {Li}},\ }\href
  {https://doi.org/10.1103/PhysRevB.105.195149} {\bibfield  {journal} {\bibinfo
   {journal} {Phys. Rev. B}\ }\textbf {\bibinfo {volume} {105}},\ \bibinfo
  {pages} {195149} (\bibinfo {year} {2022}{\natexlab{a}})}\BibitemShut
  {NoStop}%
\bibitem [{\citenamefont {Giuliani}\ and\ \citenamefont
  {Vignale}(2005)}]{Giuliani_Vignale_2005}%
  \BibitemOpen
  \bibfield  {author} {\bibinfo {author} {\bibfnamefont {G.}~\bibnamefont
  {Giuliani}}\ and\ \bibinfo {author} {\bibfnamefont {G.}~\bibnamefont
  {Vignale}},\ }\href@noop {} {\emph {\bibinfo {title} {Quantum Theory of the
  Electron Liquid}}}\ (\bibinfo  {publisher} {Cambridge University Press},\
  \bibinfo {year} {2005})\BibitemShut {NoStop}%
\bibitem [{\citenamefont {Bogoljubov}\ \emph {et~al.}(1958)\citenamefont
  {Bogoljubov}, \citenamefont {Tolmachov},\ and\ \citenamefont
  {Širkov}}]{Tolmachov1958}%
  \BibitemOpen
  \bibfield  {author} {\bibinfo {author} {\bibfnamefont {N.~N.}\ \bibnamefont
  {Bogoljubov}}, \bibinfo {author} {\bibfnamefont {V.~V.}\ \bibnamefont
  {Tolmachov}},\ and\ \bibinfo {author} {\bibfnamefont {D.~V.}\ \bibnamefont
  {Širkov}},\ }\href
  {https://doi.org/https://doi.org/10.1002/prop.19580061102} {\bibfield
  {journal} {\bibinfo  {journal} {Fortschritte der Physik}\ }\textbf {\bibinfo
  {volume} {6}},\ \bibinfo {pages} {605} (\bibinfo {year} {1958})},\ \Eprint
  {https://arxiv.org/abs/https://onlinelibrary.wiley.com/doi/pdf/10.1002/prop.19580061102}
  {https://onlinelibrary.wiley.com/doi/pdf/10.1002/prop.19580061102}
  \BibitemShut {NoStop}%
\bibitem [{\citenamefont {Morel}\ and\ \citenamefont
  {Anderson}(1962)}]{anderson1962}%
  \BibitemOpen
  \bibfield  {author} {\bibinfo {author} {\bibfnamefont {P.}~\bibnamefont
  {Morel}}\ and\ \bibinfo {author} {\bibfnamefont {P.~W.}\ \bibnamefont
  {Anderson}},\ }\href {https://doi.org/10.1103/PhysRev.125.1263} {\bibfield
  {journal} {\bibinfo  {journal} {Phys. Rev.}\ }\textbf {\bibinfo {volume}
  {125}},\ \bibinfo {pages} {1263} (\bibinfo {year} {1962})}\BibitemShut
  {NoStop}%
\bibitem [{\citenamefont {Rietschel}\ and\ \citenamefont
  {Sham}(1983)}]{sham1983prb}%
  \BibitemOpen
  \bibfield  {author} {\bibinfo {author} {\bibfnamefont {H.}~\bibnamefont
  {Rietschel}}\ and\ \bibinfo {author} {\bibfnamefont {L.~J.}\ \bibnamefont
  {Sham}},\ }\href {https://doi.org/10.1103/PhysRevB.28.5100} {\bibfield
  {journal} {\bibinfo  {journal} {Phys. Rev. B}\ }\textbf {\bibinfo {volume}
  {28}},\ \bibinfo {pages} {5100} (\bibinfo {year} {1983})}\BibitemShut
  {NoStop}%
\bibitem [{\citenamefont {Chubukov}\ \emph {et~al.}(2019)\citenamefont
  {Chubukov}, \citenamefont {Prokof'ev},\ and\ \citenamefont
  {Svistunov}}]{chubukov_prb_2019}%
  \BibitemOpen
  \bibfield  {author} {\bibinfo {author} {\bibfnamefont {A.}~\bibnamefont
  {Chubukov}}, \bibinfo {author} {\bibfnamefont {N.~V.}\ \bibnamefont
  {Prokof'ev}},\ and\ \bibinfo {author} {\bibfnamefont {B.~V.}\ \bibnamefont
  {Svistunov}},\ }\href {https://doi.org/10.1103/PhysRevB.100.064513}
  {\bibfield  {journal} {\bibinfo  {journal} {Phys. Rev. B}\ }\textbf {\bibinfo
  {volume} {100}},\ \bibinfo {pages} {064513} (\bibinfo {year}
  {2019})}\BibitemShut {NoStop}%
\bibitem [{\citenamefont {Cai}\ \emph {et~al.}(2022)\citenamefont {Cai},
  \citenamefont {Wang}, \citenamefont {Prokof'ev}, \citenamefont {Svistunov},\
  and\ \citenamefont {Chen}}]{kunchen2022prb}%
  \BibitemOpen
  \bibfield  {author} {\bibinfo {author} {\bibfnamefont {X.}~\bibnamefont
  {Cai}}, \bibinfo {author} {\bibfnamefont {T.}~\bibnamefont {Wang}}, \bibinfo
  {author} {\bibfnamefont {N.~V.}\ \bibnamefont {Prokof'ev}}, \bibinfo {author}
  {\bibfnamefont {B.~V.}\ \bibnamefont {Svistunov}},\ and\ \bibinfo {author}
  {\bibfnamefont {K.}~\bibnamefont {Chen}},\ }\href
  {https://doi.org/10.1103/PhysRevB.106.L220502} {\bibfield  {journal}
  {\bibinfo  {journal} {Phys. Rev. B}\ }\textbf {\bibinfo {volume} {106}},\
  \bibinfo {pages} {L220502} (\bibinfo {year} {2022})}\BibitemShut {NoStop}%
\bibitem [{\citenamefont {Simonato}\ \emph {et~al.}(2023)\citenamefont
  {Simonato}, \citenamefont {Katsnelson},\ and\ \citenamefont
  {R\"osner}}]{malte_prb_2023}%
  \BibitemOpen
  \bibfield  {author} {\bibinfo {author} {\bibfnamefont {M.}~\bibnamefont
  {Simonato}}, \bibinfo {author} {\bibfnamefont {M.~I.}\ \bibnamefont
  {Katsnelson}},\ and\ \bibinfo {author} {\bibfnamefont {M.}~\bibnamefont
  {R\"osner}},\ }\href {https://doi.org/10.1103/PhysRevB.108.064513} {\bibfield
   {journal} {\bibinfo  {journal} {Phys. Rev. B}\ }\textbf {\bibinfo {volume}
  {108}},\ \bibinfo {pages} {064513} (\bibinfo {year} {2023})}\BibitemShut
  {NoStop}%
\bibitem [{\citenamefont {Murakami}\ and\ \citenamefont
  {Nagaosa}(2003)}]{Murakami_2003}%
  \BibitemOpen
  \bibfield  {author} {\bibinfo {author} {\bibfnamefont {S.}~\bibnamefont
  {Murakami}}\ and\ \bibinfo {author} {\bibfnamefont {N.}~\bibnamefont
  {Nagaosa}},\ }\href {https://doi.org/10.1103/PhysRevLett.90.057002}
  {\bibfield  {journal} {\bibinfo  {journal} {Phys. Rev. Lett.}\ }\textbf
  {\bibinfo {volume} {90}},\ \bibinfo {pages} {057002} (\bibinfo {year}
  {2003})}\BibitemShut {NoStop}%
\bibitem [{\citenamefont {Shtyk}\ \emph {et~al.}(2017)\citenamefont {Shtyk},
  \citenamefont {Goldstein},\ and\ \citenamefont {Chamon}}]{chamon_2017}%
  \BibitemOpen
  \bibfield  {author} {\bibinfo {author} {\bibfnamefont {A.}~\bibnamefont
  {Shtyk}}, \bibinfo {author} {\bibfnamefont {G.}~\bibnamefont {Goldstein}},\
  and\ \bibinfo {author} {\bibfnamefont {C.}~\bibnamefont {Chamon}},\
  }\bibfield  {journal} {\bibinfo  {journal} {Physical Review B}\ }\textbf
  {\bibinfo {volume} {95}},\ \href {https://doi.org/10.1103/physrevb.95.035137}
  {10.1103/physrevb.95.035137} (\bibinfo {year} {2017})\BibitemShut {NoStop}%
\bibitem [{\citenamefont {Chandrasekaran}\ \emph {et~al.}(2020)\citenamefont
  {Chandrasekaran}, \citenamefont {Shtyk}, \citenamefont {Betouras},\ and\
  \citenamefont {Chamon}}]{chandrasekaran2020catastrophe}%
  \BibitemOpen
  \bibfield  {author} {\bibinfo {author} {\bibfnamefont {A.}~\bibnamefont
  {Chandrasekaran}}, \bibinfo {author} {\bibfnamefont {A.}~\bibnamefont
  {Shtyk}}, \bibinfo {author} {\bibfnamefont {J.~J.}\ \bibnamefont
  {Betouras}},\ and\ \bibinfo {author} {\bibfnamefont {C.}~\bibnamefont
  {Chamon}},\ }\href@noop {} {\bibfield  {journal} {\bibinfo  {journal} {Phys.
  Rev. Research}\ }\textbf {\bibinfo {volume} {2}},\ \bibinfo {pages} {013355}
  (\bibinfo {year} {2020})}\BibitemShut {NoStop}%
\bibitem [{\citenamefont {Yuan}\ \emph {et~al.}(2019)\citenamefont {Yuan},
  \citenamefont {Isobe},\ and\ \citenamefont {Fu}}]{LFu_HOVHS_2019}%
  \BibitemOpen
  \bibfield  {author} {\bibinfo {author} {\bibfnamefont {N.~F.~Q.}\
  \bibnamefont {Yuan}}, \bibinfo {author} {\bibfnamefont {H.}~\bibnamefont
  {Isobe}},\ and\ \bibinfo {author} {\bibfnamefont {L.}~\bibnamefont {Fu}},\
  }\bibfield  {journal} {\bibinfo  {journal} {Nature Communications}\ }\textbf
  {\bibinfo {volume} {10}},\ \href {https://doi.org/10.1038/s41467-019-13670-9}
  {10.1038/s41467-019-13670-9} (\bibinfo {year} {2019})\BibitemShut {NoStop}%
\bibitem [{\citenamefont {Zang}\ \emph {et~al.}(2021)\citenamefont {Zang},
  \citenamefont {Wang}, \citenamefont {Cano},\ and\ \citenamefont
  {Millis}}]{zangj_2021}%
  \BibitemOpen
  \bibfield  {author} {\bibinfo {author} {\bibfnamefont {J.}~\bibnamefont
  {Zang}}, \bibinfo {author} {\bibfnamefont {J.}~\bibnamefont {Wang}}, \bibinfo
  {author} {\bibfnamefont {J.}~\bibnamefont {Cano}},\ and\ \bibinfo {author}
  {\bibfnamefont {A.~J.}\ \bibnamefont {Millis}},\ }\href
  {https://doi.org/10.1103/PhysRevB.104.075150} {\bibfield  {journal} {\bibinfo
   {journal} {Phys. Rev. B}\ }\textbf {\bibinfo {volume} {104}},\ \bibinfo
  {pages} {075150} (\bibinfo {year} {2021})}\BibitemShut {NoStop}%
\bibitem [{\citenamefont {Classen}\ and\ \citenamefont
  {Betouras}(2024)}]{classen2024highordervanhovesingularities}%
  \BibitemOpen
  \bibfield  {author} {\bibinfo {author} {\bibfnamefont {L.}~\bibnamefont
  {Classen}}\ and\ \bibinfo {author} {\bibfnamefont {J.~J.}\ \bibnamefont
  {Betouras}},\ }\href {https://arxiv.org/abs/2405.20226} {\bibinfo {title}
  {High-order van hove singularities and their connection to flat bands}}
  (\bibinfo {year} {2024}),\ \Eprint {https://arxiv.org/abs/2405.20226}
  {arXiv:2405.20226 [cond-mat.str-el]} \BibitemShut {NoStop}%
\bibitem [{\citenamefont {Sigrist}\ and\ \citenamefont
  {Ueda}(1991)}]{sigrist_1991}%
  \BibitemOpen
  \bibfield  {author} {\bibinfo {author} {\bibfnamefont {M.}~\bibnamefont
  {Sigrist}}\ and\ \bibinfo {author} {\bibfnamefont {K.}~\bibnamefont {Ueda}},\
  }\href {https://doi.org/10.1103/RevModPhys.63.239} {\bibfield  {journal}
  {\bibinfo  {journal} {Rev. Mod. Phys.}\ }\textbf {\bibinfo {volume} {63}},\
  \bibinfo {pages} {239} (\bibinfo {year} {1991})}\BibitemShut {NoStop}%
\bibitem [{\citenamefont {Uematsu}\ \emph {et~al.}(2019)\citenamefont
  {Uematsu}, \citenamefont {Mizushima}, \citenamefont {Tsuruta}, \citenamefont
  {Fujimoto},\ and\ \citenamefont {Sauls}}]{Sauls_2019}%
  \BibitemOpen
  \bibfield  {author} {\bibinfo {author} {\bibfnamefont {H.}~\bibnamefont
  {Uematsu}}, \bibinfo {author} {\bibfnamefont {T.}~\bibnamefont {Mizushima}},
  \bibinfo {author} {\bibfnamefont {A.}~\bibnamefont {Tsuruta}}, \bibinfo
  {author} {\bibfnamefont {S.}~\bibnamefont {Fujimoto}},\ and\ \bibinfo
  {author} {\bibfnamefont {J.~A.}\ \bibnamefont {Sauls}},\ }\href
  {https://doi.org/10.1103/PhysRevLett.123.237001} {\bibfield  {journal}
  {\bibinfo  {journal} {Phys. Rev. Lett.}\ }\textbf {\bibinfo {volume} {123}},\
  \bibinfo {pages} {237001} (\bibinfo {year} {2019})}\BibitemShut {NoStop}%
\bibitem [{vsh()}]{vshape}%
  \BibitemOpen
  \href@noop {} {}\bibinfo {note} {The grid used in
  Fig.~\ref{fig:variational_results} is not dense enough at the Dirac points to
  see a clear V-shape in the density of states, but the Dirac cones are easily
  seen in the excitation spectrum as shown in the supplemental.}\BibitemShut
  {Stop}%
\bibitem [{\citenamefont {Read}\ and\ \citenamefont
  {Green}(2000{\natexlab{a}})}]{read2000prb}%
  \BibitemOpen
  \bibfield  {author} {\bibinfo {author} {\bibfnamefont {N.}~\bibnamefont
  {Read}}\ and\ \bibinfo {author} {\bibfnamefont {D.}~\bibnamefont {Green}},\
  }\href {https://doi.org/10.1103/PhysRevB.61.10267} {\bibfield  {journal}
  {\bibinfo  {journal} {Phys. Rev. B}\ }\textbf {\bibinfo {volume} {61}},\
  \bibinfo {pages} {10267} (\bibinfo {year} {2000}{\natexlab{a}})}\BibitemShut
  {NoStop}%
\bibitem [{\citenamefont {Altland}\ and\ \citenamefont
  {Zirnbauer}(1997)}]{Altland_1997}%
  \BibitemOpen
  \bibfield  {author} {\bibinfo {author} {\bibfnamefont {A.}~\bibnamefont
  {Altland}}\ and\ \bibinfo {author} {\bibfnamefont {M.~R.}\ \bibnamefont
  {Zirnbauer}},\ }\href {https://doi.org/10.1103/physrevb.55.1142} {\bibfield
  {journal} {\bibinfo  {journal} {Physical Review B}\ }\textbf {\bibinfo
  {volume} {55}},\ \bibinfo {pages} {1142–1161} (\bibinfo {year}
  {1997})}\BibitemShut {NoStop}%
\bibitem [{\citenamefont {Schnyder}\ \emph {et~al.}(2008)\citenamefont
  {Schnyder}, \citenamefont {Ryu}, \citenamefont {Furusaki},\ and\
  \citenamefont {Ludwig}}]{Schnyder_2008}%
  \BibitemOpen
  \bibfield  {author} {\bibinfo {author} {\bibfnamefont {A.~P.}\ \bibnamefont
  {Schnyder}}, \bibinfo {author} {\bibfnamefont {S.}~\bibnamefont {Ryu}},
  \bibinfo {author} {\bibfnamefont {A.}~\bibnamefont {Furusaki}},\ and\
  \bibinfo {author} {\bibfnamefont {A.~W.~W.}\ \bibnamefont {Ludwig}},\ }\href
  {https://doi.org/10.1103/PhysRevB.78.195125} {\bibfield  {journal} {\bibinfo
  {journal} {Phys. Rev. B}\ }\textbf {\bibinfo {volume} {78}},\ \bibinfo
  {pages} {195125} (\bibinfo {year} {2008})}\BibitemShut {NoStop}%
\bibitem [{\citenamefont {Ivanov}(2001)}]{ivanov_prl_2001}%
  \BibitemOpen
  \bibfield  {author} {\bibinfo {author} {\bibfnamefont {D.~A.}\ \bibnamefont
  {Ivanov}},\ }\href {https://doi.org/10.1103/PhysRevLett.86.268} {\bibfield
  {journal} {\bibinfo  {journal} {Phys. Rev. Lett.}\ }\textbf {\bibinfo
  {volume} {86}},\ \bibinfo {pages} {268} (\bibinfo {year} {2001})}\BibitemShut
  {NoStop}%
\bibitem [{\citenamefont {Read}\ and\ \citenamefont
  {Green}(2000{\natexlab{b}})}]{read_green_2000}%
  \BibitemOpen
  \bibfield  {author} {\bibinfo {author} {\bibfnamefont {N.}~\bibnamefont
  {Read}}\ and\ \bibinfo {author} {\bibfnamefont {D.}~\bibnamefont {Green}},\
  }\href {https://doi.org/10.1103/PhysRevB.61.10267} {\bibfield  {journal}
  {\bibinfo  {journal} {Phys. Rev. B}\ }\textbf {\bibinfo {volume} {61}},\
  \bibinfo {pages} {10267} (\bibinfo {year} {2000}{\natexlab{b}})}\BibitemShut
  {NoStop}%
\bibitem [{\citenamefont {Stone}\ and\ \citenamefont {Roy}(2004)}]{stone2004}%
  \BibitemOpen
  \bibfield  {author} {\bibinfo {author} {\bibfnamefont {M.}~\bibnamefont
  {Stone}}\ and\ \bibinfo {author} {\bibfnamefont {R.}~\bibnamefont {Roy}},\
  }\href {https://doi.org/10.1103/PhysRevB.69.184511} {\bibfield  {journal}
  {\bibinfo  {journal} {Phys. Rev. B}\ }\textbf {\bibinfo {volume} {69}},\
  \bibinfo {pages} {184511} (\bibinfo {year} {2004})}\BibitemShut {NoStop}%
\bibitem [{\citenamefont {{Leggett}}(1966)}]{Leggett}%
  \BibitemOpen
  \bibfield  {author} {\bibinfo {author} {\bibfnamefont {A.~J.}\ \bibnamefont
  {{Leggett}}},\ }\href {https://doi.org/10.1143/PTP.36.901} {\bibfield
  {journal} {\bibinfo  {journal} {Progress of Theoretical Physics}\ }\textbf
  {\bibinfo {volume} {36}},\ \bibinfo {pages} {901} (\bibinfo {year}
  {1966})}\BibitemShut {NoStop}%
\bibitem [{\citenamefont {Nadeem}\ \emph {et~al.}(2023)\citenamefont {Nadeem},
  \citenamefont {Fuhrer},\ and\ \citenamefont {Wang}}]{nadeem_SDE}%
  \BibitemOpen
  \bibfield  {author} {\bibinfo {author} {\bibfnamefont {M.}~\bibnamefont
  {Nadeem}}, \bibinfo {author} {\bibfnamefont {M.~S.}\ \bibnamefont {Fuhrer}},\
  and\ \bibinfo {author} {\bibfnamefont {X.}~\bibnamefont {Wang}},\ }\href@noop
  {} {\bibfield  {journal} {\bibinfo  {journal} {Nature Reviews Physics}\
  }\textbf {\bibinfo {volume} {5}},\ \bibinfo {pages} {558} (\bibinfo {year}
  {2023})}\BibitemShut {NoStop}%
\bibitem [{\citenamefont {Ando}\ \emph {et~al.}(2020)\citenamefont {Ando},
  \citenamefont {Miyasaka}, \citenamefont {Li}, \citenamefont {Ishizuka},
  \citenamefont {Arakawa}, \citenamefont {Shiota}, \citenamefont {Moriyama},
  \citenamefont {Yanase},\ and\ \citenamefont {Ono}}]{ando_SDE}%
  \BibitemOpen
  \bibfield  {author} {\bibinfo {author} {\bibfnamefont {F.}~\bibnamefont
  {Ando}}, \bibinfo {author} {\bibfnamefont {Y.}~\bibnamefont {Miyasaka}},
  \bibinfo {author} {\bibfnamefont {T.}~\bibnamefont {Li}}, \bibinfo {author}
  {\bibfnamefont {J.}~\bibnamefont {Ishizuka}}, \bibinfo {author}
  {\bibfnamefont {T.}~\bibnamefont {Arakawa}}, \bibinfo {author} {\bibfnamefont
  {Y.}~\bibnamefont {Shiota}}, \bibinfo {author} {\bibfnamefont
  {T.}~\bibnamefont {Moriyama}}, \bibinfo {author} {\bibfnamefont
  {Y.}~\bibnamefont {Yanase}},\ and\ \bibinfo {author} {\bibfnamefont
  {T.}~\bibnamefont {Ono}},\ }\href@noop {} {\bibfield  {journal} {\bibinfo
  {journal} {Nature}\ }\textbf {\bibinfo {volume} {584}},\ \bibinfo {pages}
  {373} (\bibinfo {year} {2020})}\BibitemShut {NoStop}%
\bibitem [{\citenamefont {Davydova}\ \emph {et~al.}(2022)\citenamefont
  {Davydova}, \citenamefont {Prembabu},\ and\ \citenamefont
  {Fu}}]{Davydova_JDE}%
  \BibitemOpen
  \bibfield  {author} {\bibinfo {author} {\bibfnamefont {M.}~\bibnamefont
  {Davydova}}, \bibinfo {author} {\bibfnamefont {S.}~\bibnamefont {Prembabu}},\
  and\ \bibinfo {author} {\bibfnamefont {L.}~\bibnamefont {Fu}},\ }\href@noop
  {} {\bibfield  {journal} {\bibinfo  {journal} {Science advances}\ }\textbf
  {\bibinfo {volume} {8}},\ \bibinfo {pages} {eabo0309} (\bibinfo {year}
  {2022})}\BibitemShut {NoStop}%
\bibitem [{\citenamefont {Volkov}\ \emph {et~al.}(2024)\citenamefont {Volkov},
  \citenamefont {Lantagne-Hurtubise}, \citenamefont {Tummuru}, \citenamefont
  {Plugge}, \citenamefont {Pixley},\ and\ \citenamefont {Franz}}]{Volkov_2024}%
  \BibitemOpen
  \bibfield  {author} {\bibinfo {author} {\bibfnamefont {P.~A.}\ \bibnamefont
  {Volkov}}, \bibinfo {author} {\bibfnamefont {E.}~\bibnamefont
  {Lantagne-Hurtubise}}, \bibinfo {author} {\bibfnamefont {T.}~\bibnamefont
  {Tummuru}}, \bibinfo {author} {\bibfnamefont {S.}~\bibnamefont {Plugge}},
  \bibinfo {author} {\bibfnamefont {J.~H.}\ \bibnamefont {Pixley}},\ and\
  \bibinfo {author} {\bibfnamefont {M.}~\bibnamefont {Franz}},\ }\bibfield
  {journal} {\bibinfo  {journal} {Physical Review B}\ }\textbf {\bibinfo
  {volume} {109}},\ \href {https://doi.org/10.1103/physrevb.109.094518}
  {10.1103/physrevb.109.094518} (\bibinfo {year} {2024})\BibitemShut {NoStop}%
\bibitem [{\citenamefont {Scammell}\ \emph
  {et~al.}(2022{\natexlab{b}})\citenamefont {Scammell}, \citenamefont {Li},\
  and\ \citenamefont {Scheurer}}]{scammell2022theory}%
  \BibitemOpen
  \bibfield  {author} {\bibinfo {author} {\bibfnamefont {H.~D.}\ \bibnamefont
  {Scammell}}, \bibinfo {author} {\bibfnamefont {J.}~\bibnamefont {Li}},\ and\
  \bibinfo {author} {\bibfnamefont {M.~S.}\ \bibnamefont {Scheurer}},\
  }\href@noop {} {\bibfield  {journal} {\bibinfo  {journal} {2D Mater.}\
  }\textbf {\bibinfo {volume} {9}},\ \bibinfo {pages} {025027} (\bibinfo {year}
  {2022}{\natexlab{b}})}\BibitemShut {NoStop}%
\bibitem [{\citenamefont {Narita}\ \emph {et~al.}(2022)\citenamefont {Narita},
  \citenamefont {Ishizuka}, \citenamefont {Kawarazaki}, \citenamefont {Kan},
  \citenamefont {Shiota}, \citenamefont {Moriyama}, \citenamefont {Shimakawa},
  \citenamefont {Ognev}, \citenamefont {Samardak}, \citenamefont {Yanase} \emph
  {et~al.}}]{narita_fieldfree}%
  \BibitemOpen
  \bibfield  {author} {\bibinfo {author} {\bibfnamefont {H.}~\bibnamefont
  {Narita}}, \bibinfo {author} {\bibfnamefont {J.}~\bibnamefont {Ishizuka}},
  \bibinfo {author} {\bibfnamefont {R.}~\bibnamefont {Kawarazaki}}, \bibinfo
  {author} {\bibfnamefont {D.}~\bibnamefont {Kan}}, \bibinfo {author}
  {\bibfnamefont {Y.}~\bibnamefont {Shiota}}, \bibinfo {author} {\bibfnamefont
  {T.}~\bibnamefont {Moriyama}}, \bibinfo {author} {\bibfnamefont
  {Y.}~\bibnamefont {Shimakawa}}, \bibinfo {author} {\bibfnamefont {A.~V.}\
  \bibnamefont {Ognev}}, \bibinfo {author} {\bibfnamefont {A.~S.}\ \bibnamefont
  {Samardak}}, \bibinfo {author} {\bibfnamefont {Y.}~\bibnamefont {Yanase}},
  \emph {et~al.},\ }\href@noop {} {\bibfield  {journal} {\bibinfo  {journal}
  {Nature Nanotechnology}\ }\textbf {\bibinfo {volume} {17}},\ \bibinfo {pages}
  {823} (\bibinfo {year} {2022})}\BibitemShut {NoStop}%
\bibitem [{\citenamefont {Davydova}\ \emph {et~al.}(2024)\citenamefont
  {Davydova}, \citenamefont {Geier},\ and\ \citenamefont
  {Fu}}]{davydova_nonrecip_super}%
  \BibitemOpen
  \bibfield  {author} {\bibinfo {author} {\bibfnamefont {M.}~\bibnamefont
  {Davydova}}, \bibinfo {author} {\bibfnamefont {M.}~\bibnamefont {Geier}},\
  and\ \bibinfo {author} {\bibfnamefont {L.}~\bibnamefont {Fu}},\ }\href
  {https://arxiv.org/abs/2407.01681} {\bibinfo {title} {Nonreciprocal
  superconductivity}} (\bibinfo {year} {2024}),\ \Eprint
  {https://arxiv.org/abs/2407.01681} {arXiv:2407.01681 [cond-mat.supr-con]}
  \BibitemShut {NoStop}%
\bibitem [{\citenamefont {Li}\ \emph {et~al.}(2023)\citenamefont {Li},
  \citenamefont {Kuang}, \citenamefont {Jimeno-Pozo}, \citenamefont
  {Sainz-Cruz}, \citenamefont {Zhan}, \citenamefont {Yuan},\ and\ \citenamefont
  {Guinea}}]{Li_2023}%
  \BibitemOpen
  \bibfield  {author} {\bibinfo {author} {\bibfnamefont {Z.}~\bibnamefont
  {Li}}, \bibinfo {author} {\bibfnamefont {X.}~\bibnamefont {Kuang}}, \bibinfo
  {author} {\bibfnamefont {A.}~\bibnamefont {Jimeno-Pozo}}, \bibinfo {author}
  {\bibfnamefont {H.}~\bibnamefont {Sainz-Cruz}}, \bibinfo {author}
  {\bibfnamefont {Z.}~\bibnamefont {Zhan}}, \bibinfo {author} {\bibfnamefont
  {S.}~\bibnamefont {Yuan}},\ and\ \bibinfo {author} {\bibfnamefont
  {F.}~\bibnamefont {Guinea}},\ }\href
  {https://doi.org/10.1103/PhysRevB.108.045404} {\bibfield  {journal} {\bibinfo
   {journal} {Phys. Rev. B}\ }\textbf {\bibinfo {volume} {108}},\ \bibinfo
  {pages} {045404} (\bibinfo {year} {2023})}\BibitemShut {NoStop}%
\end{thebibliography}%

% ------------------------------------------------------------------------ %
% ------------------------------------------------------------------------ %
% ------------------------------------------------------------------------ %

\onecolumngrid
	\newpage
	\makeatletter 
	
	\begin{center}
		\textbf{\large Supplementary material for \\ ``\@title ''} \\[10pt]
Daniele Guerci$^1$, Daniel Kaplan$^2$, Julian Ingham$^{3,1}$, J. H. Pixley$^{2,1}$, Andrew Millis$^{3,1}$     \\
\textit{$^1$Center for Computational Quantum Physics, Flatiron Institute, 162 5th Avenue, NY 10010, USA}\\
\textit{$^2$Department of Physics and Astronomy, Center for Materials Theory, Rutgers University, Piscataway, New Jersey 08854, USA}\\
\textit{$^3$Department of Physics, Columbia University, 538 Wst 120th Street, New York, NY 10027, USA}\\
	\end{center}
	\vspace{20pt}
	
	\setcounter{figure}{0}
	\setcounter{section}{0}
	\setcounter{equation}{0}
	
	\renewcommand{\thefigure}{S\@arabic\c@figure}
	\makeatother
	
	\onecolumngrid
	\appendix

\onecolumngrid
% \newpage
\makeatletter 

% ------------------------------------------------------------------------ %
% ------------------------------------------------------------------------ %
% ------------------------------------------------------------------------ %

These supplementary materials contain the details of analytic calculations as well as additional numerical details supporting the results presented in the main text.
Sec.~\ref{app:cm_FS} contains the details on the continuum Hamiltonian and on the bandstructure properties~\ref{app:bands}.  
In Sec.~\ref{app:screened_Coulomb} we explicitly detail the evaluation of the dressed Coulomb interaction, the properties of the pairing potential~\ref{app:pairing_pot} and of the particle-particle correlation function~\ref{app:particle_particle}. Finally, Sec.~\ref{app:bcs} details the Hartree-Fock self-consistent calculation for the superconducting ordering.

\section{Continuum model: topology and Fermi surface properties}
\label{app:cm_FS}

To characterize the single particle properties we employ the continuum model~\cite{FW_PRL_2019}: 
\begin{equation}\label{continuum_modeling}
    H_\uparrow(\br) = \begin{pmatrix}
        -\frac{(\bk-K_+)^2}{2m} + V_+(\br) + u_D & T(\br) \\ 
        T^*(\br) & -\frac{(\bk-K_-)^2}{2m} + V_-(\br) - u_D
    \end{pmatrix},
\end{equation}
where we employed the gauge where $\bk$ of top and bottom layers are measured with respect to $K_\pm=4\pi/(3a_{\rm M})(-\sqrt{3}/2,\pm1/2)$, the interlayer tunneling potential $T(\br)=w(1+e^{i\bm g_1\cdot\br}+e^{i\bm g_2\cdot\br})$, the displacement field $u_D$ and finally the intralayer potential $U_{\pm}(\br)=2v\sum_{j=1,3,5}\cos(\bm b_j\cdot\br\pm\phi)$. 
We used the parameters $(m,w,v,\phi,a_{0})=(0.43m_e,-18{\rm meV},9{\rm meV},128^\circ,3.32{\rm \AA})$~\cite{wang2023topological}. 
Finally, we have introduced the reciprocal lattice vectors $\bm g_{1/2}=(4\pi/\sqrt{3}a_{\rm M})(\pm1/2,\sqrt{3}/2)$ with $a_{\rm M}$ moir\'e lattice constant and $\bm b_{1}=4\pi/\sqrt{3}a_{\rm M}$, $\bm b_{3}=\omega\bm b_{1}$ and $\bm b_{5}=\omega^*\bm b_{1}$ with $\omega=\exp(2\pi i/3)$ and also $\bm q_j=(4\pi/3a_{\rm M})e^{i\pi/2+2\pi (j-1)/3}$ with $j=1,2,3$. The Hamiltonian for valley $K'$ is obtained by time reversal symmetry $H_{\downarrow}(\br)=H^*_{\uparrow}(\br)$.

\begin{figure}
    \centering
    \includegraphics[width=1\linewidth]{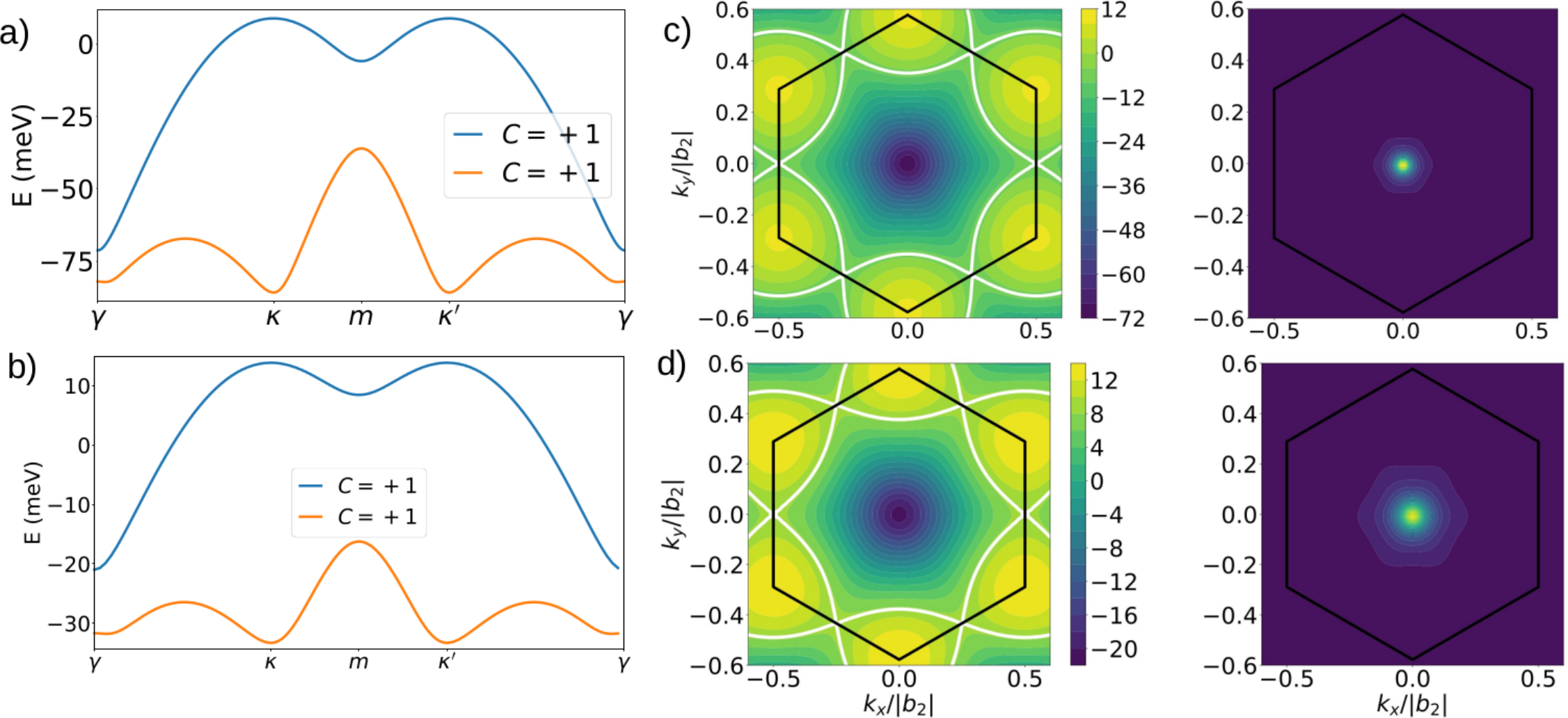}
    \caption{Panels a) and b) show the bandstructure of $\rm{tWSe}_2$ for twist angles $\theta=3.65^\circ$ and $\theta=5^\circ$, respectively.  
    Figs. c) and d) show the energy contour including the Fermi surface at the van Hove filling (solid white line) and the Berry curvature for twist angles $5^\circ$ and $3.65^\circ$, respectively. Despite an overall renormalization of the kinetic energy due to the smaller twist angle, the energy contour shows similar qualitative features with quantitative differences concerning the degree of nesting of the van Hove Fermi surface, the bandwidth and the Berry curvature concentration. Continuum model parameters are taken from Ref.~\cite{wang2023topological}.}
    \label{fig:bands}
\end{figure}

\subsection{Bandstructure properties}
\label{app:bands}

The model is invariant under $C_{3z}$ and time reversal symmetry $\mathcal T$. Furthermore, at vanishing displacement field, we have three dimensional inversion symmetry $I$ and $C_{2y}$~\cite{wang2023topological,Jiabin_2024,kolar2024hofstadterspectrumchernbands}. We observe that $I$ is an approximate symmetry of the model and it is weakly broken by higher harmonics and higher order corrections in the dispersion~\cite{PhysRevLett.132.036501,Zhang_2024}. We observe that the absence of $I$ strongly disfavours fully polarized spin triplet states $S^z=\pm1$ implying that the leading superconducting instabilities belong to the $S^z=0$ triplet and singlet channels.

\begin{figure}
    \centering
    \includegraphics[width=\linewidth]{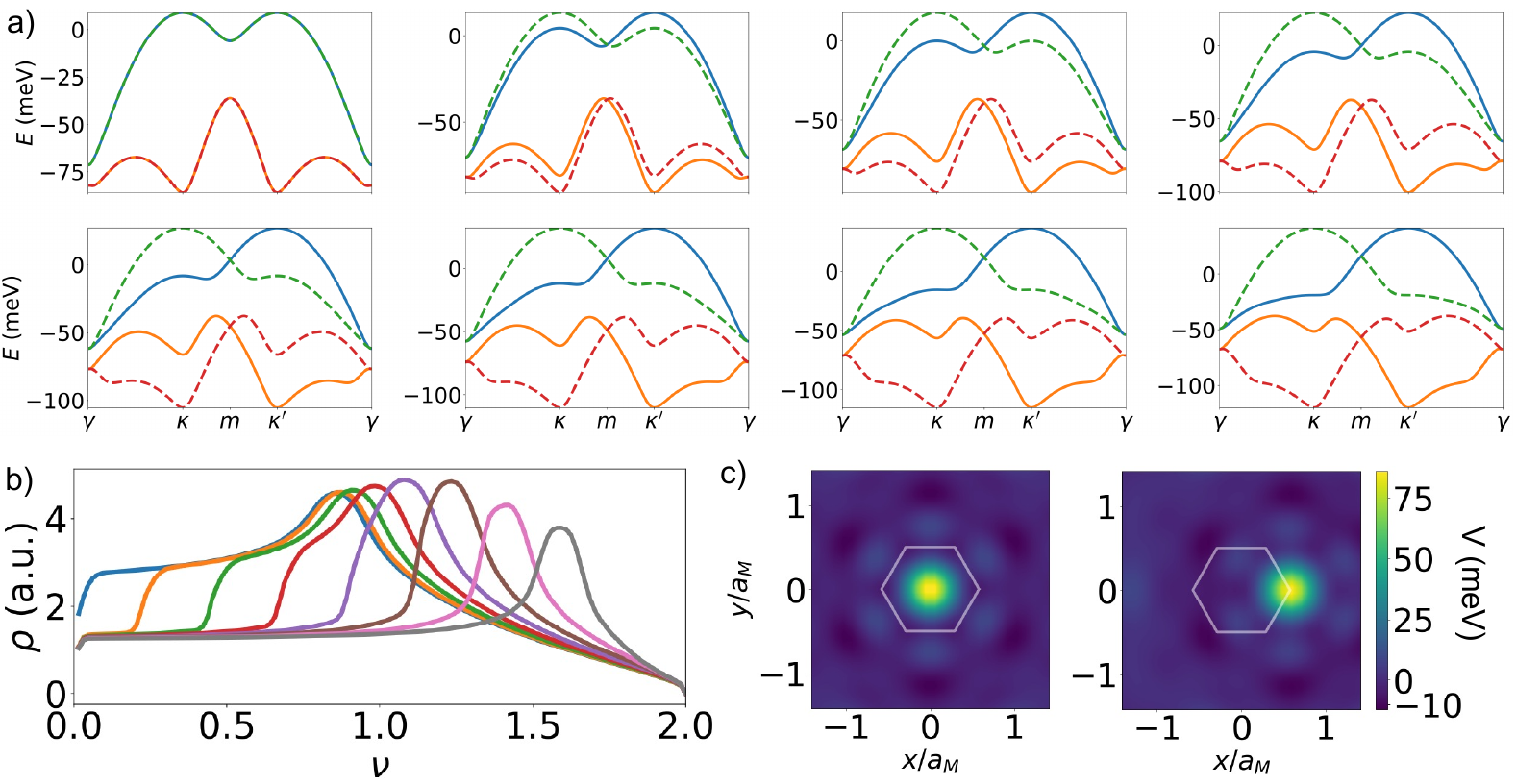}
    \vspace{0.05cm}
    \caption{a) Dispersion relation for the values $u_D=0,10,20,30,40,50,60,70$meV from left to right. We fix the twist angle to $\theta=5^\circ$, solid and dashed lines correspond to spin down and up, respectively. b) Density of states as a function of the filling factor, different colors corresponds to different $u_D=0,10,\cdots,70$meV. c) Real-space structure of the dressed Coulomb potential at the van Hove filling for $u_D=0$ and setting $\br'=0,(\bm a_1-\bm a_2)/3$. We employed the continuum model parameters in Ref.~\cite{wang2023topological}, $\epsilon=20$ and $d_{\rm sc}=24$nm.}
    \label{fig:ek_dispersion}
\end{figure}

\begin{figure}
    \centering
    \includegraphics[width=\linewidth]{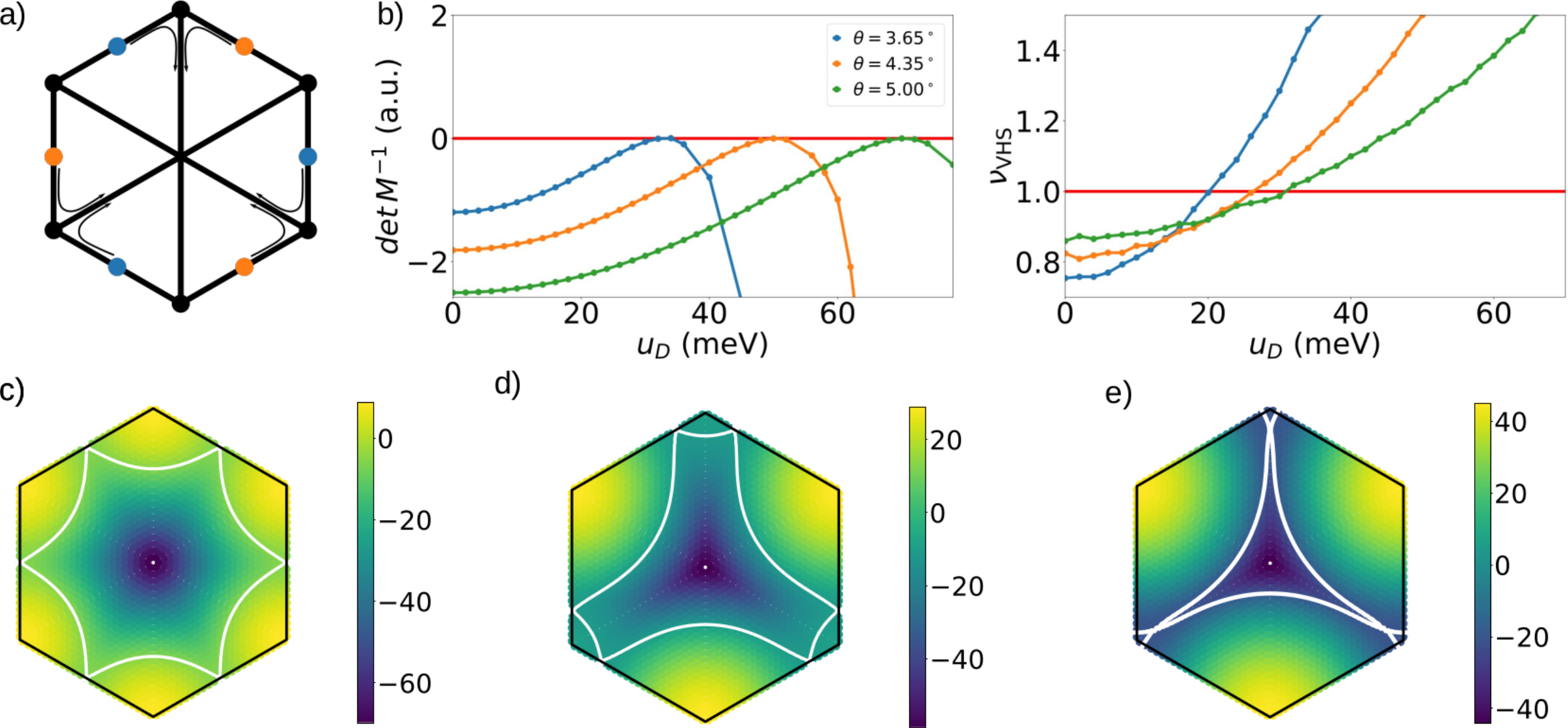}
    \caption{a) Sketch of the evolution of the van Hove singularity points for spin $\uparrow$ as a function the displacement field. The VHS moves from $m_j$ to $\kappa$, gives rise to an higher order VHS at $u_{D,c}$ and, finally, from $\kappa$ the VHS points moves towards $\gamma$ along the $\gamma\kappa$ line. b) Evolution of ${\rm det}M^{-1}$ at the van Hove point and of the filling factor where the VHS is realized for different values of $u_D$ for twist angles $3.65^\circ,4.35^\circ$ and $5^\circ$. Panels c), d) and e) shows the dispersion $\epsilon_{\bk\uparrow}$ with color code expressed in unit of meV and the Fermi surface at the van Hove singularity filling for the values of displacement field $u_D=0$ (c), $u_D=44$meV (d) and $u_D=78$meV (c). The spin $\downarrow$ sector is obtained by time reversal symmetry $\mathcal T$. }
    \label{fig:vhs_evolution}
\end{figure}

The bands at zero displacement field an twist angle $3.65^\circ,\, 5^\circ$ are given in Figs.~\ref{fig:bands}a) and b). The bands are topological up to a large displacement field $u_D$ which depends on the twist angle $\theta$, i.e. the smaller $\theta$, the smaller is the critical displacement field needed to induced the topological transition. 
We highlight a strong particle-hole asymmetry in the topmost band~\ref{fig:bands}, below the van Hove filling the bands are flat while becomes more dispersive around $\gamma$ with bandwidth $W=80,35$meV for twist angles $5^\circ,3.65^\circ$, respectively. At $u_D=0$, $I$ implies $\epsilon_{\bk\sigma}=\epsilon_{-\bk\sigma}$ which has consequences on the properties of the Fermi surface. Specifically, the topmost band are characterized by a van Hove singularity at $m_j$ for $u_D=0$ shown as solid white lines in Fig.~\ref{fig:bands}c) and d) where the different parabola centered around $\kappa$ and $\kappa'$ of the moir\'e Brillouin zone meet. 
The position at $u_D=0$ of the VHS can be also understood noticing that due to inversion symmetry $\epsilon_{-\bk\sigma}=\epsilon_{\bk\sigma}$ implying that linear terms in $\bk$ are not allowed and the dispersion around $m_1$ reads: 
\begin{equation}\label{kp_dispersion}
\xi_1(\bk)=\epsilon(m_1+\bk)-\epsilon(m_1)\approx a k_x^2 - b k^2_y,
\end{equation}
the dispersion around the other $m_j$ points is obtained employing the $C_{3z}$ symmetry, $\xi_{j}(\bk)=\xi_{1}(C^{-j+1}_{3z}\bk)$. The coefficients $a$ and $b$~\eqref{kp_dispersion} can be readily obtained applying non degenerate perturbation theory at $m_1$: 
\begin{equation}\label{inverse_mass_tensor}
\xi_{j}(\bk)=\bk^2 - 4\sum_{ab}k_ak_b\sum_{n>0}\frac{\mel{u}{J^a}{u_n}\mel{u_n}{J^b}{u}}{\epsilon_{n}-\epsilon}=k_aM^{-1}_{ab} k_b,
\end{equation}
where $\ket{u_n}$ and $\epsilon_{n}$ are eigenstates and eigenvalues at the high symmetry point $m_j$. Additionally, we have introduced the operator: 
\begin{equation}\label{current}
    J^a_{\bQ,\bQ'}=(m_j-\bQ)_a\delta_{\bm Q,\bm Q'},
\end{equation}
diagonal matrix in $\bQ=n\bm g_1+m\bm g_2$ sites of the reciprocal moir\'e lattice and $a=x,y$ spacial direction. We emphasize that Eqs.~\eqref{inverse_mass_tensor} and~\eqref{current} generalize straightforwardly to the case where the VHS is away from the $m_j$ points. The evolution of the bands as a function of displacement field is shown in Fig.~\ref{fig:ek_dispersion} for twist angle $5^\circ$.  We emphasize that in the experimentally relevant regime the topmost band is topological with Chern number $C=\pm1$ for valley $K/K'$.
Displacement field breaks $I$ implying  $\epsilon_{\bk\uparrow}\neq \epsilon_{\bk\downarrow}$ and thanks to time reversal symmetry $\mathcal T$ we have $\epsilon_{\bk\uparrow}=\epsilon_{-\bk\downarrow}$. Increasing the displacement field modifies the critical Fermi surface moving the VHS away from the $m_j$ points. The skecth in Fig.~\ref{fig:vhs_evolution}a) shows the trajectory of the VHS points which first approach $\kappa/\kappa'$ (depending on the spin projection) where a higher order van Hove singularity takes place and, then, they move towards $\gamma$ along the $\gamma\kappa$ lines. 
The evolution of ${\rm det}M^{-1}$~\eqref{inverse_mass_tensor} is shown in Fig.~\ref{fig:vhs_evolution}b), we also show the filling factor $\nu$ where the VHS takes place as a function of $u_D$. Notice that the higher order VHS takes place at $u_{D,c}$ where ${\rm det} M^{-1}=0$. Finally, panels c), d) and e) show the dispersion and the Fermi surface for the of $u_D=0,44,78$meV.  

% The evolution of the van Hove singularity as a function of the displacement field $u_D$ is given in Fig.~\ref{fig:Fermi_surface_BC}b) and c), due to $C_{2y}$ the positive $u_D<0$ is obtained from $u_D>0$ sending $\langle\hat l^z\rangle\to-\langle\hat l^z\rangle$. Notice that in Fig.~\ref{fig:Fermi_surface_BC}b) the color code denotes the degree of layer polarization of the low density Fermi liquid while in Fig.~\ref{fig:Fermi_surface_BC}c) the density of state.


\section{Screened Coulomb potential and linearised gap equation}
\label{app:screened_Coulomb}

In the random phase approximation, we replace $V_0$ with the screened Coulomb potential $V$:
\begin{equation}\label{coulomb_interaction_screened}
    {\hat H}_{\rm int}=\frac{1}{2A}\sum_{\bk_1\cdots\bk_4}\sum_{\bm g\bm g'}\sum_{\sigma\sigma'}\delta_{\bk_1+\bk_2-\bk_3-\bk_4,\Delta\bm Q}V_{\bm g\bm g'}(\bk_1-\bk_4) F^{\bk_1,\bk_4}_{\bm g\sigma}F^{\bk_2,\bk_3}_{-\bm g'+\Delta\bm Q\sigma'}c^\dagger_{\bk_1\sigma}c^\dagger_{\bk_2\sigma'}c_{\bk_3\sigma'}c_{\bk_4\sigma},
\end{equation}
where $V$ is obtained by summing up the series of bubble diagrams $V=[1-V_0\Pi]^{-1}V_0$~\cite{Cea_2021}. We have $V_{-\bm g'-\bm g}(-\bq)=V_{\bm g\bm g'}(\bq)$ and $V^*_{\bm g\bm g'}(\bq)=V_{\bm g'\bm g }(\bq)$ implying that $V$ is a hermitian matrix in the reciprocal lattice vectors $\bm g,\bm g'$. Due to the moir\'e translational symmetry the real space Coulomb interaction reads: 
\begin{equation}
     V(\br,\br')=\frac{1}{A}\sum_{\bq}\sum_{\bm g,\bm g'}V_{\bm g,\bm g'}(\bq) e^{i\bq\cdot(\br-\br')}e^{-i\bm g\cdot\br}e^{i\bm g'\cdot\br'},
\end{equation}
which as a result of the properties of the Coulomb potential satisfies $V(\br,\br')^*=V(\br,\br')$ and $V(\br,\br')=V(\br',\br)$. 
% We show in Fig.~\ref{} the real space evolution of the potential $V(\br_0,\br)$ with $\br_0=0,(\bm a_1-\bm a_2)/3$ corresponding to the MM and XM high-symmetry stacking. 
The screened Coulomb potential is displayed in the maintext. 
Following Ref.~\cite{Cea_2021,Ghazaryan_2021,Li_2023,Long_2024}, we find the linearised gap equation: 
\begin{eqnarray}\label{gap_equation}
\Delta^{l_1l_2}_{\sigma\sigma'}(\br,\br') &=& -V(\br,\br')\beta^{-1}\sum_{i\omega}\int d^2\bm x d^2\bm x'\sum_{l_3l_4} G^{l_1l_3}_{\sigma}(\br,\bm x,\omega) G^{l_2l_4}_{\sigma'}(\br',\bm x',-\omega)\Delta^{l_3l_4}_{\sigma\sigma'}(\bm x,\bm x'),
\end{eqnarray}
where $\Delta^{l_1l_2}_{\sigma\sigma'}(\br,\br')$ is the pairing function for two electrons of spin $\sigma,\sigma'$, $V$ the dressed Coulomb potential with spatial pattern displayed in Fig.~\ref{fig:ek_dispersion}, position $(l_1\br,l_2\br')$ with $l_{1/2}$ layer degree of freedom, $G$ the single particle Green's function: 
\begin{equation}\label{greensfunction_G}
    G_\sigma^{\ell\ell'}(\br,\br',i\omega)=\sum_{\bk}\frac{\psi_{\bk\sigma\ell}(\br)\psi^*_{\bk\sigma\ell'}(\br')}{i\omega-\xi_{\bk\sigma}},
\end{equation}
where $\xi_{\bk\sigma}=\epsilon_{\bk\sigma}-\mu$ and $\bm \psi_{\bk\sigma}(\br)=[\psi_{\bk\sigma 1}(\br),\psi_{\bk\sigma 2}(\br)]^T$ is the wavefunction layer spinor of the topmost moir\'e band. Projecting the interaction~\eqref{coulomb_interaction_screened} in the particle-particle channel with vanishing center of mass momentum and employing Eq.~\eqref{greensfunction_G}, we obtain the linearised gap equation: 
\begin{equation}
    \Delta_{ss'}(\bk)=-\frac{T}{A}\sum_{\bk'}\sum_{s_1s_2}\sum_{i\epsilon} V_{s's,s_1s_2}(\bk,\bk') G_{s_1}(\bk',i\epsilon) G_{s_2}(-\bk',-i\epsilon)\Delta_{s_1s_2}(\bk').
\end{equation}
We have introduced the pairing interaction: 
\begin{equation}
    \begin{split}
    \label{scattering_vertex}
    V_{s_1s_2,s_3s_4}(\bk,\bk')&=\delta_{s_1s_4}\delta_{s_2s_3}\sum_{\bm g\bm g'}V_{\bm g,\bm g'}(\bk-\bk')\Lambda^{\bk,\bk'}_{-\bm gs_3}\Lambda^{-\bk,-\bk'}_{\bm g's_4}.
    \end{split}
\end{equation}
Performing the Matsubara sum we obtain: 
\begin{equation}
    I_{s_1s_2}(\bk)=T\sum_{i\epsilon}G_{s_1}(\bk,i\epsilon) G_{s_2}(-\bk,-i\epsilon)=\frac{1}{2(\xi_{\bk s_1}+\xi_{-\bk s_2})}\left[\tanh\frac{\beta\xi_{\bk s_1}}{2}+\tanh\frac{\beta\xi_{-\bk s_2}}{2}\right],
\end{equation}
and redefining $\Delta_{ss'}(\bk)= \phi_{ss'}(\bk)/\sqrt{I_{ss'}(\bk)}$ we arrive at the integral eigenvalue equation
\begin{equation}
    \lambda \phi_{ss'}(\bk)=\sum_{\bk'}\sum_{s_1s_2}K_{ss',s_1s_2}(\bk,\bk') \phi_{s_1s_2}(\bk'),
\end{equation}
where 
\begin{equation}
    K_{s_1s_2,s_3s_4}(\bk,\bk')=-\sqrt{I_{s_1s_2}(\bk)}V_{s_2s_1,s_3s_4}(\bk,\bk')\sqrt{I_{s_3s_4}(\bk')}/A. 
\end{equation}
To solve the gap equations we fix the time reversal gauge, by explicitly enforcing the relation between $\ket{\bk\uparrow}$ and $\ket{-\bk\downarrow}$, through: 
\begin{equation}\label{TRS_gauge}
    z_{-\bm g\uparrow}(-\bk)=z^*_{\bm g\downarrow}(\bk),
\end{equation}
where $z_{\bm g\sigma}(\bk)$ is the Fourier amplitude $z_{\bm g\sigma}(\bk)=\braket{\bm g}{u_{\bk\sigma}}$. The time reversal symmetry gauge is then imposed building the basis such that for any $\bk$ the spin $\downarrow$ is computed from the $\uparrow$ sector through Eq.~\eqref{TRS_gauge}.  
\begin{figure}
    \centering
    \includegraphics[width=0.8\linewidth]{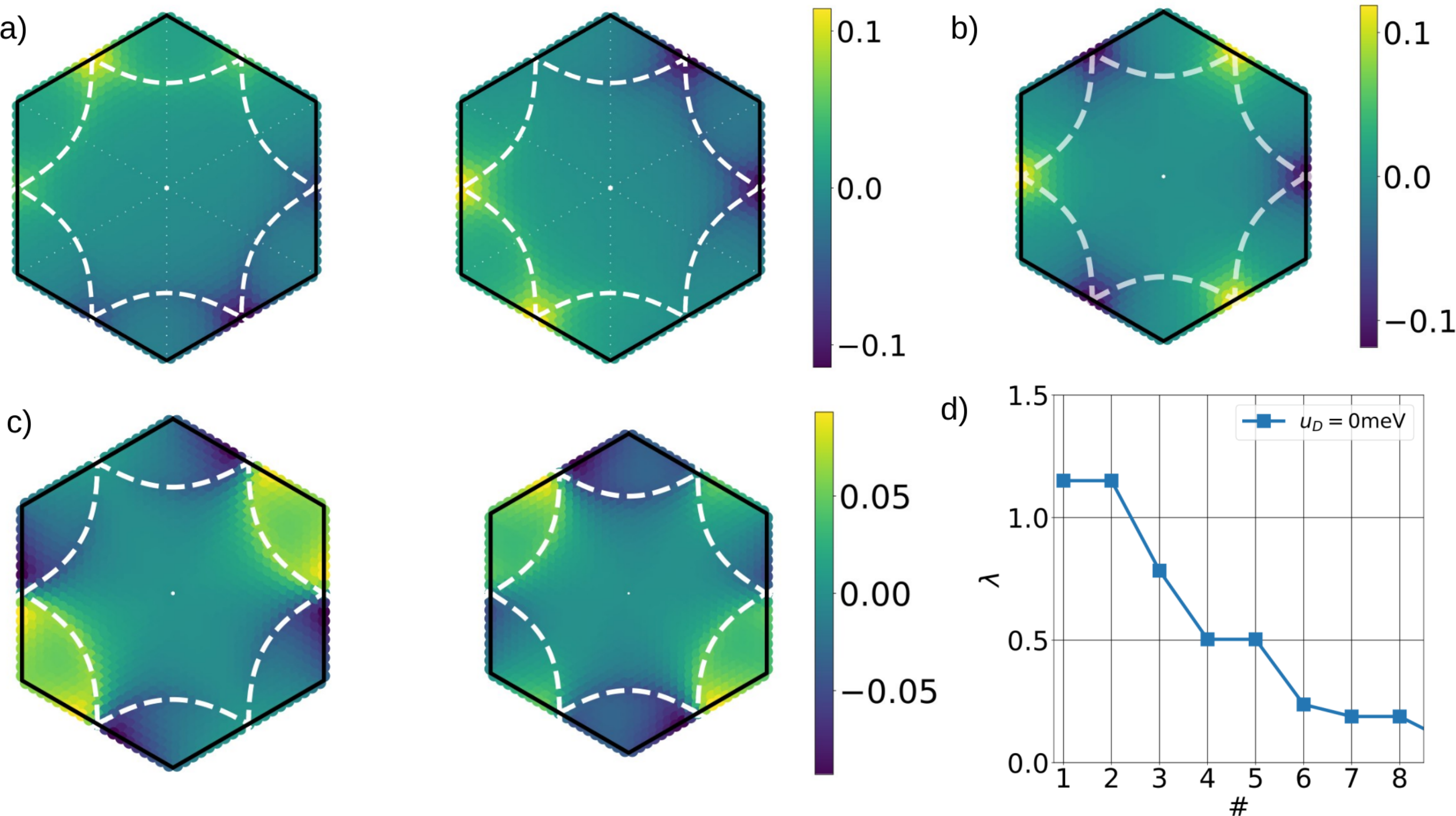}
    \caption{Panels a), b) and c) show the leading superconducting instabilities at $u_D=0$ belonging to the irreducible representations $E_u$, $A_{1u}$ and $E_g$, respectively. Panel d) shows the eigenvalue spectrum of the pairing operator. The largest eigenvalues are two-fold degenerate and belongs to $E_u$, subleading instabilities are $A_{1u}$ and $E_{g}$.  
    At finite displacement field (orange line) the degeneracy persists protected by $C_{3z}$ and $\mathcal T$, here eigenstates do not have a well defined parity. Calculations have been performed setting the twist angle $5^\circ$ at vanishing displacement field $u_D=0$, dielectric constant $\varepsilon=20$, and for $k_BT/E_k=9\times10^{-3}$ with $E_k=\hbar^2(4\pi/3a_{\rm M})^2/(2m^*_e)=107.3$meV kinetic energy scale.}
    \label{fig:superconductivity_zero_displacement}
\end{figure}

\subsection{Symmetries and properties of the pairing potential}
\label{app:pairing_pot}

\begin{table}[]
\centering
\begin{tabular}{|c||c|c|c|c|c|c|}
\hline
 $D_{3d}$ & $E$ & $2C_3$ & $3C_2$ & $I$ & $2S_3$ & $ 3\sigma_v$ \\
\hline\hline
$A_{1g}$ &  $1$  & $1$  & $1$ & 1 & 1 & 1 \\
\hline
$A_{2g}$ & $1$  &  $1$ & $-1$ & 1& 1 &  $-1$ \\
\hline
$E_g$ & $2$  &  $-1$ & $0$ & 2 &$-1$ &  0 \\
\hline
$A_{1u}$ &  $1$  & $1$  & $1$ &$-1$ &$-1$ &$-1$ \\
\hline
$A_{2u}$ & $1$  &  $1$ & $-1$ &$-1$ &$-1$ & 1\\
\hline
$E_{u}$ & $2$  &  $-1$ & $0$ &$-2$ & 1 & 0 \\
\hline
\end{tabular}\vspace{0.5cm}
\caption{Character table of $D_{3d}$ at zero displacement field. Four one dimensional representations $A_{1/2g/u}$ and two two-dimensional representation $E_{g/u}$ which $g/u$ refer to their parity under the three-dimensional inversion $I$. Due to the Ising spin-orbit couplings it is important to remark that the table refers to a spin space point group. Including higher order harmonics or higher order $\bk\cdot\bp$ corrections lower the point group to $D_3$. Finally, a finite displacement field lower the point group to $C_{3}$. We highlight that $C_{3z}$ does not commute with time reversal symmetry mixing eigenstates with opposite eigenvalues $\omega,\omega^*$ protecting two-fold degeneracies in the particle-particle spectrum of $K(\bk,\bk')$. } 
\label{tab:character_table}
\end{table}

% \begin{table}[]
% \centering
% \begin{tabular}{|c||c|c|c|}
% \hline
%  $C_{3}$ & $E$ & $C_3$ & $C^2_3$ \\
% \hline\hline
% $A$ &  $1$  & $1$  & $1$  \\
% \hline
% $E$ & $1$  &  $\omega$ & $\omega^*$ \\
% \hline
% $E$ & $1$  &  $\omega^*$ & $\omega$  \\
% \hline
% \end{tabular}\vspace{0.5cm}
% \caption{Character table of $D_{3d}$ at zero displacement field. Four one dimensional representations $A_{1/2g/u}$ and two two-dimensional representation $E_{g/u}$ which $g/u$ refer to their parity under the three-dimensional inversion $I$. Due to the Ising spin-orbit couplings it is important to remark that the table refers to a spin space point group. Including higher order harmonics or higher order $\bk\cdot\bp$ corrections lower the point group to $D_3$. Finally, a finite displacement field lower the point group to $C_{3}$. We highlight that $C_{3z}$ does not commute with time reversal symmetry mixing eigenstates with opposite eigenvalues $\omega,\omega^*$ protecting two-fold degeneracies in the particle-particle spectrum of $K(\bk,\bk')$. } 
% \label{tab:character_table}
% \end{table}

The projected screened Coulomb interaction takes the form in Eq.~\eqref{scattering_vertex}. Within the time reversal symmetric gauge we decomposed the form factors as $\Lambda^{\bk,\bk'}_{\bm g\sigma}=X^{\bk,\bk'}_{\bm g}+i\sigma Y^{\bk\bk'}_{\bm g}$ and we obtained: 
\begin{equation}
    V_{s_1s_2,s_3s_4}(\bk,\bk')=\sum_{\mu\nu=0,z}V_{\mu\nu}(\bk,\bk')\sigma^\mu_{s_1s_4}\sigma^\nu_{s_2s_3},
\end{equation}
where: 
\begin{eqnarray}\label{decomposition_interaction}
    &V_{00}(\bk,\bk')=\sum_{\bm g\bm g'}V_{\bm g\bm g'}(\bk-\bk')X^{\bk,\bk'}_{-\bm g}X^{-\bk,-\bk'}_{\bm g'},\\
    &V_{z0}(\bk,\bk')=i\sum_{\bm g\bm g'}V_{\bm g\bm g'}(\bk-\bk')Y^{\bk,\bk'}_{-\bm g}X^{-\bk,-\bk'}_{\bm g'},\\
    &V_{0z}(\bk,\bk')=i\sum_{\bm g\bm g'}V_{\bm g\bm g'}(\bk-\bk')X^{\bk,\bk'}_{-\bm g}Y^{-\bk,-\bk'}_{\bm g'},\\
    &V_{zz}(\bk,\bk')=-\sum_{\bm g\bm g'}V_{\bm g\bm g'}(\bk-\bk')Y^{\bk,\bk'}_{-\bm g}Y^{-\bk,-\bk'}_{\bm g'},
\end{eqnarray}
and $X^{\bk,\bk'}_{\bm g}=(\Lambda^{\bk,\bk'}_{\bm g\uparrow}+\Lambda^{\bk,\bk'}_{\bm g\downarrow})/2$, $Y^{\bk,\bk'}_{\bm g}=-i(\Lambda^{\bk,\bk'}_{\bm g\uparrow}-\Lambda^{\bk,\bk'}_{\bm g\downarrow})/2$. 
As a result of the time reversal symmetry we have: 
\begin{equation}
    X^{\bk,\bk'}_{\bm g}=X^{-\bk',-\bk}_{\bm g},\quad Y^{\bk,\bk'}_{\bm g}=-Y^{-\bk',-\bk}_{\bm g},
\end{equation}
which directly implies 
\begin{eqnarray}
    V_{00}(\bk,\bk')=V_{00}(\bk',\bk),\quad V_{zz}(\bk,\bk')=V_{zz}(\bk',\bk).
\end{eqnarray}
Finally, we have the mixed terms $V_{0z}(\bk,\bk')$ and $V_{z0}(\bk,\bk')$ which satisfies the following conditions: 
\begin{equation}
    V_{z0}(\bk,\bk')=-V_{0z}(\bk',\bk).
\end{equation}
An additional symmetry which is present also for finite displacement field is $C_{3z}$. The threefold rotational symmetry acts on the Hamiltonian as $C_{3z}H_\sigma(\bk)C^{-1}_{3z}=H_\sigma(C_{3z}\bk)$ leading to the relation $C_{3z}\ket{u_{\bk\sigma}}=\ket{u_{C_{3z}\bk\sigma}}$ with direct implications on the matrix elements $\Lambda^{C_{3z}\bk,C_{3z}\bk'}_{\bm g\sigma}=\Lambda^{\bk,\bk'}_{C^{-1}_{3z}\bm g\sigma}$. This result combined with the symmetry properties of the screened Coulomb potential $V_{C_{3z}\bm g,C_{3z}\bm g'}(\bq)=V_{\bm g,\bm g'}(C_{3z}\bq)$ leads to the relation: 
\begin{equation}
    V_{\mu\nu}(C_{3z}\bk,C_{3z}\bk')=V_{\mu\nu}(\bk,\bk'),
\end{equation}
the scattering vertex is invariant under the threefold rotation symmetry. This property directly implies that the eigenspectrum of the pairing Kernel operator can be classified according to the symmetry properties under $C_{3z}$. The presence of time reversal symmetry $\mathcal T$ protects two-fold degeneracies as $\mathcal T$ is off-diagonal in subspaces of eigenstates transforming non trivially under $C_{3z}$. 

Finally, we discuss the role of the extra symmetries valid only at vanishing displacement field. The first one is inversion $I$ implying that $\Lambda^{\bk,\bk'}_{\bm g\sigma}=\Lambda^{-\bk,-\bk'}_{-\bm g\sigma}$. Additionally, due to inversion symmetry we find $V_{\bm g\bm g'}(\bq)=V_{\bm g'\bm g}(\bq)$. Combining the latter identity with time reversal symmetry we find $X^{\bk,\bk'}_{\bm g}=\Re\Lambda^{\bk,\bk'}_{\bm g\uparrow}$ and $Y^{\bk,\bk'}_{\bm g}=\Im\Lambda^{\bk,\bk'}_{\bm g\uparrow}$. Furthermore, we have: 
\begin{equation}
    V_{00}(-\bk,-\bk')=\sum_{\bm g\bm g'}V_{\bm g,\bm g'}(-\bk+\bk')X^{-\bk,-\bk'}_{-\bm g}X^{\bk,\bk'}_{\bm g'}=\sum_{\bm g\bm g'}V_{-\bm g',-\bm g}(-\bk+\bk')X^{-\bk,-\bk'}_{\bm g'}X^{\bk,\bk'}_{-\bm g}=V_{00}(\bk,\bk'),
\end{equation}
similarly we find $V_{zz}(-\bk,-\bk')=V_{zz}(\bk,\bk')$. This latter relation implies that, in the symmetric inversion case $V_{00/zz}$ cannot mix the triplet and singlet sectors, as expected from the different parity under inversion of the two pairing channels. 
Finally, we have 
\begin{equation}
    V_{0z}(-\bk,-\bk')=V_{z0}(\bk,\bk').
\end{equation}
$C_{2y}$ sends $\uparrow\to\downarrow$ and $\bk\to(-k_x,k_y)$ imposing the relation $\Lambda^{C_{2y}\bk,C_{2y}\bk'}_{\bm g\uparrow}=\Lambda^{\bk,\bk'}_{C_{2y}\bm g\downarrow}$ which implies $X^{C_{2y}\bk,C_{2y}\bk'}_{\bm g}=X^{\bk,\bk'}_{C_{2y}\bm g}$ and $Y^{C_{2y}\bk,C_{2y}\bk'}_{\bm g}=-Y^{\bk,\bk'}_{C_{2y}\bm g}$. As a result, we find: 
\begin{eqnarray}
    V_{00/zz}(C_{2y}\bk,C_{2y}\bk')=V_{00/zz}(\bk,\bk'),\quad V_{0z/z0}(C_{2y}\bk,C_{2y}\bk')=-V_{0z/0z}(\bk,\bk').
\end{eqnarray}
Interestingly, the components $V_{0z}$ and $V_{z0}$ are odd under $C_{2y}$ and must necessarily shows nodes in the mini Brillouin zone. This is also expected since these two components mix triplet and singlet with opposite angular momenta and display vortices. 
% Finally, we conclude with a technical remark related to the current implementation of the code. Within the current implementation of the model we have $\bm g'=-(\bm b_1-\bm b_2)-\bm g$.

\subsection{Properties of the particle-particle Kernel}
\label{app:particle_particle}

In this section we discuss the symmetries of the pairing kernel. We start from the $u_D=0$ limit where the kernel enjoys the point group $D_{3d}$. Then, we move to the more general case of a finite displacement field where the symmetries reduce to $C_{3z}$ and $T$.  

\subsubsection{Vanishing displacement field}

To start with we notice that in the presence of inversion symmetry $I$ the dressed Coulomb potential satisfies the relation $V_{\bm g,\bm g'}(\bq)=V_{\bm g',\bm g}(\bq)$ which is directly inherited from $\Pi_{\bm g\bm g'}(\bq)$. Furthermore, in this limit $\xi_{\bk\sigma}=\xi_{\bk}$ is spin independent which implies: 
\begin{equation}
    I_{ss'}(\bk)=\frac{1}{2\xi_{\bk}}\tanh\frac{\beta\xi_{\bk}}{2}=F_{\bk}.
\end{equation}
Without, loss of generality we write:
\begin{equation}
    \lambda \phi_{ss'}(\bk)=-\frac{1}{A}\sum_{\bk'}\sqrt{F_{\bk}}V_{ss'}(\bk,\bk')\sqrt{F_{\bk'}} \phi_{ss'}(\bk'),
\end{equation}
where we have explicitly imposed the constraint imposed by the spin conservation. Focusing on the spin $S^z=0$ sector, we decompose the amplitude $\phi_{ss'}(\bk)$ as: 
\begin{equation}
    \phi(\bk)=\begin{bmatrix}
        0 & \psi(\bk)+d^z(\bk) \\ 
        -\psi(\bk)+d^z(\bk) & 0 
    \end{bmatrix}, 
\end{equation}
with $\psi(\bk)$ even function of $\bk$ and $d^z(\bk)$ odd. For $\psi(\bk)$ we have: 
\begin{equation}
    \begin{split}
    \lambda\psi(\bk) = &-\frac{1}{2A}\sum_{\bk'}\sqrt{F_{\bk}}\left[V_{\uparrow\downarrow}(\bk,\bk')+V_{\downarrow\uparrow}(\bk,\bk')\right]\sqrt{F_{\bk'}}\psi(\bk')\\
    &-\frac{1}{2A}\sum_{\bk'}\sqrt{F_{\bk}}\left[V_{\uparrow\downarrow}(\bk,\bk')-V_{\downarrow\uparrow}(\bk,\bk')\right]\sqrt{F_{\bk'}}d^z(\bk'),
    \end{split}
\end{equation}
Employing the decomposition~\eqref{decomposition_interaction}, we find: 
\begin{equation}
    \frac{V_{\uparrow\downarrow}(\bk,\bk')+V_{\downarrow\uparrow}(\bk,\bk')}{2}=V_{00}(\bk,\bk')-V_{zz}(\bk,\bk'),\quad \frac{V_{\uparrow\downarrow}(\bk,\bk')-V_{\downarrow\uparrow}(\bk,\bk')}{2}=V_{z0}(\bk,\bk')-V_{0z}(\bk,\bk').
\end{equation}
Using the inversion symmetry of the model $X^{\bk,\bk}_{\bm g}=X^{-\bk,-\bk'}_{-\bm g}$ and $Y^{\bk,\bk}_{\bm g}=Y^{-\bk,-\bk'}_{-\bm g}$
\begin{eqnarray}
    V_{z0}(\bk,\bk')-V_{z0}(\bk,\bk')&=i\sum_{\bm g\bm g'}V_{\bm g\bm g'}(\bk-\bk')\left[Y^{\bk,\bk'}_{-\bm g'}X^{-\bk,-\bk'}_{\bm g'}-X^{\bk,\bk'}_{-\bm g'}Y^{-\bk,-\bk'}_{\bm g'}\right]\\
    &=i\sum_{\bm g\bm g'}Y^{\bk,\bk'}_{-\bm g'}X^{-\bk,-\bk'}_{\bm g'}\left[V_{\bm g\bm g'}(\bk-\bk')-V_{\bm g'\bm g}(\bk-\bk')\right]=0.
\end{eqnarray}
As expected from the opposite inversion eigenvalues in table~\ref{tab:character_table}, we find that the singlet $\psi(\bk)$ and triplet $d^z(\bk)$ are decoupled: 
\begin{equation}
    \lambda\psi(\bk) =-\frac{1}{A}\sum_{\bk'}\sqrt{F_{\bk}}\left[V_{00}(\bk,\bk')-V_{zz}(\bk,\bk')\right]\sqrt{F_{\bk'}}\psi(\bk'). 
\end{equation}
Similarly, in the triplet sector we find: 
\begin{equation}
    \lambda d^z(\bk) =-\frac{1}{A}\sum_{\bk'}\sqrt{F_{\bk}}\left[V_{00}(\bk,\bk')-V_{zz}(\bk,\bk')\right]\sqrt{F_{\bk'}} d^z(\bk'). 
\end{equation}
We now compute numerically the spectral properties of the kernel $K_{\bk,\bk'}=-\sqrt{F_{\bk}}\left[V_{00}(\bk,\bk')-V_{zz}(\bk,\bk')\right]\sqrt{F_{\bk'}}/A$. In the limit of vanishing displacement field $u_D=0$, the point group of the continuum theory is $D_{3d}$~\ref{tab:character_table} characterized by the irreducible representations $A_{1g/u},A_{2g/u}$ and the two-dimensional $E_{g/u}$ with $g/u$ referring to the parity under the three-dimensional inversion $I$. Triplet and singlet sectors are thus decoupled as they transform with opposite parities under inversion $I$. This is explicitly manifested in Fig.~\ref{fig:superconductivity_zero_displacement}a), b) and c) where the different eigenstates have different inversion symmetry eigenvalues.  
The leading superconducting instability belongs is a non-trivial two-dimensional irrep in the $E_u$ (p-wave like) sector consisting of $S^z=0$ triplet. The subleading superconducting instabilities in the $S^z=0$ sector are $A_{1u}$ (f-wave like) and $E_g$ (d-wave like) with eigenvalues shown in Fig.~\ref{fig:superconductivity_zero_displacement}d) for a representative value of the temperature and $u_D=0$.   
In the two-dimensional degenerate subspace $E_u$ we can express the order parameter $d^z(\bk)$ as: 
\begin{equation}
    d^z(\bk) = \sum_{\alpha}\eta_{\alpha} d^z_{\alpha}(\bk), 
\end{equation}
where $\bm \eta=(\eta_1,\eta_2)^T$ in the subspace spanned by $d^z_{1/2}(\bk)$ basis functions of the $E_u$ irreducible representation. By definition $d^z_{1/2}(-\bk)=-d^z_{1/2}(\bk)$ and: 
\begin{equation}
     \int \frac{d^2\bk}{\Omega_{\rm mBZ}} d^{z*}_{a}(\bk)d^{z}_b(\bk)=\delta_{ab}.   
\end{equation}
Under $C_{3z}$ we have: 
\begin{eqnarray}
    C_{3z}d^z_1(\bk)=d^z_1(C_{3z}^{-1}\bk)=-\frac{d_1(\bk)}{2}-\frac{\sqrt{3}d_2(\bk)}{2},\quad C_{3z}d^z_2(\bk)=d^z_2(C_{3z}^{-1}\bk)=-\frac{d_2(\bk)}{2}+\frac{\sqrt{3}d_1(\bk)}{2}. 
\end{eqnarray}
As a result we have: 
\begin{eqnarray}
    d^z(C_{3z}\bk)=\left[C_{3z}\bm\eta\right]\cdot\bm d^z(\bk),
\end{eqnarray}
where in the basis $(d^z_1,d^z_2)$ the threefold rotation around z reads: 
\begin{eqnarray}
    C_{3z}=\begin{pmatrix}
        -1/2 & -\sqrt{3}/2 \\ 
        \sqrt{3}/2 & -1/2
    \end{pmatrix}.
\end{eqnarray}
We can also employ the basis $d^z_{\pm}=d^z_1\pm i d^z_2$ where the threefold rotational symmetry takes the form $C_{3z}=\text{diag}[\omega,\omega^*]$. To quadratic order in $\bm\eta$ we find: 
\begin{equation}
    \mathcal F_{(2)}(\bm \eta)=A\bm\eta^\dagger\cdot\bm\eta,
\end{equation}
where $\bm\eta=[\eta_1,\eta_2]^T$ is related to the basis $\eta_{\pm}$ by the unitary transformation $$U=\frac{1}{\sqrt{2}}\begin{pmatrix}
    1 & i \\ 
    1 & -i
\end{pmatrix},\quad \begin{pmatrix}
    \eta_{+} \\ 
    \eta_{-}
\end{pmatrix}=U\begin{pmatrix}
    \eta_{1} \\ 
    \eta_{2}
\end{pmatrix}.$$
$C_{2y}$ matrix presentation is: 
\begin{equation}
    C_{2y}=\begin{pmatrix}
        -1/2 & -\sqrt{3}/2 \\ 
        -\sqrt{3}/2 & 1/2
    \end{pmatrix}.
\end{equation}
$C_{3z}$ invariant contributions include $|\eta_\pm|^4$, $|\eta_+|^2|\eta_-|^2$. In the basis $(\eta_1,\eta_2)^T$ the resulting fourth order contribution to the free energy reads: 
\begin{equation}
    \mathcal F_{(4)}(\bm\eta)=B(\bm\eta^\dagger\cdot\bm\eta)^2+C(\eta_1^*\eta_2-\eta_1\eta^*_2)^2,
\end{equation}
leading to the free energy in the main text which is also consistent with Ref.~\cite{Sauls_2019}.
 Remarkably, the same expression holds also at finite displacement field as it simply follows from $\mathcal T$ and $C_{3z}$ symmetry.

\subsection{Finite displacement field}
\label{app:away}
\begin{figure}
\centering
\includegraphics[width=0.8\linewidth]{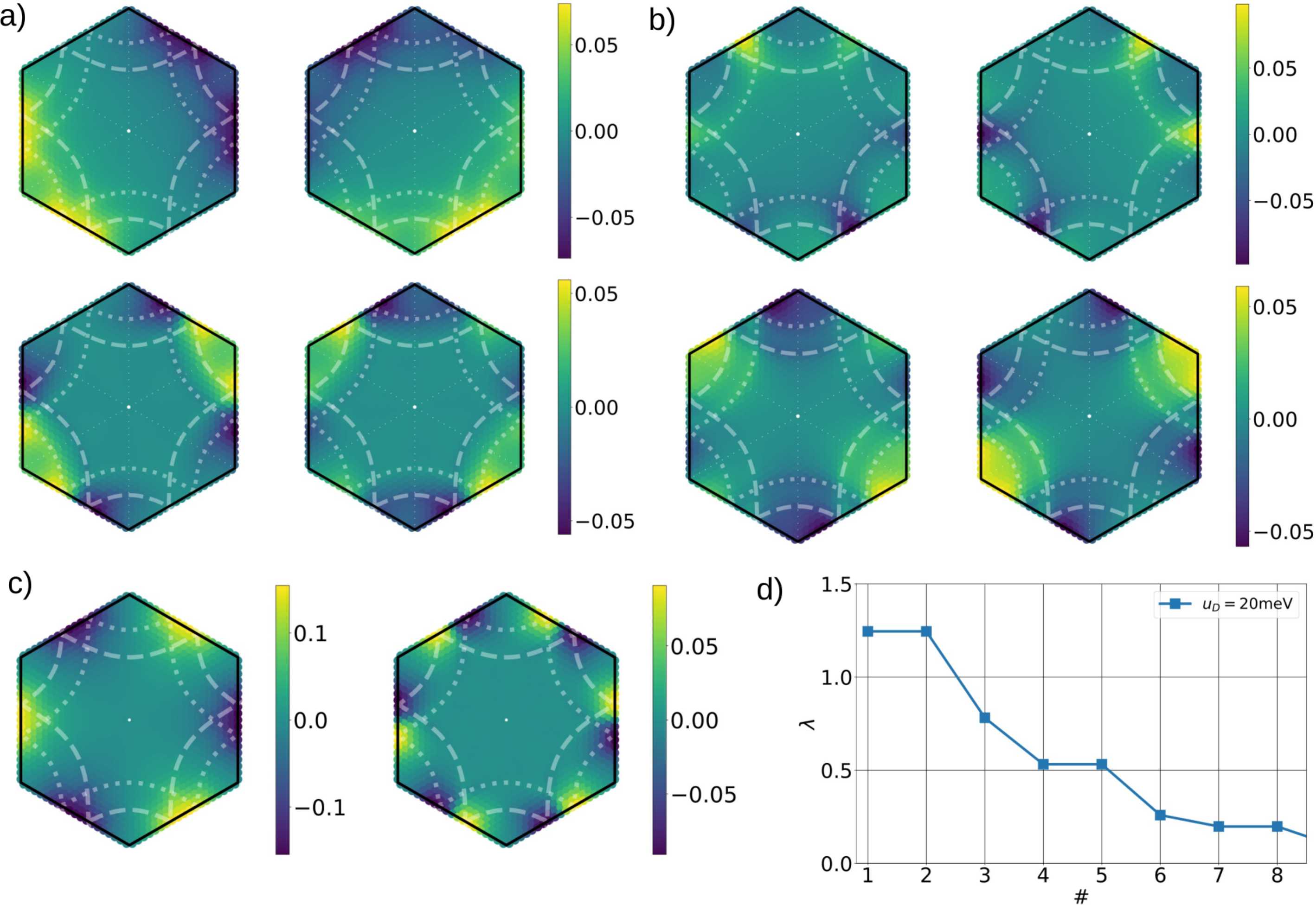}
\caption{Spectrum of the pairing Kernel for finite displacement field $u_D=20$meV. Panels a), b) and c) show the momentum space structure of the $E^-$, $E^+$ and $A^-$ instabilities, respectively. Dotted and dashed lines show the spin $\uparrow$ and $\downarrow$ Fermi surface whose degeneracy is lifted by the applied displacement field $(\xi_{\bk\uparrow}\neq \xi_{\bk\downarrow})$. We emphasize that the instabilities do not have a well defined eigenvalue under inversion symmetry are composed by an even $\psi(\bk)$ and off component $d^z(\bk)$ under inversion. 
The eigenvalues are displayed in panel d), the leading instability is in the two-fold subspace $E^-$ connected to $E_u$ at $u_D=0$.  }
\label{fig:superconductivity_finite_displacement}
\end{figure}

\begin{figure}
    \centering
    \includegraphics[width=0.7\linewidth]{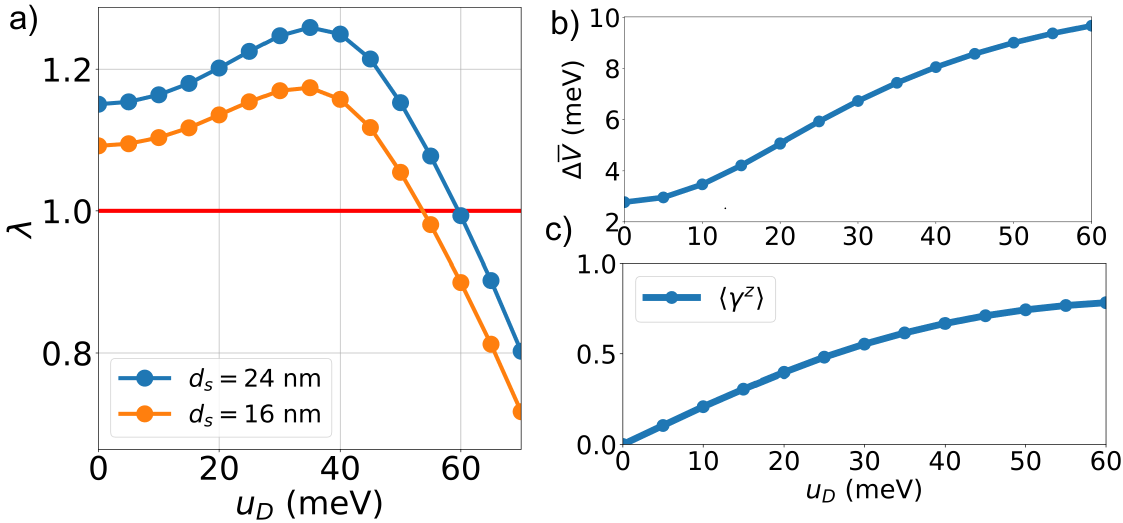}
    \vspace{0.1cm}
    \caption{a) Leading superconducting instability eigenvalue for $d_{\rm sc}=24,16$nm along the van Hove filling. Reducing $d_{\rm sc}$ suppresses the superconducting instability. Panels b) and c) show the evolution of $\Delta \overline{V}$ and $\langle\gamma^z\rangle$, respectively, at the van Hove filling for increasing value of $u_D$.}
    \label{fig:supp4}
\end{figure}

In the limit of finite displacement field $I$ and $C_{2y}$ are broken and, as a result, the singlet and triplet channels mix. Fig.~\ref{fig:superconductivity_finite_displacement} show the spectrum of the pairing kernel in the $S^z=0$. Specifically, Fig.~\ref{fig:superconductivity_finite_displacement}a) and b) show the two-dimensional irreducible representation $E^-$ and $E^+$, top panels the odd ($d^z(\bk)$) and bottom the even ($\psi(\bk)$) components of the eigenstates. 
Finally, Fig.~\ref{fig:superconductivity_finite_displacement}c) we have the $A^-_1$ instability, here left and right panels show odd and even components.  
The mixing originates from $\xi_{\bk\uparrow}\neq\xi_{\bk\downarrow}$ as well as from the contribution of the form factors $\Lambda^{\bk,\bk'}_{\bm g\sigma}$ encoding the non trivial topological properties of the bands. We find: 
\begin{equation}
    \begin{split}
        &\lambda \begin{pmatrix}
            \psi(\bk) \\
             d^z(\bk)
        \end{pmatrix}=-\frac{1}{2A}\sum_{\bk'}\left[\left(\sqrt{F_{\bk\uparrow}F_{\bk'\uparrow}}+\sqrt{F_{\bk\downarrow}F_{\bk'\downarrow}}\right)\overline V(\bk,\bk')+\left(\sqrt{F_{\bk\uparrow}F_{\bk'\uparrow}}-\sqrt{F_{\bk\downarrow}F_{\bk'\downarrow}}\right)\delta V(\bk,\bk')\right] \begin{pmatrix}
            \psi(\bk') \\
             d^z(\bk')
        \end{pmatrix}\\
        &-\frac{1}{2A}\sum_{\bk'}\left[\left(\sqrt{F_{\bk\uparrow}F_{\bk'\uparrow}}-\sqrt{F_{\bk\downarrow}F_{\bk'\downarrow}}\right)\overline V(\bk,\bk')+\left(\sqrt{F_{\bk\uparrow}F_{\bk'\uparrow}}+\sqrt{F_{\bk\downarrow}F_{\bk'\downarrow}}\right)\delta V(\bk,\bk')\right]\sigma^x \begin{pmatrix}
            \psi(\bk') \\
             d^z(\bk')
        \end{pmatrix},
    \end{split}
\end{equation}
where we have introduced $\overline{V}=V_{00}-V_{zz}$ and $\delta V=V_{z0}-V_{0z}$. Fig.~\ref{fig:supp4}a) shows the evolution of the leading eigenvalue of the pairing Kernel $K(\bk,\bk')$ for two different values of the distance from the gates $d_{\rm sc}=24,16$nm along the van Hove filling.  Fig.~\ref{fig:supp4}b) shows the evolution of the quantity: 
\begin{equation}
    \Delta \overline{V} =\max \overline{V}(\bk,\bk)-\int\frac{d^2\bk}{\Omega_{\rm mBZ}}\overline{V}(\bk,\bk),
\end{equation}
and Fig.~\ref{fig:supp4}c) the average layer polarization $\langle\gamma^z\rangle$ where the average is taken over the Fermi sea as a function of the displacement field $u_D$. We notice that:
\begin{equation}
    \overline{V}(\bk,\bk)=\frac{1}{2}\sum_{\bm g,\bm g'}\sum_{\sigma}\left[\Lambda^{\bk,\bk}_{\bm g\sigma}\right]^* V_{\bm g,\bm g'}(0)\Lambda^{\bk,\bk}_{\bm g'\sigma},
\end{equation}
An increase in layer polarization amplifies the diagonal component of the dressed Coulomb potential, thereby increasing the repulsion between carriers and, eventually, reducing the tendency to superconductivity.

\section{Detials on the variational approach and on the topological properties of the superconductor}
\label{app:bcs}

\begin{figure}
    \centering
    \includegraphics[width=0.8\linewidth]{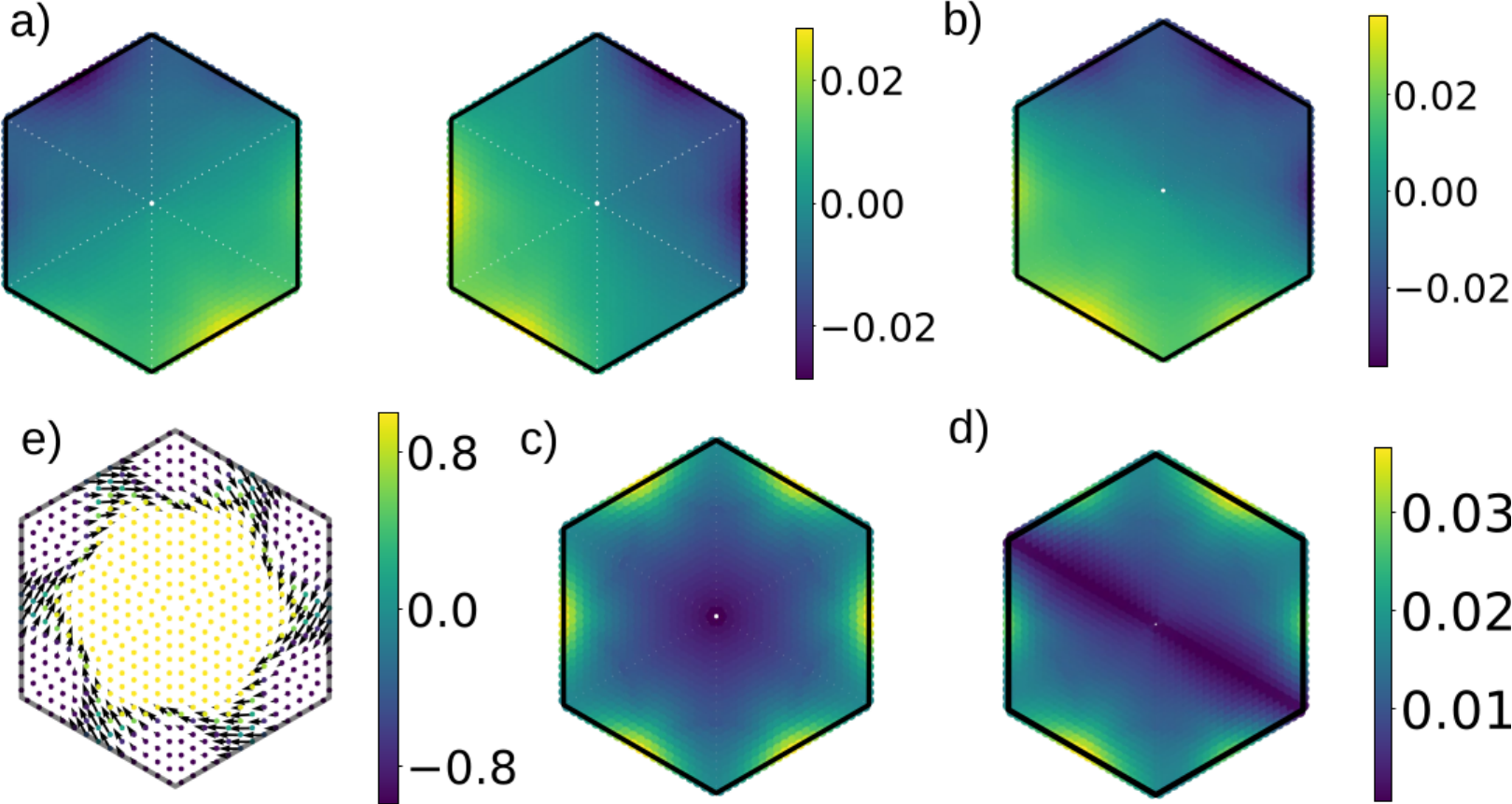}
    \vspace{0.05cm}
    \caption{Panels a) and b) show the real and imaginary part of $d^z(\bk)$ in the chiral and nematic ($d^z$ purely real) superconductors, respectively, in the mini Brillouin zone. Panels c) and d) display $|d^z(\bk)|$. Finally, 
    The states are saddle point of the energy $E_{\Psi}$~\eqref{variational_energy_app} and are obtained employing $\epsilon=20$, $\theta=5^\circ$, $k_BT=0.5$meV and $u_D=0$. }
    \label{fig:figure_variational_zero_D}
\end{figure}
\begin{figure}
    \centering
    \includegraphics[width=0.7\linewidth]{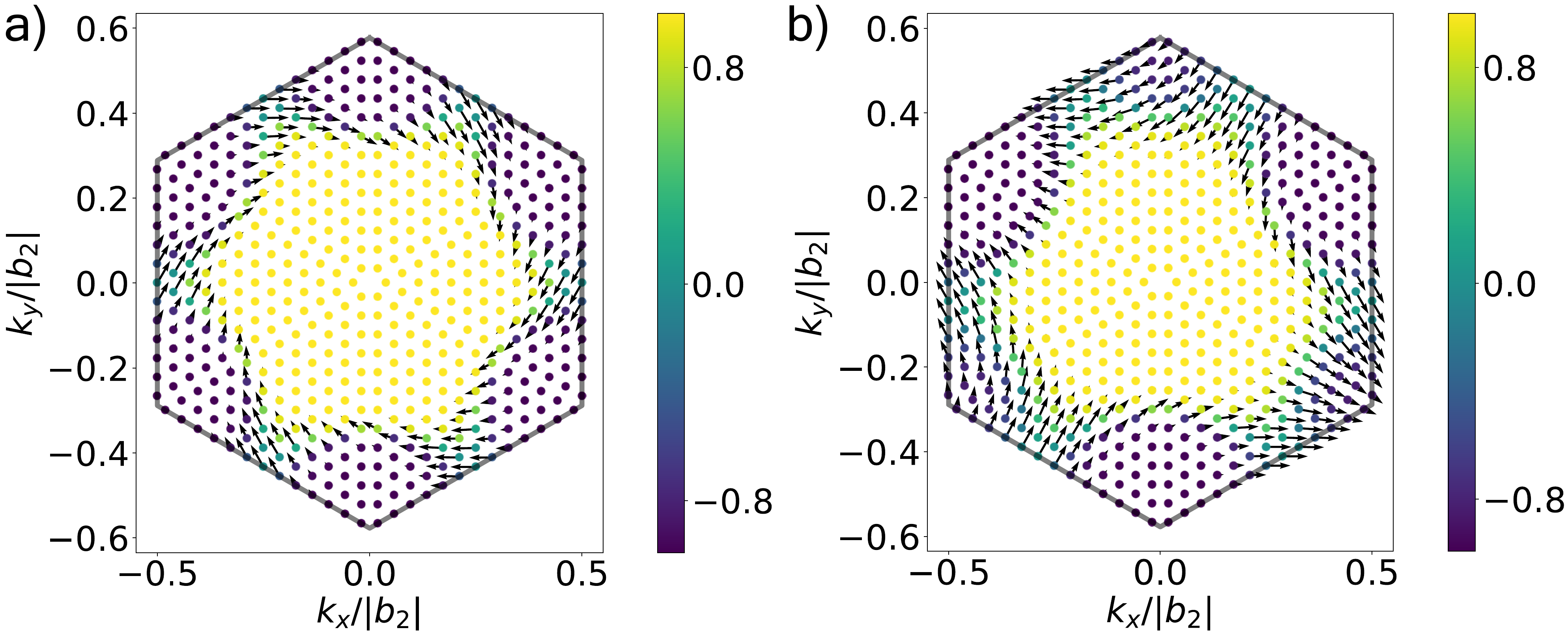}
        \caption{Pseudospin $\bm n(\bk)$ ($|\bm n|=1$) of the BdG Hamiltonian in the mini Brillouin zone for the chiral state for $u_D=0$ (panel a) and $u_D=15$meV (panel b). The color code shows the out of plane component $n_z(\bk)$. }
    \label{fig:nk_skyrmion}
\end{figure}
\begin{figure}
    \centering
    \includegraphics[width=0.7\linewidth]{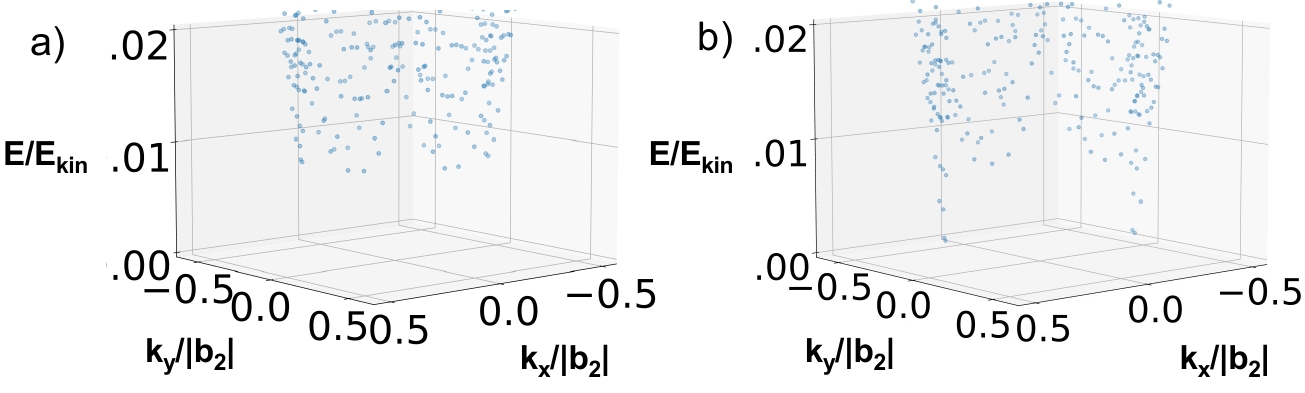}
        \caption{Positive energy BdG spectrum for the chiral (a) and nematic (b) states. The chiral state exhibits a full gap, while the nematic state features low-energy Dirac nodes along the high-symmetry line connecting $\kappa$ with $-\kappa$.}
    \label{fig:ek_BdG}
\end{figure}

To determine the ordering in the two-fold degenerate manifold we perform a variational calculation finding the best BCS wavefunction $\ket{\Psi}$ optimizing the energy of the Hamiltonian $\hat H_{0}+\hat H_{\rm eff}$. 
The energy of the mean-field ansatz is given by: 
\begin{equation}\label{variational_energy_app}
    % E_\Psi=\sum_{\bk}\sum_{\sigma}\left[\epsilon_{\bk\sigma}+\delta\epsilon_{\bk\sigma}\right] \langle c^\dagger_{\bk\sigma}c_{\bk\sigma}\rangle+\frac{1}{2}\sum_{\bk}\sum_{\sigma\sigma'}\langle c^\dagger_{\bk\sigma}\Delta_{\sigma\sigma'}(\bk)c^\dagger_{-\bk\sigma'}\rangle,
    E_\Psi=\sum_{\bk}\sum_{\sigma}\left[\epsilon_{\bk\sigma}+\delta\epsilon_{\bk\sigma}\right] \langle c^\dagger_{\bk\sigma}c_{\bk\sigma}\rangle+\frac{1}{2}\sum_{\bk}\sum_{\sigma\sigma'}\langle c^\dagger_{\bk\sigma}\Delta_{\sigma\sigma'}(\bk)c^\dagger_{-\bk\sigma'}\rangle,
\end{equation}
where the second contribution takes into account the condensate energy. 
% and the variation $\delta\epsilon$ describes the Hartree corrections to the quasiparticle dispersion: 
% \begin{equation}
%     \delta\epsilon_{\bk\sigma}=\frac{1}{A}\sum_{\sigma'}V_{\sigma\sigma'}(\bk,\bk)\langle c^\dagger_{-\bk\sigma'}c_{-\bk\sigma'}\rangle.
% \end{equation}
The Bogoliubov de Gennes (BdG) Hamiltonian of the $S^z=0$ superconductor in the basis $\Psi_{\bk}=[c_{\bk\uparrow},c_{\bk\downarrow},c^\dagger_{-\bk\uparrow},c^\dagger_{-\bk\downarrow}]^T$ reads: 
\begin{equation}\label{BdG_Hamiltonian}
    \hat{\mathcal H}_{\rm BdG}=\frac{1}{2}\sum_{\bk}\Psi^\dagger_{\bk}\begin{pmatrix}
        \hat \xi_{\bk} & \hat \Delta(\bk) \\ 
        \hat \Delta^\dagger(\bk) & -\hat \xi_{-\bk}
    \end{pmatrix}\Psi_{\bk},
\end{equation}
where $\hat \xi_{\bk}={\rm diag}[\xi_{\bk\uparrow},\xi_{\bk\downarrow}]$ and the self-consistent pairing gap introduced in the Hamiltonian~\eqref{BdG_Hamiltonian} which gives: 
\begin{equation}\label{self_consistent_gap}
    \delta\epsilon_{\bk\sigma}=\frac{1}{A}\sum_{\sigma'}V_{\sigma\sigma'}(\bk,\bk)\langle c^\dagger_{-\bk\sigma'} c_{-\bk\sigma'}\rangle,\quad \Delta_{\sigma\sigma'}(\bk)=-\frac{1}{A}\sum_{\bk'}V_{\sigma\sigma'}(\bk,\bk')\langle c_{\bk'\sigma}c_{-\bk'\sigma'}\rangle.
\end{equation}
We initialize the solution from two different ansatz: the first corresponding the nematic state $\bm\eta=|\bm \eta|(\cos\varphi,\sin\varphi)$ obtained imposing $\bm\eta\in\mathbb R$. The second describes a chiral state where we allow the coefficients $\bm\eta$ to develop a complex amplitude. Fig.~\ref{fig:figure_variational_zero_D} shows the order parameter developed at self-consistency in the two different cases. One clearly see that the chiral state is characterized by a full gap in the spectrum while the nematic state has a line of nodes. The nematic phase has been obtained by enforcing that $\eta_1,\eta_2$ are both real, the self-consistent solution selects a linear combination of the two components with $\varphi=\arctan\eta_2/\eta_1=\pi n/3$ with $n\in\mathbb Z$. The chiral superconductor is characterized by the particle-hole symmetry $\mathcal C=i\tau^y\sigma^y\mathcal K$, $\mathcal C\mathcal H_{\rm BdG}(\bk)\mathcal C^{-1}=-H_{\rm BdG}(-\bk)$,  with $\tau^x$ in the particle-hole basis consistent with a D superconductor of the Altland-Zirnbauer class~\cite{Altland_1997,Schnyder_2008}. This can be readily understood noticing that the BdG Hamiltonain decouples in two blocks $[c_{\bk\uparrow},c^\dagger_{-\bk\downarrow}]^T$ and $[c_{\bk\downarrow},c^\dagger_{-\bk\uparrow}]^T$. Within each subspace, the effectively spinless Hamiltonian reduces to 
\begin{eqnarray}
    \mathcal H_{\rm BdG}(\bk)=\sum_{s=\pm}\frac{\tau^0+s\tau^z}{2}\left[\xi_{\bk}\sigma^z+\Re d^z(\bk)\sigma^x-\Im d^z(\bk)\sigma^y\right]
\end{eqnarray}
leading to a pseudospin vector $\bm n(\bk)=[\Re d^z(\bk),-\Im d^z(\bk),\xi_{\bk}]/\sqrt{\xi^2_{\bk}+|d^z(\bk)|^2}$. The vector field $\bm n(\bk)$ is shown in Fig.~\ref{fig:nk_skyrmion}a) where the color code show the $z$ component of $\bm n(\bk)$ and displays a net winding with Chern number $C=1$. 
On the other hand, the nematic state is nodal and, therefore, characterized by a set of nodes along the $\gamma\kappa'$ line where the gap closes implying that it is not possible to define a versor $\bm n(\bk)$. Figs.~\ref{fig:ek_BdG}a) and b) shows the BdG spectrum of the chiral and nematic states at zero displacement field. 

At a finite displacement field, the pairing function $\Delta(\bk)$ does not have a well defined inversion symmetry as it is characterized by the mixing of spin $S^z=0$ (intervalley) triplet and singlet channels. The resulting BdG Hamiltonian reads: 
\begin{eqnarray}
    \mathcal H_{\rm BdG}(\bk) = \sum_{s=\pm}\frac{\tau^0+s\tau^z}{2}h_s(\bk),\quad h_\pm(\bk)=\xi_{\bk\uparrow/\downarrow}+\left[d^z(\bk)\mp\psi(\bk)\right]\sigma^ ++h.c..
\end{eqnarray}
The winding of the pseudospin vector $\bm n_+(\bk)$ is displayed in Fig.~\ref{fig:nk_skyrmion}b). To quantify the mixing between singlet and triplet we define the quantities: 
\begin{eqnarray}
    P_t=\frac{1}{N}\int_{\rm mBZ}\frac{d^2\bk}{\Omega_{\rm BZ}}|d^z(\bk)|^2,\quad P_s=\frac{1}{N}\int_{\rm mBZ}\frac{d^2\bk}{\Omega_{\rm BZ}}|\psi(\bk)|^2,\quad \mathcal N=\int_{\rm mBZ}\frac{d^2\bk}{\Omega_{\rm BZ}}\left[|d^z(\bk)|^2+|\psi(\bk)|^2\right].
\end{eqnarray}

\end{document}